\pdfoutput=1
\documentclass[11pt,twoside,a4paper,cmspaper,final,collab]{cms-tdr}
\def\svnVersion{3be35c3-D}\def\svnDate{2021/04/20}\def\cmsCernNoTag{CERN-EP-2021-016}\def\cmsCernDate{\today}\def\cmsMessage{"Published in the European Physical Journal C as \href{http://dx.doi.org/10.1140/epjc/s10052-021-09200-x}{\doi{10.1140/epjc/s10052-021-09200-x}.}"}
\begin{document}\cmsNoteHeader{HIG-19-001}

\newcommand{\ggH}{\ensuremath{\Pg\Pg\PH}}
\newcommand{\VBF}{\ensuremath{\mathrm{VBF}}}
\newcommand{\qqH}{\ensuremath{\PQq\PAQq\PH}}
\newcommand{\WH}{\ensuremath{\PW\PH}}
\newcommand{\ZH}{\ensuremath{\PZ\PH}}
\newcommand{\VH}{\ensuremath{\PV\PH}}
\newcommand{\VVV}{\ensuremath{\PV\PV\PV}}
\newcommand{\ttH}{\ensuremath{\ttbar\PH}}
\newcommand{\tH}{\ensuremath{\PQt\PH}}
\newcommand{\bbH}{\ensuremath{\bbbar\PH}}
\newcommand{\tHq}{\ensuremath{\PQt\PH\PQq}}
\newcommand{\tHW}{\ensuremath{\PQt\PH\PW}}
\newcommand{\ZZ}{\ensuremath{\PZ\PZ}}
\newcommand{\qqZZ}{\ensuremath{\PQq\PAQq\to\PZ\PZ}}
\newcommand{\ggZZ}{\ensuremath{\Pg\Pg\to\PZ\PZ}}
\newcommand{\ttZ}{\ensuremath{\ttbar\PZ}}
\newcommand{\ttZZ}{\ensuremath{\ttbar\PZ\PZ}}
\newcommand{\ttWW}{\ensuremath{\ttbar\PW\PW}}
\newcommand{\HZZfl}{\ensuremath{\PH\to\PZ\PZ\to4\ell}}
\newcommand{\HZZ}{\ensuremath{\PH\to\PZ\PZ}}
\newcommand{\Hllll}{\ensuremath{\PH\to4\ell}}
\newcommand{\mH}{\ensuremath{m_{\PH}}}
\newcommand{\mllll}{\ensuremath{m_{4\ell}}}
\newcommand{\KD}{\ensuremath{{\mathcal D}^{\text{kin}}_{\text{bkg}}} }
\newcommand{\DbkgVBFdec}{\ensuremath{{\mathcal{D}}^{\mathrm{VBF}+\text{dec}}_{\text{bkg}}}\xspace}
\newcommand{\DbkgVHdec}{\ensuremath{{\mathcal{D}}^{\VH+\text{dec}}_{\text{bkg}}}\xspace}
\newcommand{\DMeVbfjj}{\ensuremath{{\mathcal D}_{\text{2jet}}^{\VBF}}\xspace}
\newcommand{\DMeVbfj}{\ensuremath{{\mathcal D}_{\text{1jet}}^{\VBF}}\xspace}
\newcommand{\mll}{\ensuremath{m_{\ell\ell}}}
\newcommand{\mlplm}{\ensuremath{m_{\ell^{+}\ell^{-}}}}
\newcommand{\DMeWh}{\ensuremath{{\mathcal D}_{\text{2jet}}^{\WH}}\xspace}
\newcommand{\DMeZh}{\ensuremath{{\mathcal D}_{\text{2jet}}^{\ZH}}\xspace}
\newcommand{\DMeVh}{\ensuremath{{\mathcal D}_{\text{2jet}}^{\VH}}\xspace}
\newcommand{\Zee}{\ensuremath{\PZ\to\Pep\Pem}}
\newcommand{\Zll}{\ensuremath{\PZ\to\ell^{+}\ell^{-}}}
\newcommand{\muV}{\ensuremath{\mu_{\PV}}}
\newcommand{\muF}{\ensuremath{\mu_{\mathrm{f}}}}
\newcommand{\muVlong}{\ensuremath{\mu_{\mathrm{VBF},\PV\PH}}}
\newcommand{\muFlong}{\ensuremath{\mu_{\Pg\Pg\PH,\,\ttH,\bbH,\tH}} }
\newcommand{\usedLumiABC}{137\fbinv}
\newcommand{\valMuF}{\ensuremath{0.96^{+0.14}_{-0.12}}}
\newcommand{\valMuV}{\ensuremath{0.82^{+0.36}_{-0.31}}}
\newcommand{\sigmaObs}{\ensuremath{(\sigma\mathcal{B})_{{\text{obs}}}}}
\newcommand{\sigmaSM}{\ensuremath{(\sigma\mathcal{B})_{\mathrm{SM}}}}
\newcommand{\stxsggH}{\mathtt{ggH}}
\newcommand{\stxsqqH}{\mathtt{qqH}}
\newcommand{\stxsVHlep}{\mathtt{VH\text{-}lep}}
\newcommand{\stxsttH}{\mathtt{ttH}}
\newcommand{\stageoneggHbinone}{\mathtt{ggH\text{-}0j/p_T[0,10]}}
\newcommand{\stageoneggHbintwo}{\mathtt{ggH\text{-}0j/p_T[10,200]}}
\newcommand{\stageoneggHbinthree}{\mathtt{ggH\text{-}1j/p_T[0,60]}}
\newcommand{\stageoneggHbinfour}{\mathtt{ggH\text{-}1j/p_T[60,120]}}
\newcommand{\stageoneggHbinfive}{\mathtt{ggH\text{-}1j/p_T[120,200]}}
\newcommand{\stageoneggHbinsix}{\mathtt{ggH\text{-}2j/p_T[0,60]}}
\newcommand{\stageoneggHbinseven}{\mathtt{ggH\text{-}2j/p_T[60,120]}}
\newcommand{\stageoneggHbineight}{\mathtt{ggH\text{-}2j/p_T[120,200]}}
\newcommand{\stageoneggHbinnine}{\mathtt{ggH\text{-}2j/m_{jj}>350}}
\newcommand{\stageoneggHbinten}{\mathtt{ggH/p_T>200}}
\newcommand{\stageoneqqHbinone}{\mathtt{qqH\text{-}rest}}
\newcommand{\stageoneqqHbintwo}{\mathtt{qqH\text{-}2j/m_{jj}[60,120]}}
\newcommand{\stageoneqqHbinthree}{\mathtt{qqH\text{-}2j/m_{jj}[350,700]}}
\newcommand{\stageoneqqHbinfour}{\mathtt{qqH\text{-}2j/m_{jj}>700}}
\newcommand{\stageoneqqHbinfive}{\mathtt{qqH\text{-}3j/m_{jj}>350}}
\newcommand{\stageoneqqHbinsix}{\mathtt{qqH\text{-}2j/p_T>200}}
\newcommand{\stageoneVHbinone}{\mathtt{VH\text{-}lep/p_T^\PH[0,150]}}
\newcommand{\stageoneVHbintwo}{\mathtt{VH\text{-}lep/p_T^\PH>150}}

\providecommand{\cmsTable}[1]{\resizebox{\textwidth}{!}{#1}}
\newlength\cmsTabSkip\setlength{\cmsTabSkip}{1ex}

\ifthenelse{\boolean{cms@external}}{\providecommand{\cmsLeft}{upper\xspace}}{\providecommand{\cmsLeft}{left\xspace}}
\ifthenelse{\boolean{cms@external}}{\providecommand{\cmsRight}{lower\xspace}}{\providecommand{\cmsRight}{right\xspace}}
\ifthenelse{\boolean{cms@external}}{\providecommand{\cmsUpperLeft}{upper\xspace}}{\providecommand{\cmsUpperLeft}{upper left\xspace}}
\ifthenelse{\boolean{cms@external}}{\providecommand{\cmsUpperRight}{center\xspace}}{\providecommand{\cmsUpperRight}{upper right\xspace}}

\cmsNoteHeader{HIG-19-001}
\title{Measurements of production cross sections of the Higgs boson in the four-lepton final state in proton-proton collisions at \texorpdfstring{$\sqrt{s} = 13\TeV$}{sqrt(s) = 13 TeV}}
\titlerunning{Measurements of production cross sections of the Higgs boson in the four-lepton final state\ldots}
\date{\today}

\abstract{
   Production cross sections of the Higgs boson are measured in the $\PH\to\PZ\PZ\to4\ell$ ($\ell=\Pe,\Pgm$) decay channel.    A data sample of proton-proton collisions at a center-of-mass energy of 13\TeV, collected by the CMS detector at the LHC and corresponding to an integrated luminosity of 137\fbinv is used.    The signal strength modifier $\mu$, defined as the ratio of the Higgs boson production rate in the $4\ell$ channel to the standard model (SM) expectation, is measured to be $\mu=0.94 \pm 0.07 \stat ^{+0.09}_{-0.08} \syst $ at a fixed value of $m_{\PH} = 125.38\GeV$.    The signal strength modifiers for the individual Higgs boson production modes are also reported.    The inclusive fiducial cross section for the $\PH\to4\ell$ process is measured to be $2.84^{+0.23}_{-0.22} \stat ^{+0.26}_{-0.21} \syst \unit{fb}$,    which is compatible with the SM prediction of $2.84 \pm 0.15 \unit{fb}$ for the same fiducial region.    Differential cross sections as a function of the transverse momentum and rapidity of the Higgs boson, the number of associated jets, and the transverse momentum of the leading associated jet are measured.    A new set of cross section measurements in mutually exclusive categories targeted to identify production mechanisms and kinematical features of the events is presented.    The results are in agreement with the SM predictions.
}

\hypersetup{%
pdfauthor={CMS Collaboration},%
pdftitle={Measurements of production cross sections of the Higgs boson in the four-lepton final state in proton-proton collisions at sqrt(s)=13 TeV},%
pdfsubject={CMS},%
pdfkeywords={CMS, Higgs}}

\maketitle

\section{Introduction}
\label{sec:intro}

The discovery of the Higgs boson (\PH) in 2012 by the ATLAS and CMS collaborations~\cite{Aad:2012tfa,Chatrchyan:2012ufa,Chatrchyan:2013lba} has been a major step towards the understanding of the electroweak symmetry breaking mechanism~\cite{Englert:1964et,Higgs:1964ia,Higgs:1964pj,Guralnik:1964eu,Higgs:1966ev,Kibble:1967sv}.
Further studies by the two experiments~\cite{ATLASPropertiesRun1,CMSPropertiesRun1,ATLASCMSMassRun1,ATLASCMSPropertiesRun1} have shown that the properties of the new particle are consistent with the standard model (SM) expectations for the \PH boson.

The $\HZZfl$ decay channel ($\ell=\Pe,\Pgm$) has a large signal-to-background ratio thanks to a low background rate and the complete reconstruction of the final state decay products,
capitalizing on the excellent lepton momentum resolution of the CMS detector.
The measurements performed using this decay channel with the LHC Run 1 data set at center-of-mass energies of 7 and 8\TeV, and the Run 2 data set at 13\TeV include the determination of
the mass, the spin and the parity of the \PH boson~\cite{ATLASH4lLegacyRun1,CMSH4lLegacyRun1,CMSH4lSpinParity,CMSH4lAnomalousCouplings,CMSH4l2016,ATLASH4l2016},
its width~\cite{CMSH4lWidth,CMSH4lLifetime,ATLASH4lWidth,ATLASH4lWidth2016}, the inclusive and differential fiducial cross sections~\cite{ATLASH4lFiducial8TeV,CMSH4lFiducial8TeV, CMSH4l2016,ATLASH4lFiducial2016, ATLASH4lLegacyRun2, ATLASH4lFiducialRun2}, and the tensor structure of the \PH boson interaction with a pair of neutral gauge bosons in both on-shell and off-shell regions~\cite{CMSH4lAnomalousCouplings,CMSH4lLifetime, CMSH4lAnomalousCouplings2016,ATLASH4l2016,CMSHVVAnomalousCouplings2016}.

This paper presents the measurement of production cross sections in granular kinematic regions of the \PH boson in the $\HZZfl$ decay channel.
A data sample of proton-proton ($\Pp\Pp$) collisions at a center-of-mass energy of $\sqrt{s}=13\TeV$, collected by the CMS detector at the LHC and corresponding to an integrated luminosity of $\usedLumiABC$ is used.
The inclusive signal strength modifier, defined as the ratio of the \PH boson production rate in the $4\ell$ channel to the SM expectation,
and signal strength modifiers for the individual \PH boson production modes are measured. The measurements of the inclusive and differential fiducial cross sections are also presented,
and their compatibility with the SM predictions is tested.
The present analysis is similar to that previously performed by the CMS Collaboration~\cite{CMSH4l2016}, but is based on a larger data sample.

In addition, measurements of the \PH boson cross sections within the simplified template cross section (STXS) framework~\cite{Bendavid:2018nar,deFlorian:2016spz,Berger:2019wnu} are also presented.
The main goals of the STXS framework are to increase the reinterpretability of the precision \PH boson measurements and to minimize the theory dependence.
This is achieved by defining exclusive kinematic regions in the \PH boson production phase space.
The results presented within the STXS framework nonetheless depend on the SM simulation used to model the experimental acceptance of the signal processes, which could be modified in beyond the SM (BSM) scenarios.
These kinematic regions, referred to as bins, are defined in different stages corresponding to increasing degrees of granularity.
This paper presents results in the STXS stage 0 where the bins correspond closely to the different \PH boson production mechanisms.
Previous measurements of cross sections in stage 0 production bins in the $\Hllll$ decay channel were already presented by the CMS Collaboration~\cite{CMSH4l2016}.
In the STXS framework, additional stages are defined by further splitting of the bins enhancing the sensitivity to possible signature of BSM physics at high transverse momentum of the \PH boson.
Measurements of stage 0, stage 1, and stage 1.1 cross sections in the $\Hllll$ decay channel were recently published by the ATLAS Collaboration~\cite{ATLASH4lLegacyRun2}.
The most recent refinement of STXS binning is referred to as STXS stage 1.2.
This paper presents a first set of the cross section measurements in the STXS stage 1.2 bins in the $\Hllll$ decay channel.

The paper is organized as follows.
A brief introduction of the CMS detector is given in Section~\ref{sec:detector}.
The data, as well as the simulated signal and background samples, are described in Section~\ref{sec:datasets}.
The event reconstruction and selection, the kinematic discriminants, and the categorization of the \PH boson candidate events are described in Sections~\ref{sec:objects},~\ref{sec:observables}, and~\ref{sec:categorization}, respectively.
The background estimation is detailed in Section~\ref{sec:bkgd} while the signal modeling is described in Section~\ref{sec:signal}.
The experimental and theoretical systematic uncertainties are described in Section~\ref{sec:systematics} and the results are presented in Section~\ref{sec:results}.
Concluding remarks are given in Section~\ref{sec:summary}.

\section{The CMS detector}
\label{sec:detector}

The central feature of the CMS apparatus is a superconducting solenoid of 6\unit{m} internal diameter,
providing a magnetic field of 3.8\unit{T}. Within the solenoid volume are a silicon pixel and strip tracker,
a lead tungstate crystal electromagnetic calorimeter (ECAL), and a brass and scintillator hadron calorimeter (HCAL),
each composed of a barrel and two endcap sections.
Forward calorimeters extend the pseudorapidity $\eta$ coverage provided by the barrel and endcap detectors.
Muons are detected in gas-ionization chambers embedded in the steel flux-return yoke outside the solenoid.

Events of interest are selected using a two-tiered trigger system.
The first level, composed of custom hardware processors,
uses information from the calorimeters and muon detectors to select events at a rate of around 100\unit{kHz} within a fixed latency of about 4\unit{\mus}\cite{Sirunyan:2020zal}.
The second level, known as the high-level trigger,
consists of a farm of processors running a version of the full event reconstruction software optimized for fast processing,
and reduces the event rate to around 1\unit{kHz} before data storage~\cite{Khachatryan:2016bia}.

{\tolerance=800 The candidate vertex with the largest value of summed physics-object squared transverse momentum $\pt^2$ is taken to be the primary $\Pp\Pp$ interaction vertex (PV).
The physics objects are the jets, clustered using the jet finding algorithm~\cite{Cacciari:2008gp,Cacciari:2011ma} with the tracks assigned to candidate vertices as inputs,
and the associated missing transverse momentum, taken as the negative vector sum of the \pt of those jets.\par}

The electron momentum is estimated by combining the energy measurement in the ECAL with the momentum measurement in the tracker.
The momentum resolution for electrons with $\pt \approx 45\GeV$ from $\PZ\to\Pe\Pe$ decays ranges from 1.7\% to 4.5\%.
It is generally better in the barrel region than in the endcaps,
and also depends on the bremsstrahlung energy emitted by the electron as it traverses the material in front of the ECAL~\cite{EGM-17-001}.
The ECAL consists of 75\,848 lead tungstate crystals,
which provide coverage of $\abs{\eta} < 1.48 $ in the barrel region and $1.48 < \abs{\eta} < 3.00$ in the two endcap regions (EE).
Preshower detectors consisting of two planes of silicon sensors interleaved with a total of $3 X_0$ of lead are located in front of each EE detector.

Muons are measured in the pseudorapidity range $\abs{\eta} < 2.4$, with detection planes made using three technologies: drift tubes,
cathode strip chambers, and resistive plate chambers.
The single muon trigger efficiency exceeds 90\% over the full $\eta$ range,
and the efficiency to reconstruct and identify muons is greater than 96\%.
Matching muons to tracks measured in the silicon tracker results in a relative transverse momentum resolution,
for muons with \pt up to 100\GeV, of 1\% in the barrel and 3\% in the endcaps.
The \pt resolution in the barrel is better than 7\% for muons with \pt up to 1\TeV~\cite{Sirunyan:2018}.

A more detailed description of the CMS detector, together with a definition of the coordinate system used and the relevant kinematic variables, can be found in Ref.~\cite{Chatrchyan:2008zzk}.

\section{Data and simulated samples}
\label{sec:datasets}

This analysis is based on the $\Pp\Pp$ collision data collected by the CMS detector at the LHC in 2016, 2017, and 2018
with integrated luminosities of 35.9, 41.5, and 59.7\fbinv, respectively~\cite{CMS-PAS-LUM-17-001,CMS-PAS-LUM-17-004,CMS-PAS-LUM-18-002}.
The collision events are selected by high-level trigger algorithms that require the presence of leptons passing loose identification and isolation requirements.
The main triggers select either a pair of electrons or muons, or an electron and a muon.
The minimal transverse momentum of the leading and subleading leptons changed throughout the years to account for the different data-taking conditions and is summarized in Table~\ref{tab:HLT}.

\begin{table}[htb]
	\centering
		\topcaption{
        The minimal \pt of the leading/subleading leptons for the main di-electron (\Pe/\Pe), di-muon (\Pgm/\Pgm), and electron-muon (\Pe/\Pgm, \Pgm/\Pe) high-level trigger algorithms used in the $\Hllll$ analysis in 2016, 2017, and 2018.
        \label{tab:HLT}
    }
    \begin{tabular}{cccc}
   & \Pe/\Pe ({\GeVns}) & \Pgm/\Pgm ({\GeVns})& \Pe/\Pgm, \Pgm/\Pe ({\GeVns})\\
   \hline
   2016 & 17/12 & 17/8 & 17/8, 8/23 \\
   2017 & 23/12 & 17/8 & 23/8, 12/23 \\
   2018 & 23/12 & 17/8 & 23/8, 12/23
\end{tabular}
\end{table}

To maximize the coverage of the $\Hllll$ phase space, triggers requiring three leptons with relaxed transverse momenta thresholds and no isolation requirement are also used, as are isolated single-electron and single-muon triggers.
The overall trigger efficiency for simulated signal events that pass the full event selection (described in Section~\ref{sec:objects}) is larger than 99\%.
The trigger efficiency is derived from data using a sample of $4\ell$ events collected by the single-lepton triggers and a method based on the ``tag and probe'' technique.
One of the four leptons is matched to a candidate reconstructed by the single-lepton trigger and the remaining three leptons in the event are used as probes.
The probe leptons are combined in an attempt to reconstruct any of the triggers used in the analysis.
The efficiency in data is found to be in agreement with the expectation from the simulation.

{\tolerance=800 Monte Carlo (MC) simulation samples for the signals and the relevant background processes are used to evaluate the signal shape, estimate backgrounds, optimize the event selection, and evaluate the acceptance and systematic uncertainties.
The SM \PH boson signals are simulated at next-to-leading order (NLO) in perturbative QCD (pQCD) with
the \POWHEG~2.0~\cite{Nason:2004rx,Frixione:2007vw,Alioli:2010xd} generator for the five main production processes:
gluon fusion ($\ggH$)~\cite{Alioli:2008tz},
vector boson fusion ($\VBF$)~\cite{Nason:2009ai},
associated production with a vector boson ($\VH$, where \PV is a \PW or a \PZ boson)~\cite{Luisoni2013},
and associated production with a pair of top quarks ($\ttH$)~\cite{Hartanto:2015uka}.
The $\ZH$ production occurs in two ways, $\PQq\PAQq\to\ZH$ and a much smaller contribution from $\Pg\Pg\to\ZH$, which is simulated at leading order (LO) using \textsc{jhugen}~7.3.0~\cite{Gao:2010qx, Bolognesi:2012mm,Anderson:2013afp,Gritsan:2016hjl,Gritsan:2020pib}.
In addition to the five main production processes, the contributions due to \PH boson production in association with a single top quark ($\tH$) and either a quark ($\tHq$) or a $\PW$ boson ($\tHW$)
are simulated at LO using \textsc{jhugen}~7.0.2 and \MGvATNLO~2.2.2~\cite{amcatnlo}, respectively.
The associated production with a pair of bottom quarks ($\bbH$) is simulated at LO with \textsc{jhugen}~7.0.2.
In all cases, the decay of the \PH boson to four leptons is modeled with \textsc{jhugen}~7.0.2.
The theoretical predictions used for the various production and decay modes can be found in Refs.~\cite{Anastasiou:2015ema,Anastasiou2016,Ciccolini:2007jr,Ciccolini:2007ec,Bolzoni:2010xr,Bolzoni:2011cu,Brein:2003wg,Ciccolini:2003jy,Beenakker:2001rj,Beenakker:2002nc,Dawson:2002tg,Dawson:2003zu,Yu:2014cka,Frixione:2014qaa,Demartin:2015uha,Demartin:2016axk,Denner:2011mq,Djouadi:1997yw,hdecay2,Bredenstein:2006rh,Bredenstein:2006ha,Boselli:2015aha,Actis:2008ts} and are summarized in Ref.~\cite{deFlorian:2016spz}.\par}

{\tolerance=5000 The $\PZ\PZ$ background contribution from quark-antiquark annihilation is simulated at NLO pQCD with \POWHEG~2.0~\cite{Melia:2011tj}, while the $\ggZZ$ process is generated at LO with \MCFM~7.0.1~\cite{MCFM}.
The $\PW\PZ$ background and the triboson backgrounds $\PZ\PZ\PZ$, $\PW\PZ\PZ$, and $\PW\PW\PZ$ are modeled at NLO using \MGvATNLO~2.4.2.
The smaller $\ttZ$, $\ttWW$, and $\ttZZ$ background processes are simulated at LO with \MGvATNLO~2.4.2.
The events containing $\PZ$ bosons with associated jets ($\PZ+$jets) are simulated at NLO with \MGvATNLO~2.4.2 and the $\ttbar$ background is simulated at NNLO with \POWHEG~2.0.
The reducible background determination does not rely on the MC but is based on data, as described in Section~\ref{sec:redbkgd}.\par}

All signal and background event generators are interfaced with \PYTHIA 8.230~\cite{Sjostrand:2014zea} using the CUETP8M1 tune~\cite{Khachatryan:2015pea} for the 2016 data-taking period and the CP5 tune~\cite{Sirunyan:2019dfx} for the 2017 and 2018 data-taking periods,
to simulate the multi-parton interaction and hadronization effects. The NNPDF3.0 set of parton distribution functions (PDFs) is used~\cite{Ball:2014uwa}.
The generated events are processed through a detailed simulation of the CMS detector based on \GEANTfour~\cite{Agostinelli:2002hh,GEANT} and are reconstructed with the same algorithms that are used for data.
The simulated events include overlapping $\Pp\Pp$ interactions (pileup) and have been reweighted so that the distribution of the number of interactions per LHC bunch crossing in simulation matches that observed in data.

\section{Event reconstruction and selection}
\label{sec:objects}

The particle-flow (PF) algorithm~\cite{CMS-PRF-14-001} aims to reconstruct and identify each individual particle (PF candidate) in an event,
with an optimized combination of information from the various elements of the CMS detector.
The energy of photons is obtained from the ECAL measurement.
The energy of electrons is determined from a combination of the electron momentum at the PV as determined by the tracker,
the energy of the corresponding ECAL cluster, and the energy sum of all bremsstrahlung photons spatially compatible with originating from the electron track.
The energy of muons is obtained from the curvature of the corresponding track.
The energy of charged hadrons is determined from a combination of their momentum measured in the tracker and the matching ECAL and HCAL energy deposits,
corrected for the response function of the calorimeters to hadronic showers.
Finally, the energy of neutral hadrons is obtained from the corresponding ECAL and HCAL energies.

The missing transverse momentum vector \ptvecmiss is computed as the negative vector sum of the transverse momenta of all the PF candidates in an event,
and its magnitude is denoted as \ptmiss~\cite{Sirunyan:2019kia}.
The \ptvecmiss is modified to account for corrections to the energy scale of the reconstructed jets in the event.

Muons with $\pt^{\Pgm} > 5\GeV$ are reconstructed within the geometrical acceptance, corresponding to the region $\abs{\eta^{\Pgm}} < 2.4$, by combining information from the silicon tracker and the muon system~\cite{Sirunyan:2018}.
The matching between the inner and outer tracks proceeds either outside-in, starting from a track in the muon system, or inside-out, starting from a track in the silicon tracker.
Inner tracks that match segments in only one or two stations of the muon system are also considered because they may belong to very low-\pt muons that do not have sufficient energy to penetrate the entire muon system.
The muons are selected among the reconstructed muon track candidates by applying minimal requirements on the track in both the muon system and the inner tracker system, and taking into account the compatibility with small energy deposits in the calorimeters.

To discriminate between prompt muons from \PZ boson decay and those arising from electroweak (EW) decays of hadrons within jets,
an isolation requirement of ${\mathcal I}^{\Pgm}<0.35$ is imposed, where the relative isolation is defined as
\begin{linenomath}
\ifthenelse{\boolean{cms@external}}
{
\begin{multline}
\label{eqn:pfiso}
{\mathcal I}^{\Pgm} \equiv \Big( \sum \pt^\text{charged} + \max\big[ 0, \sum \pt^\text{neutral} +\\
\sum \pt^{\Pgg} - \pt^{\Pgm,\mathrm{PU}} \big] \Big) / \pt^{\Pgm} .
\end{multline}
}
{
\begin{equation}
\label{eqn:pfiso}
{\mathcal I}^{\Pgm} \equiv \Big( \sum \pt^\text{charged} + \max\big[ 0, \sum \pt^\text{neutral} +
\sum \pt^{\Pgg} - \pt^{\Pgm,\mathrm{PU}} \big] \Big) / \pt^{\Pgm} .
\end{equation}
}
\end{linenomath}
{\tolerance=800 In Eq.~(\ref{eqn:pfiso}), $\sum \pt^\text{charged}$ is the scalar sum of the
transverse momenta of charged hadrons originating from
the chosen PV of the event.
The quantities $\sum \pt^\text{neutral}$ and $\sum \pt^{\Pgg}$ are the
scalar sums of the transverse momenta for neutral hadrons and photons, respectively.
The isolation sums involved are all restricted to a volume bound by a
cone of angular radius  $\Delta R=0.3$ around the muon direction at the
PV, where the angular  distance between two particles
$i$ and $j$ is $\Delta R(i,j) = \sqrt{\smash[b]{(\eta^i-\eta^j)^{2} + (\phi^i-\phi^j)^{2}}}$.
Since the isolation variable is particularly sensitive to energy deposits from pileup interactions,
a $\pt^{\Pgm,\mathrm{PU}}$ contribution is subtracted, defined as $\pt^{\Pgm,\mathrm{PU}}\equiv0.5\sum_{i}\pt^\text{\textit{i},PU}$, where $i$ runs over the charged hadron PF candidates not originating from
the PV, and the factor of 0.5 corrects for the different fraction of charged and neutral particles in the cone~\cite{PUmitigationCMS}.\par}

{\tolerance=1200 Electrons with $\pt^{\Pe} > 7\GeV$ are reconstructed within the
geometrical acceptance, corresponding to the pseudorapidity region $\abs{\eta^{\Pe}} < 2.5$~\cite{EGM-17-001}.
Electrons are identified using a multivariate discriminant which includes observables sensitive to
the presence of bremsstrahlung along the electron trajectory, the geometrical and momentum-energy matching between the
electron trajectory and the associated cluster in the ECAL, the shape of the electromagnetic shower in the ECAL,
and variables that discriminate against electrons originating from photon conversions.
Instead of an additional isolation requirement, similar to the one for muons, the electron multivariate discriminant also includes the isolation sums described above ($\sum \pt^{\text{charged}}$, $\sum \pt^{\text{neutral}}$, and $\sum \pt^{\Pgg}$) but computed around the electron direction.
The inclusion of isolation sums helps suppressing electrons originating from electroweak decays of hadrons within jets~\cite{DPS-2018} and has a better performance than a simple requirement on the relative isolation observable.
The package \textsc{xgboost}~\cite{Chen:2016btl} is used for the training and optimization of the multivariate discriminant employed for electron identification and isolation.
The training is performed with simulated events that are not used at any other stage of the analysis.
Events are divided into six regions defined by two transverse momentum ranges (7--10$\GeV$ and $>$10$\GeV$) and three pseudorapidity regions:
central barrel ($\abs{\eta^{\Pe}}<0.8$), outer barrel ($0.8<\abs{\eta^{\Pe}}<1.479$), and endcaps ($1.479<\abs{\eta^{\Pe}}<2.5$).
Separate training is performed for the three different data-taking periods and selection requirements are determined such that the signal efficiency remains the same for all three periods.\par}

The effect of the final-state radiation (FSR) from leptons is recovered as follows.
Bremsstrahlung photons already associated to electrons in the reconstruction step are not considered in this procedure.
Photons reconstructed by the PF algorithm within $\abs{\eta^\Pgg}<2.4$ are considered as FSR candidates if they satisfy the conditions $\pt^{\Pgg} > 2 \GeV$ and ${\mathcal I}^{\Pgg} < 1.8$,
where the photon relative isolation ${\mathcal I}^{\Pgg}$ is defined as for the muon in Eq.~(\ref{eqn:pfiso}).
Every such photon is associated to the closest selected lepton in the event.
Photons that do not satisfy the requirements $\Delta R(\Pgg,\ell)/(\pt^{\Pgg})^2<0.012 \GeV^{-2}$ and $\Delta R(\Pgg,\ell) < 0.5$ are discarded.
The lowest-$\Delta R(\Pgg,\ell)/(\pt^{\Pgg})^2$ photon candidate of every lepton, if any, is retained.
The photons thus identified are excluded from the isolation computation of the muons selected in the event.

In order to suppress muons from in-flight decays of hadrons and electrons from photon conversions,
leptons are rejected if the ratio of their impact parameter in three dimensions, computed with respect to the PV position,
to their uncertainty is greater or equal to four.

The momentum scale and resolution of electrons and muons are calibrated in bins of $\pt^{\ell}$ and $\eta^{\ell}$
using the decay products of known dilepton resonances as described in Refs.~\cite{EGM-17-001,Sirunyan:2018}.

A ``tag and probe'' technique~\cite{CMS:2011aa} based on samples of $\PZ$ boson events in data and simulation
is used to measure the efficiency of the reconstruction and selection
for prompt electrons and muons in several bins of $\pt^{\ell}$ and $\eta^{\ell}$.
The difference in the efficiencies measured in simulation and data is used to rescale the yields of selected events in the simulated samples.

For each event, hadronic jets are clustered from the reconstructed particles using the infrared- and collinear-safe anti-\kt algorithm~\cite{Cacciari:2008gp, Cacciari:2011ma} with a distance parameter of 0.4.
The jet momentum is determined as the vectorial sum of all particle momenta in the jet, and is found from simulation to be within 5 to 10\% of the true momentum over the whole \pt spectrum and detector acceptance.
Additional $\Pp\Pp$ interactions within the same or nearby bunch crossings can contribute extra tracks and calorimetric energy depositions to the jet.
To mitigate this effect, tracks identified as originating from pileup vertices are discarded and an offset correction is applied to correct for the remaining contributions.
Jet energy corrections are derived from simulation to match that of particle level jets on average.
In situ measurements of the momentum balance in dijet, photon + jet, $\PZ +$~jet, and multijet events are used to account for any residual differences in jet energy scale in data and simulation~\cite{Khachatryan:2016kdb}.
Jet energies in simulation are smeared to match the resolution in data.
The jet energy resolution amounts typically to 16\% at 30\GeV, 8\% at 100\GeV, and 4\% at 1\TeV.
Additional selection criteria are applied to remove jets potentially dominated by anomalous contributions from various subdetector components or reconstruction failures.
To be considered in the analysis, jets must satisfy the conditions $\pt^{\text{jet}}>30\GeV$
and $\abs{\eta^{\text{jet}}} < 4.7$, and be separated from all selected lepton candidates and any selected FSR photons by $\Delta R(\ell/\cPgg,\text{jet})>0.4$.
Jets are also required to pass the tight identification criteria and the tight working point of pileup jet identification described in Ref.~\cite{PUmitigationCMS}.

For event categorization, jets are tagged as {\PQb} jets using the DeepCSV algorithm~\cite{Sirunyan:2017ezt}, which combines information about impact parameter significance, secondary vertex, and jet kinematics.
Data to simulation scale factors for the {\PQb} tagging efficiency are applied as a function of jet \pt, $\eta$, and flavor.

The event selection is designed to extract signal candidates from events containing at least four well-identified and isolated leptons, each originating from the PV and possibly accompanied by an FSR photon candidate.
In what follows, unless otherwise stated, FSR photons are included in invariant mass computations.

First, $\PZ$ candidates are formed with pairs of leptons of the same flavor and opposite-charge ($\Pep\Pem$, $\PGmp\PGmm$) that pass the requirement $12 < \mlplm  < 120\GeV$.
They are then combined into $\PZ\PZ$ candidates, wherein we denote as $\PZ_1$ the $\PZ$ candidate with an invariant mass closest to the nominal $\PZ$ boson mass~\cite{Zyla:2020zbs}, and as $\PZ_2$ the other one.
The flavors of the involved leptons define three mutually exclusive subchannels: $4\Pe$, $4\Pgm$, and $2\Pe 2\Pgm$.

To be considered for the analysis, $\PZ\PZ$ candidates have to pass a set of kinematic requirements that improve the sensitivity to \PH boson decays.
The $\PZ_1$ invariant mass must be larger than 40\GeV.
All leptons must be separated in angular space by at least $\Delta R(\ell_i, \ell_j) > 0.02$.
At least two leptons are required to have $\pt > 10\GeV$ and at least one is required to have $\pt > 20\GeV$.
In the $4\Pgm$ and $4\Pe$ subchannels, where an alternative $\PZ_a \PZ_b$ candidate can be built out of the same four leptons, we discard candidates with $m_{\PZ_b}<12\GeV$ if $\PZ_a$ is closer to the nominal $\PZ$ boson mass than $\PZ_1$ is. This rejects events that contain an on-shell $\PZ$ and a low-mass dilepton resonance.
To further suppress events with leptons originating from hadron decays in jet fragmentation or from the decay of low-mass resonances, all four opposite-charge lepton pairs that can be built with the four leptons (irrespective of flavor) are required to satisfy the condition $m_{\ell^{+}\ell'^{-}} > 4\GeV$, where selected FSR photons are disregarded in the invariant mass computation.
Finally, the four-lepton invariant mass $\mllll$ must be larger than 70\GeV, which defines the mass range of interest for the subsequent steps of the analysis.

In events where more than one $\PZ\PZ$ candidate passes the above selection, the candidate with the highest value of $\KD$ (defined in Section~\ref{sec:observables}) is retained, except if two candidates consist of the same four leptons, in which case the candidate with  the $\PZ_1$ mass closest to the nominal $\PZ$ boson mass is retained.

\section{Kinematic discriminants}
\label{sec:observables}

The full kinematic information from each event using either the \PH boson decay products and/or the associated particles in the \PH boson production is extracted by means of matrix element calculations and is used to form several kinematic discriminants.
These computations rely on the \textsc{MELA} package~\cite{Gao:2010qx,Bolognesi:2012mm,Anderson:2013afp,Gritsan:2020pib} and exploit the \textsc{jhugen} matrix elements for the signal and the \MCFM matrix elements for the background.
Both the \PH boson decay kinematics and the kinematics of the associated production of $\PH+1~$jet, $\PH+2$~jets, $\VBF$, $\ZH$, and $\WH$ are explored.
The full event kinematics is described by decay observables $\vec\Omega^{\PH\to4\ell}$ or observables describing the associated production $\vec\Omega^{\PH+\text{jj}}$,
which may or may not include the $\PH\to4\ell$ decay kinematic information depending on the use case.
The definition of these observables can be found in Refs.~\cite{Gao:2010qx,Bolognesi:2012mm,Anderson:2013afp}.

Two types of kinematic discriminants are exploited in the $\Hllll$ analysis.
First we construct the three categorization discriminants in order to classify signal events into exclusive categories as defined in Section~\ref{subsec:Reco_Categories}.
Categorization discriminants are designed to increase the purity of the targeted production mechanism in a dedicated event category.
In addition, we define another set of three kinematic discriminants that are taken as an observable in the two-dimensional likelihood fits carried out to extract the results, as described in Section~\ref{sec:results}.
These kinematic discriminants are designed to separate the targeted \PH boson production mechanism from its dominant background.

Categorization discriminants are calculated following the prescription in Refs.~\cite{CMSH4l2016,CMSHighMassPaper,CMSH4lLifetime}.
The discriminants sensitive to the $\VBF$ signal topology with two associated jets,
the $\VBF$ signal topology with one associated jet, and the $\VH$ (either $\ZH$ or $\WH$) signal topology with two associated jets are:
\begin{linenomath}
\ifthenelse{\boolean{cms@external}}
{
\begin{equation}
\label{eqn:prodmela}
\begin{aligned}
\DMeVbfjj &=
\left[1+
\frac{ \mathcal{P}_{\PH\text{jj}} (\vec\Omega^{\PH+\text{jj}} | m_{4\ell}) }
{\mathcal{P}_{\VBF}  (\vec\Omega^{\PH+\text{jj}} | m_{4\ell})  }
\right]^{-1} \\
\DMeVbfj &=
\left[1+
\frac{ \mathcal{P}_{\PH\text{j}} (\vec\Omega^{\PH+\text{j}} | m_{4\ell}) }
{\int d\eta^{\text{j}}\mathcal{P}_{\VBF}  (\vec\Omega^{\PH+\text{jj}} | m_{4\ell}) }
\right]^{-1} \\
\DMeWh &=
\left[1+
\frac{ \mathcal{P}_{\PH\text{jj}} (\vec\Omega^{\PH+\text{jj}} | m_{4\ell}) }
{\mathcal{P}_{\WH}  (\vec\Omega^{\PH+\text{jj}} | m_{4\ell})  }
\right]^{-1} \\
\DMeZh &=
\left[1+
\frac{ \mathcal{P}_{\PH\text{jj}} (\vec\Omega^{\PH+\text{jj}} | m_{4\ell}) }
{\mathcal{P}_{\ZH}  (\vec\Omega^{\PH+\text{jj}} | m_{4\ell})  }
\right]^{-1},
\end{aligned}
\end{equation}
}
{
\begin{equation}
\label{eqn:prodmela}
\begin{aligned}
\DMeVbfjj &=
\left[1+
\frac{ \mathcal{P}_{\PH\text{jj}} (\vec\Omega^{\PH+\text{jj}} | m_{4\ell}) }
{\mathcal{P}_{\VBF}  (\vec\Omega^{\PH+\text{jj}} | m_{4\ell})  }
\right]^{-1}
\,\,\,
\DMeVbfj &&=
\left[1+
\frac{ \mathcal{P}_{\PH\text{j}} (\vec\Omega^{\PH+\text{j}} | m_{4\ell}) }
{\int d\eta^{\text{j}}\mathcal{P}_{\VBF}  (\vec\Omega^{\PH+\text{jj}} | m_{4\ell}) }
\right]^{-1}
\,\\
\DMeWh &=
\left[1+
\frac{ \mathcal{P}_{\PH\text{jj}} (\vec\Omega^{\PH+\text{jj}} | m_{4\ell}) }
{\mathcal{P}_{\WH}  (\vec\Omega^{\PH+\text{jj}} | m_{4\ell})  }
\right]^{-1}
\,\,\,
\DMeZh &&=
\left[1+
\frac{ \mathcal{P}_{\PH\text{jj}} (\vec\Omega^{\PH+\text{jj}} | m_{4\ell}) }
{\mathcal{P}_{\ZH}  (\vec\Omega^{\PH+\text{jj}} | m_{4\ell})  }
\right]^{-1},
\end{aligned}
\end{equation}
}
\end{linenomath}
where $\mathcal{P}_{\VBF}$, $\mathcal{P}_{\PH\text{jj}}$, $\mathcal{P}_{\PH\text{j}}$, and $\mathcal{P}_{\VH}$ are the probabilities for the $\VBF$ process,
the $\ggH$ process in association with two jets (combination of $\Pg\Pg/\PQq\Pg/\PQq\PQq^\prime$ parton collisions producing $\PH+2$~jets),
the $\ggH$ process in association with one jet ($\PH+1$~jet),
and the $\VH$ process, respectively.
The quantity $\int d\eta^{\text{j}}\mathcal{P}_{\VBF}$ is the integral of the two-jet $\VBF$ matrix element probability over the $\eta^j$ values of the unobserved jet, with the constraint that the total transverse momentum of the $\PH+2$~jets system is zero.
The discriminant $\DMeVh$, used for event categorization, is defined as the maximum value of the two discriminants, $\DMeVh = \max(\DMeZh,\DMeWh)$.

A set of three discriminants used in the likelihood fits is calculated as in Refs.~\cite{CMSH4lAnomalousCouplings,CMSH4l2016}.
The discriminant sensitive to the $\Pg\Pg/\PQq\PAQq\to4\ell$ process exploits the kinematics of the four-lepton decay system.
It is used in most of the event categories described in Section~\ref{sec:categorization} to separate signal from background and is defined as:
\begin{linenomath}
\begin{equation}
\label{eq:ggmela}
\mathcal{D}^{\text{kin}}_{\text{bkg}} =
\left[1+
\frac{ \mathcal{P}^{\PQq \PAQq }_\text{bkg} (\vec\Omega^{\PH\to4\ell} | m_{4\ell})  }
{ \mathcal{P}^{\cPg\cPg}_\text{sig}(\vec\Omega^{\PH\to4\ell} | m_{4\ell}) }
\right]^{-1},
\end{equation}
\end{linenomath}
where $\mathcal{P}^{\cPg\cPg}_\text{sig}$ is the probability for the signal
and $\mathcal{P}^{\PQq \PAQq }_\text{bkg}$ is the probability for the dominant $\PQq\PAQq\to4\ell$ background process, calculated using the LO matrix elements.
In the $\VBF$-2jet-tagged and $\VH$-hadronic-tagged event categories (defined in Section~\ref{subsec:Reco_Categories}), the background includes the QCD production of $\PZ\PZ/\PZ\Pgg^{*}/\Pgg^{*}\Pgg^{*}\to4\ell$ in association with two jets, the EW background from the vector boson scattering (VBS), as well as the triboson ($\PV\PV\PV$) production process.
We therefore use dedicated production-dependent discriminants based on the kinematics of the four-lepton decay and information from the associated jets (noted with $\VBF+$~dec or $\VH+$~dec), defined as:
\begin{linenomath}
\ifthenelse{\boolean{cms@external}}
{
\begin{multline}
\label{eq:vbfmela}
\DbkgVBFdec =
\Bigg[1+\\
\frac{c^{\VBF}(\mllll) [\mathcal{P}^{\text{EW}}_\text{bkg}(\vec\Omega^{\PH+\text{jj}}| m_{4\ell})+\mathcal{P}^{\mathrm{QCD}}_\text{bkg}(\vec\Omega^{\PH+\text{jj}}| m_{4\ell})]}{\mathcal{P}^{\text{EW}}_\mathrm{sig}(\vec\Omega^{\PH+\text{jj}}| m_{4\ell})}
\Bigg]^{-1}
\end{multline}
}
{
\begin{equation}
\label{eq:vbfmela}
\DbkgVBFdec =
\left[1+
\frac{c^{\VBF}(\mllll) [\mathcal{P}^{\text{EW}}_\text{bkg}(\vec\Omega^{\PH+\text{jj}}| m_{4\ell})+\mathcal{P}^{\mathrm{QCD}}_\text{bkg}(\vec\Omega^{\PH+\text{jj}}| m_{4\ell})]}{\mathcal{P}^{\text{EW}}_\mathrm{sig}(\vec\Omega^{\PH+\text{jj}}| m_{4\ell})}
\right]^{-1}
\end{equation}
}
\end{linenomath}
\begin{linenomath}
\ifthenelse{\boolean{cms@external}}
{
\begin{multline}
\label{eq:vhmela}
\DbkgVHdec =
\Bigg[1+\\
\frac{c^{\VH}(\mllll) [\mathcal{P}^{\text{EW}}_\text{bkg}(\vec\Omega^{\PH+\text{jj}}| m_{4\ell})+\mathcal{P}^{\mathrm{QCD}}_\text{bkg}(\vec\Omega^{\PH+\text{jj}}| m_{4\ell})]}{\mathcal{P}^{\text{EW}}_\mathrm{sig}(\vec\Omega^{\PH+\text{jj}}| m_{4\ell})}
\Bigg]^{-1},
\end{multline}
}
{
\begin{equation}
\label{eq:vhmela}
\DbkgVHdec =
\left[1+
\frac{c^{\VH}(\mllll) [\mathcal{P}^{\text{EW}}_\text{bkg}(\vec\Omega^{\PH+\text{jj}}| m_{4\ell})+\mathcal{P}^{\mathrm{QCD}}_\text{bkg}(\vec\Omega^{\PH+\text{jj}}| m_{4\ell})]}{\mathcal{P}^{\text{EW}}_\mathrm{sig}(\vec\Omega^{\PH+\text{jj}}| m_{4\ell})}
\right]^{-1},
\end{equation}
}
\end{linenomath}
where $\mathcal{P}^{\text{EW}}_\mathrm{sig}$ is the probability for the $\VBF$ and $\VH$ signal,
$\mathcal{P}^{\text{EW}}_\text{bkg}$ is the probability for the VBS and VVV background processes,
and $\mathcal{P}^{\mathrm{QCD}}_\text{bkg}$ is the probability for $\PZ\PZ/\PZ\Pgg^{*}/\Pgg^{*}\Pgg^{*}\to4\ell$ QCD production in association with two jets.
The quantity $c^{p}(\mllll)$ for category $p$ is the $\mllll$-dependent parameter that allows to change the relative normalization of the EW probabilities, separately for the $\VBF$ and $\VH$ topologies.
For each slice of $\mllll$, the distributions of the signal and background discriminants are plotted,
and the $c^{p}(\mllll)$ value is determined in such a way that the two distributions cross at 0.5.
This procedure allows rescaling of the distributions for the linear-scale binning of the templates used in the likelihood fits described in Section~\ref{sec:results}.

\section{Event categorization}
\label{sec:categorization}

In order to improve the sensitivity to the \PH boson production mechanisms,
the selected events are classified into mutually exclusive categories based on the features of the reconstructed objects associated with the $\Hllll$ candidates.
Event categorization is organized in two steps with increasing granularity of the categories.
First step is primarily designed to separate the $\ggH$, $\VBF$, $\VH$, and $\ttH$ processes.
There is little sensitivity to $\bbH$ or $\tH$, even though these production modes are considered explicitly in the analysis.
The reconstructed event categories from the first step are further subdivided (as discussed in Section~\ref{subsec:Reco_Categories}) in order to study each production mechanism in more detail.
This subdivision is carried out by matching the recommended binning of the framework of STXS described in the following section.

\subsection{STXS production bins}
\label{subsec:STXS_Categories}

The STXS framework has been adopted by the LHC experiments as a common framework for studies of the \PH boson.
It has been developed to define fine-grained measurements of the \PH boson production modes in various kinematic regions, and to reduce the theoretical uncertainties that are folded into the measurements.
It also allows for the use of advanced categorization techniques and provides a common scheme for combining measurements in different decay channels or by different experiments.
The regions of phase space defined by this framework are referred to as production bins and are determined by using generator-level information for \PH bosons with rapidity $\abs{y_\PH}<2.5$.
Generator-level jets are defined as anti-\kt jets with a distance parameter of 0.4 and a \pt threshold of 30\GeV; no requirement is placed on the generator-level leptons.

The STXS framework has been designed to complement the Run I measurements of the production signal strength modifiers and fiducial differential cross sections of the \PH boson by combining their advantages.
The sensitivity to theoretical uncertainties in the signal strength modifier results is suppressed by excluding dominant theoretical uncertainties causing production bin migration effects from the STXS measurements.
They are included only when comparing the results with the theoretical predictions.
In contrast to the fiducial differential cross section measurements,
in the STXS framework measurements are optimized for sensitivity by means of event categories and matrix element discriminants.
To account for the evolving experimental sensitivity, different stages of production bins with increasing granularity are developed.

The stage 0 production bins are called $\stxsggH$, $\mathtt{qqH}$, $\stxsVHlep$, and $\stxsttH$ and are designed to closely match the main \PH boson production mechanisms.
The $\stxsqqH$ bin includes the EW production of the \PH boson in association with two quarks from either $\VBF$ or $\VH$ events with hadronic decays of the vector boson \PV.
The $\stxsVHlep$ production bin includes $\VH$ events with leptonic decays of the vector boson \PV.
The low rate $\bbH$ and $\tH$ production processes are considered together with the $\stxsggH$ and $\stxsttH$ production bins, respectively.
In this analysis, a modified version of the stage 0 production bins is also studied, where instead of $\stxsVHlep$ and $\stxsqqH$ bins we define the $\mathtt{WH}$, $\mathtt{ZH}$,
and $\mathtt{VBF}$ bins that map the \PH boson production mechanisms without the splitting of the $\VH$ events in leptonic and hadronic decays.

\begin{figure*}[htbp]
	\centering
		\includegraphics[width=1.3\textwidth, angle=90]{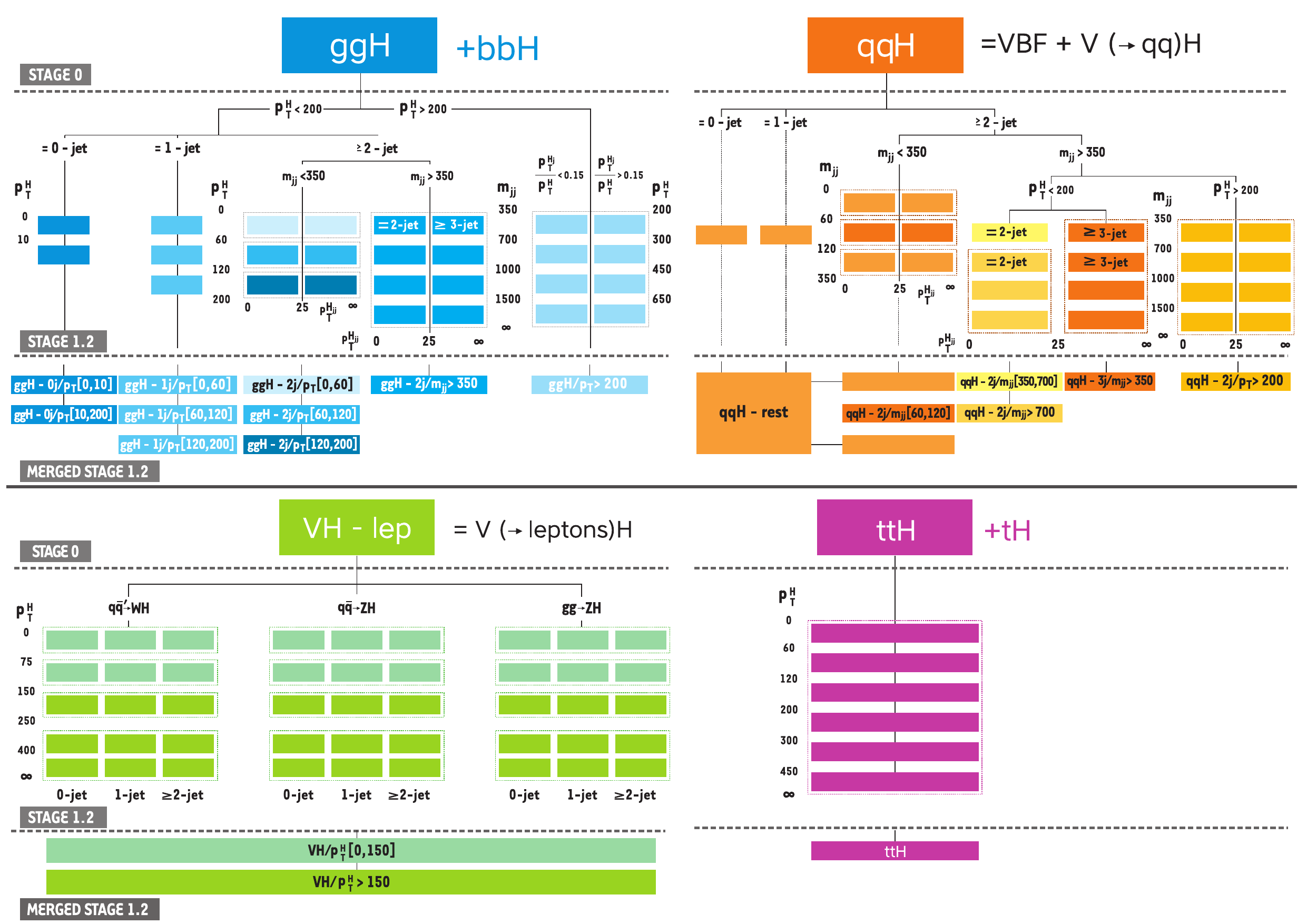}
		\caption{Binning of the gluon fusion production process,
		the electroweak production process (combines $\VBF$ and $\VH$ with hadronic \PV decay),
		the $\VH$ production process with leptonic \PV decay (combining $\WH$, $\ZH$, and gluon fusion $\ZH$ production),
		and the $\ttH$ production process in the merged stage 1.2 of the STXS framework used in the $\HZZfl$ analysis.
			\label{fig:stage1p2_reduced}}
\end{figure*}

Stage 1 of the STXS framework was designed by further splitting the bins from the stage 0,
one of the main motives being the enhanced sensitivity to possible signatures of BSM physics.
This is achieved by dividing stage 0 bins with additional requirements on generator-level quantities that include the transverse momentum of the \PH boson ($\pt^{\PH}$),
the number of associated jets ($N^{\text{j}}$), the dijet invariant mass ($m_{\text{jj}}$), the transverse momentum of the \PH boson and the leading jet ($\pt^{\PH\text{j}}$),
and the transverse momentum of the \PH boson and the two leading jets ($\pt^{\PH\text{jj}}$).
These bins were designed in order to maximize sensitivity to new physics while also taking into account the current experimental sensitivity limited mostly by the amount of collected data.
The most recent set of bins defined in the STXS framework is referred to as stage 1.2.
This paper presents a first set of cross section measurements in the $\Hllll$ channel for the stage 1.2 of the STXS framework.
However, several stage 1.2 production bins are merged as the full set of bins cannot be measured with the current data sample.
The merging scheme results in 19 production bins; it is illustrated in Fig.~\ref{fig:stage1p2_reduced} and discussed in more detail below.

{\tolerance=800 The $\ggH$ production process is split into events with $\pt^{\PH}<200\GeV$ and $\pt^{\PH}>200\GeV$.
The events with $\pt^{\PH}>200\GeV$ are placed into one single production bin called $\stageoneggHbinten$.
The events with $\pt^{\PH}<200\GeV$ are split in events with zero, one, and two or more jets.
The events with zero or one jets are split into the following production bins according to the \PH boson \pt : $\stageoneggHbinone$,
$\stageoneggHbintwo$, $\stageoneggHbinthree$, $\stageoneggHbinfour$, and $\stageoneggHbinfive$.
The events with two or more jets are split according to the dijet invariant mass as follows.
The events with $m_{\text{jj}}<350\GeV$ are split into three production bins according to the \PH boson \pt: $\stageoneggHbinsix$, $\stageoneggHbinseven$, and $\stageoneggHbineight$.
The events with $m_{\text{jj}}>350\GeV$ are all placed into one production bin $\stageoneggHbinnine$, which merges four bins originally suggested in stage 1.2 of the STXS framework.\par}

The merging scheme of the electroweak $\stxsqqH$ production bins is as follows.
The events with zero jets, one jet, or with two or more jets with $m_{\text{jj}}<60\GeV$ or $120<m_{\text{jj}}<350\GeV$ correspond to production bins with insufficient statistics; they are all merged into one bin called $\stageoneqqHbinone$.
The events with two or more jets and $60<m_{\text{jj}}<120\GeV$ are placed in the $\stageoneqqHbintwo$ bin.
The events with two or more jets and $m_{\text{jj}}>350\GeV$ are split into events with $\pt^{\PH}<200\GeV$ and $\pt^{\PH}>200\GeV$.
The events with $\pt^{\PH}>200\GeV$ are placed into one single production bin called $\stageoneqqHbinsix$.
The events with $\pt^{\PH}<200\GeV$ and $\pt^{\text{\PH jj}}<25\GeV$ are split into two production bins, $\stageoneqqHbinthree$ and $\stageoneqqHbinfour$, and otherwise if $\pt^{\text{\PH jj}}>25\GeV$ are merged in a single bin called $\stageoneqqHbinfive$.

{\tolerance=800 The three production processes $\PQq\PAQq^{'}\to\WH$, $\Pg\Pg\to\ZH$, and $\PQq\PAQq\to\ZH$ are combined to build $\stxsVHlep$ reduced stage 1.2 production bins.
Several proposed production bins are merged into just two bins according to the \pt of the \PH boson: $\stageoneVHbinone$ and $\stageoneVHbintwo$.\par}

In stage 1.2 of the STXS framework the $\stxsttH$ stage 0 production bin is split in five different bins according to the \pt of the \PH boson.
Because of the very low expected yields all these bins are merged into a single bin that includes the $\tH$ production process as well.

{\tolerance=1200 Finally, in stage 1.2 the $\bbH$ production process, which has small cross section, is classified in the $\stageoneggHbintwo$ production bin.\par}

The first measurement of STXS stage 1.2 cross sections was recently performed by the CMS Collaboration~\cite{HIG-19-015}.

\subsection{Reconstructed event categories}
\label{subsec:Reco_Categories}

In order to be sensitive to different production bins, the $\PZ\PZ$ candidates that pass the event selection described in Section~\ref{sec:objects} are classified into several dedicated reconstructed event categories.
The category definitions are based on the multiplicity of jets, {\PQb}-tagged jets, and additional leptons in the event.
Additional leptons are not involved in the $\ZZ$ candidate selection but, if present, should satisfy the identification, vertex compatibility, and isolation requirements.
Requirements on the categorization discriminants described in Section~\ref{sec:observables}, the invariant mass of the two leading jets, and the transverse momentum of the $\PZ\PZ$ candidate are also exploited.

The event categorization is carried out in two steps.
In the first step, the $\PZ\PZ$ candidates are split into seven initial categories to target the main \PH boson production mechanisms corresponding to the stage 0 production bins.
The first step of the categorization closely follows the analysis strategy from the previous publication~\cite{CMSH4l2016}.
To ensure exclusive categories, an event is considered for the subsequent category only if it does not satisfy the requirements of the previous one.

In the first categorization step, the following criteria are applied:
\begin{itemize}
\item The {$\VBF$-2jet-tagged} category requires exactly 4 leptons. In addition there must be either 2 or 3 jets
of which at most 1 is {\PQb}-tagged, or at least 4 jets and no {\PQb}-tagged jets. Finally, $\DMeVbfjj>0.5$ is required.
\item The {$\VH$-hadronic-tagged} category requires exactly 4 leptons. In addition there must be 2 or 3 jets with no {\PQb}-tagging requirements,
or at least 4 jets and no {\PQb}-tagged jets. Finally, $\DMeVh > 0.5$ is required.
\item The {$\VH$-leptonic-tagged} category requires no more than 3 jets and no {\PQb}-tagged jets in the event,
and exactly 1 additional lepton or 1 additional pair of opposite sign, same flavor leptons. This category also includes events
with no jets and at least 1 additional lepton.
\item The {$\ttH$-hadronic-tagged} category requires at least 4 jets, of which at least 1 is {\PQb}-tagged, and no additional leptons.
\item The {$\ttH$-leptonic-tagged} category requires at least 1 additional lepton in the event.
\item The {$\VBF$-1jet-tagged} category requires exactly 4 leptons, exactly 1 jet and $\DMeVbfj>0.7$.
\item The {untagged} category consists of the remaining events.
\end{itemize}

Reconstructed events are further subdivided in the second step of the categorization that is designed to closely match the merged stage 1.2 production bins described in the previous section.
In the second categorization step, the {untagged}, {$\VBF$-2jet-tagged}, {$\VH$-hadronic-tagged},
and {$\VH$-leptonic-tagged} categories are further split exploiting additional variables like the invariant mass of the two leading jets and the transverse momentum of the $\PZ\PZ$ candidate.
A total number of twenty-two reconstructed event categories is defined and details of the categorization are presented in Table~\ref{tab:categorisation}.

\begin{table*}[htb]
	\centering
		\topcaption{Event categorization criteria of the $\Hllll$ analysis targeting stage 1.2 STXS production bins.
     Events from the first step of the categorization are further classified based on the kinematical properties listed in the table.
     A dash indicates no requirement.
		\label{tab:categorisation}}
		\cmsTable
		{
		\begin{tabular}{ccccc}
			Reconstructed event category &  1$^\mathrm{st}$ categorization step & Number of jets & Kinematical requirements (\GeV) & Targeted production bin \\
			\hline
			Untagged-0j-$\pt^{4\ell}[0,10]$ & Untagged & 0 & $0<\pt^{4\ell}<10$ & $\stageoneggHbinone$ \\
			Untagged-0j-$\pt^{4\ell}[10,200]$ & Untagged & 0 & $10<\pt^{4\ell}<200$ & $\stageoneggHbintwo$ \\
      Untagged-1j-$\pt^{4\ell}[0,60]$ & Untagged & 1 & $0<\pt^{4\ell}<60$ & $\stageoneggHbinthree$ \\
      Untagged-1j-$\pt^{4\ell}[60,120]$ & Untagged & 1 & $60<\pt^{4\ell}<120$ & $\stageoneggHbinfour$ \\
      Untagged-1j-$\pt^{4\ell}[120,200]$ & Untagged & 1 & $120<\pt^{4\ell}<200$ & $\stageoneggHbinfive$ \\
      Untagged-2j-$\pt^{4\ell}[0,60]$ & Untagged & 2 & $0<\pt^{4\ell}<60$, $m_{\text{jj}}<350$ & $\stageoneggHbinsix$ \\
      Untagged-2j-$\pt^{4\ell}[60,120]$ & Untagged & 2 & $60<\pt^{4\ell}<120$, $m_{\text{jj}}<350$ & $\stageoneggHbinseven$ \\
      Untagged-2j-$\pt^{4\ell}[120,200]$ & Untagged & 2 & $120<\pt^{4\ell}<200$, $m_{\text{jj}}<350$ & $\stageoneggHbineight$ \\
      Untagged-$\pt^{4\ell}>200$ & Untagged &\NA& $\pt^{4\ell}>200$ & $\stageoneggHbinten$ \\
      Untagged-2j-$m_{\mathrm{jj}}>350$ & Untagged & 2 & $m_{\text{jj}}>350$ & $\stageoneggHbinnine$ \\[\cmsTabSkip]
      $\VBF$-1jet-tagged & $\VBF$-1jet-tagged &\NA&\NA& $\stageoneqqHbinone$ \\[\cmsTabSkip]
      $\VBF$-2jet-tagged-$m_{\mathrm{jj}}[350,700]$ & $\VBF$-2jet-tagged &\NA& $\pt^{4\ell}<200$, $\pt^{4\ell\text{jj}}<25$, $350<m_{\text{jj}}<700$ & $\stageoneqqHbinthree$ \\
      $\VBF$-2jet-tagged-$m_{\mathrm{jj}}>700$ & $\VBF$-2jet-tagged &\NA& $\pt^{4\ell}<200$, $\pt^{4\ell\text{jj}}<25$, $m_{\text{jj}}>700$ & $\stageoneqqHbinfour$ \\
      $\VBF$-3jet-tagged-$m_{\mathrm{jj}}>350$ & $\VBF$-2jet-tagged &\NA& $\pt^{4\ell}<200$, $\pt^{4\ell\text{jj}}>25$, $m_{\text{jj}}>350$ & $\stageoneqqHbinfive$ \\
      $\VBF$-2jet-tagged-$\pt^{4\ell}>200$ & $\VBF$-2jet-tagged &\NA& $\pt^{4\ell}>200$, $m_{\text{jj}}>350$ & $\stageoneqqHbinsix$ \\
      $\VBF$-rest & $\VBF$-2jet-tagged &\NA& $m_{\text{jj}}<350$ & $\stageoneqqHbinone$ \\[\cmsTabSkip]
      $\VH$-hadronic-tagged-$m_{\mathrm{jj}}[60,120]$ & $\VH$-hadronic-tagged &\NA& $60<m_{\text{jj}}<120$ & $\stageoneqqHbintwo$ \\
      $\VH$-rest & $\VH$-hadronic-tagged &\NA& $m_{\text{jj}}<60$ or $m_{\text{jj}}>120$ & $\stageoneqqHbinone$ \\[\cmsTabSkip]
      $\VH$-leptonic-tagged-$\pt^{4\ell}[0,150]$ & $\VH$-leptonic-tagged &\NA& $\pt^{4\ell}<150$ & $\stageoneVHbinone$ \\
      $\VH$-leptonic-tagged-$\pt^{4\ell}>150$ & $\VH$-leptonic-tagged &\NA& $\pt^{4\ell}>150$ & $\stageoneVHbintwo$ \\[\cmsTabSkip]
      $\ttH$-leptonic-tagged & $\ttH$-leptonic-tagged &\NA&\NA& $\stxsttH$ \\[\cmsTabSkip]
      $\ttH$-hadronic-tagged & $\ttH$-hadronic-tagged &\NA&\NA& $\stxsttH$ \\
	\end{tabular}}
\end{table*}

\section{Background estimation}
\label{sec:bkgd}

\subsection{Irreducible backgrounds}
\label{sec:irrbkgd}

The irreducible background to the \PH boson signal in the $4\ell$ channel, which comes from the
production of $\PZ\PZ$ via $\Pq\Paq$ annihilation or gluon fusion, is estimated using simulation.
The fully differential cross section for the $\qqZZ$ process is computed at NNLO~\cite{Grazzini2015407},
and the NNLO/NLO K factor as a function of $m_{\PZ\PZ}$ is applied to the \POWHEG sample.
This K factor varies from 1.0 to 1.2 and is 1.1 at $m_{\PZ\PZ}=125\GeV$.
Additional NLO electroweak corrections that depend on the initial state quark flavor and kinematics
are also applied in the region $m_{\PZ \PZ}>2m_{\PZ}$ following the prescription in Ref.~\cite{Bierweiler:2013dja}.

The production of $\PZ\PZ$ via gluon fusion contributes at NNLO in pQCD.
It has been shown that the soft collinear approximation is able to describe the cross section for this process and the interference term at NNLO\@~\cite{Bonvini:1304.3053}.
Further calculations also show that the K factors are very similar at NLO for signal and background~\cite{Melnikov:2015laa} and at NNLO for signal and interference terms~\cite{Li:2015jva}.
Therefore, the same K factor is used for signal and background~\cite{Passarino:1312.2397v1}.
The NNLO K factor for the signal is obtained as a function of $m_{\PZ\PZ}$ using the \textsc{hnnlo}~v2 program~\cite{Catani:2007vq,Grazzini:2008tf,Grazzini:2013mca} by calculating the NNLO and LO $\Pg\Pg\to\PH\to2\ell2\ell^\prime$ cross sections for the $\PH$ boson decay width of 4.07\MeV and taking their ratios.
The NNLO/LO K factor for {\ggZZ} varies from $\approx$2.0 to 2.6 and is 2.27 at $m_{\PZ\PZ}=125\GeV$;
a systematic uncertainty of 10\% is assigned to it when applied to the background process.

The triboson background processes $\PZ\PZ\PZ$, $\PW\PZ\PZ$, and $\PW\PW\PZ$, as well as $\ttZ$, $\ttWW$, and $\ttZZ$ are also considered.
These rare backgrounds are all estimated from simulation and are jointly referred to as the EW backgrounds.

Simulated samples are used to obtain shapes of the four-lepton invariant mass that are later used to build the likelihood function.
For each irreducible background contribution, events are divided in three final states ($4\Pgm$, $4\Pe$, and $2\Pe 2\Pgm$) and 22 event sub-categories defined in Section~\ref{subsec:STXS_Categories}.
To extract the shape of the $m_{4\ell}$ distribution, expected yields are fitted to empirical functional forms built from a third order Bernstein polynomial.
In sub-categories with not enough statistics to perform a fit, the shape is extracted from the inclusive distribution in the corresponding final state.

\subsection{Reducible backgrounds}
\label{sec:redbkgd}

Additional backgrounds to the \PH boson signal in the $4\ell$ channel arise from processes in which decays of heavy-flavor hadrons,
in-flight decays of light mesons within jets, or (for electrons) charged hadrons overlapping with $\pi^0$ decays are misidentified as leptons.
The main processes leading to these backgrounds are $\PZ\text{+jets}$,  $\ttbar\text{+jets}$,
$\PZ\gamma\text{+jets}$, $\PW\PW\text{+jets}$, and $\PW\PZ\text{+jets}$ production.
We denote these reducible backgrounds as "$\PZ\text{+X}$" since they are dominated by the $\PZ\text{+jets}$ process.
The contribution from the reducible background is estimated with two independent methods, each with dedicated control regions in data.
The control regions are defined by the presence of both a lepton pair satisfying all the requirements of a $\PZ_1$ candidate and two additional opposite sign (OS) or same sign (SS) leptons;
the two additional leptons satisfy identification requirements looser than those used in the analysis.
These four leptons are then required to pass the analysis $\PZ\PZ$ candidate selection.
The event yield in the signal region is obtained by weighting the control region events by the lepton misidentification probability $f_{\Pe}$ ($f_{\Pgm}$),
defined as the fraction of non-signal electrons (muons) that are identified by the analysis selection criteria.
A detailed description of both methods can be found in Ref.~\cite{CMSH4l2016}.

The lepton misidentification rates $f_{\Pe}$ and $f_{\Pgm}$ are measured as a function of $\pt^\ell$ and $\eta^\ell$ by means of a sample that includes a $\PZ_1$ candidate consisting of a pair of leptons,
both passing the selection requirements used in the analysis, and exactly one additional lepton passing the relaxed selection.
Furthermore, the \ptmiss is required to be less than 25\GeV  in order to suppress contamination from $\PW\PZ$ and $\ttbar$ processes.

For the OS method, the mass of the $\PZ_1$ candidate is required to satisfy the condition $\abs{\PZ_1 - m_{\PZ}} < 7\GeV$  in order to reduce the contribution of (asymmetric) photon conversions, which is estimated separately.
In the SS method, the contribution of photon conversions to the misidentification rate is estimated with dedicated samples.

The predicted yields of the reducible background from the two methods are in agreement within their uncertainties for each final state ($4\Pgm$, $4\Pe$, and $2\Pe 2\Pgm$).
The final yield used in the analysis is a weighted average of the two independent estimates.
To extract the shape of the $m_{4\ell}$ distribution for the reducible background a maximum-likelihood fit is performed in each of the 22 event sub-categories defined in Section~\ref{subsec:STXS_Categories}.
For each sub-category, the expected "$\PZ\text{+X}$" yields from the OS and SS methods are binned as a function of $m_{4\ell}$ and fitted to empirical functional forms built from Landau distributions~\cite{Landau:1944if}.
In sub-categories with not enough statistics to perform a fit, the shape is extracted from the inclusive distribution in the corresponding final state.

{\tolerance=800 The dominant systematic uncertainty on the reducible background estimation arises from the difference in the composition of the sample from which the misidentification rate is computed and that of the control regions.
It is determined from the MC simulation and is found to be around 30\%, depending on the final state.
Additional sources of systematic uncertainty arise from the limited number of events in the control regions as well as in the region where the misidentification rates are computed.\par}

\section{Signal modeling}
\label{sec:signal}

{\tolerance=800 In order to generate an accurate signal model, the \pt spectrum of the \PH boson, $\pt^{\PH}$, was tuned in the
\POWHEG simulation of the dominant gluon fusion production mode to better match the predictions from full phase space
calculations implemented in the \textsc{HRes}~generator~\cite{deFlorian:2012mx, Grazzini:2013mca, Bagnaschi:2011tu}.\par}

In order to take advantage of the accuracy of the \textsc{nnlops}~\cite{Hamilton:2013fea} simulation available for the $\ggH$ process, an event reweighting procedure is used.
Events originating from the $\ggH$ process are subdivided into classes with 0, 1, 2, and $\geq$3 jets; the jets with $\pt > 30\GeV$ are clustered from all stable particles using the anti-\kt algorithm with a distance parameter of 0.4,
excluding the decay products of the \PH boson or associated vector bosons.
The weights are obtained as the ratios of the $\pt^{\PH}$ distributions from the \textsc{nnlops} and the \POWHEG generators for each event class;
the sum of the weights in each sample is normalized to the inclusive cross section.

\begin{figure}[!htb]
	\centering
		\includegraphics[width=0.49\textwidth]{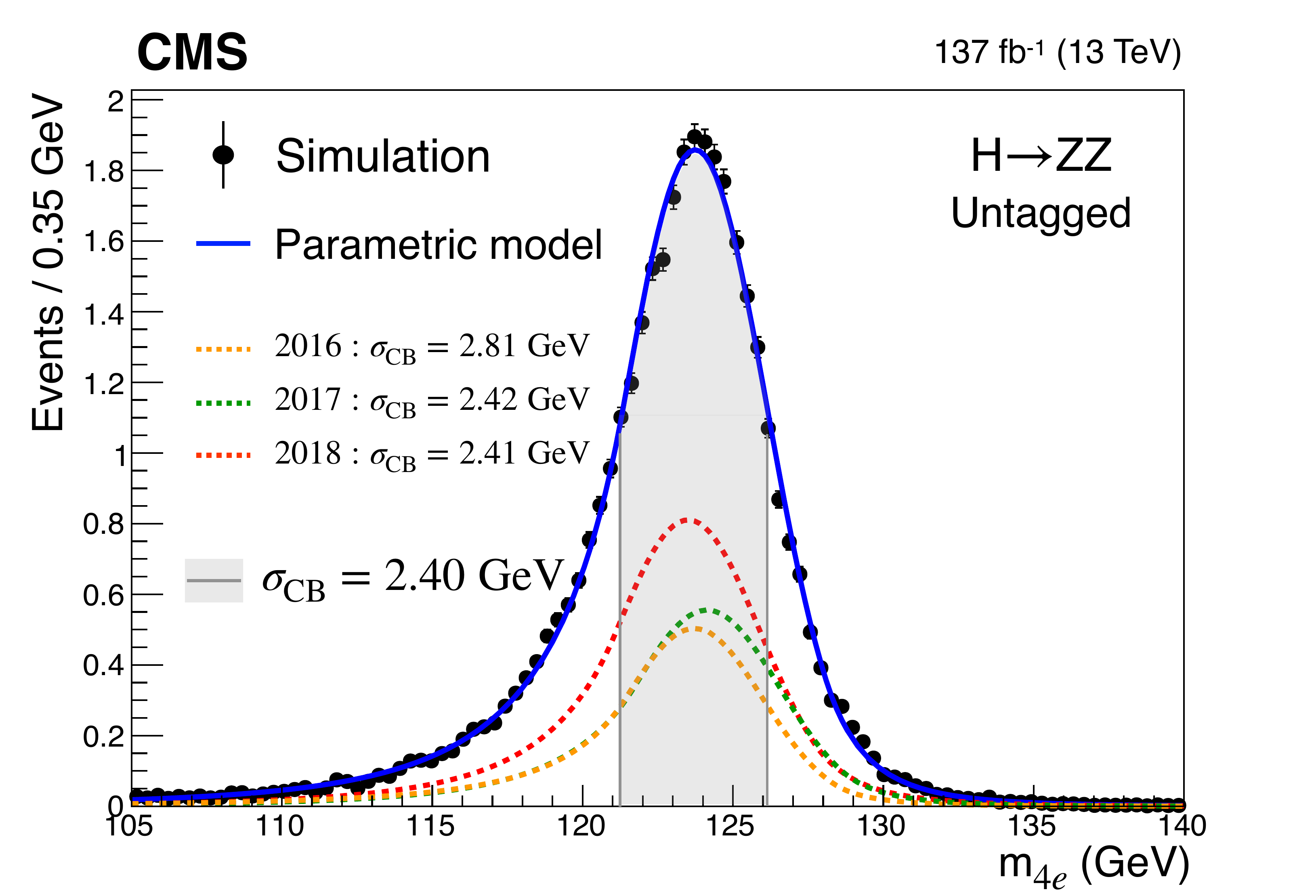}
    \includegraphics[width=0.49\textwidth]{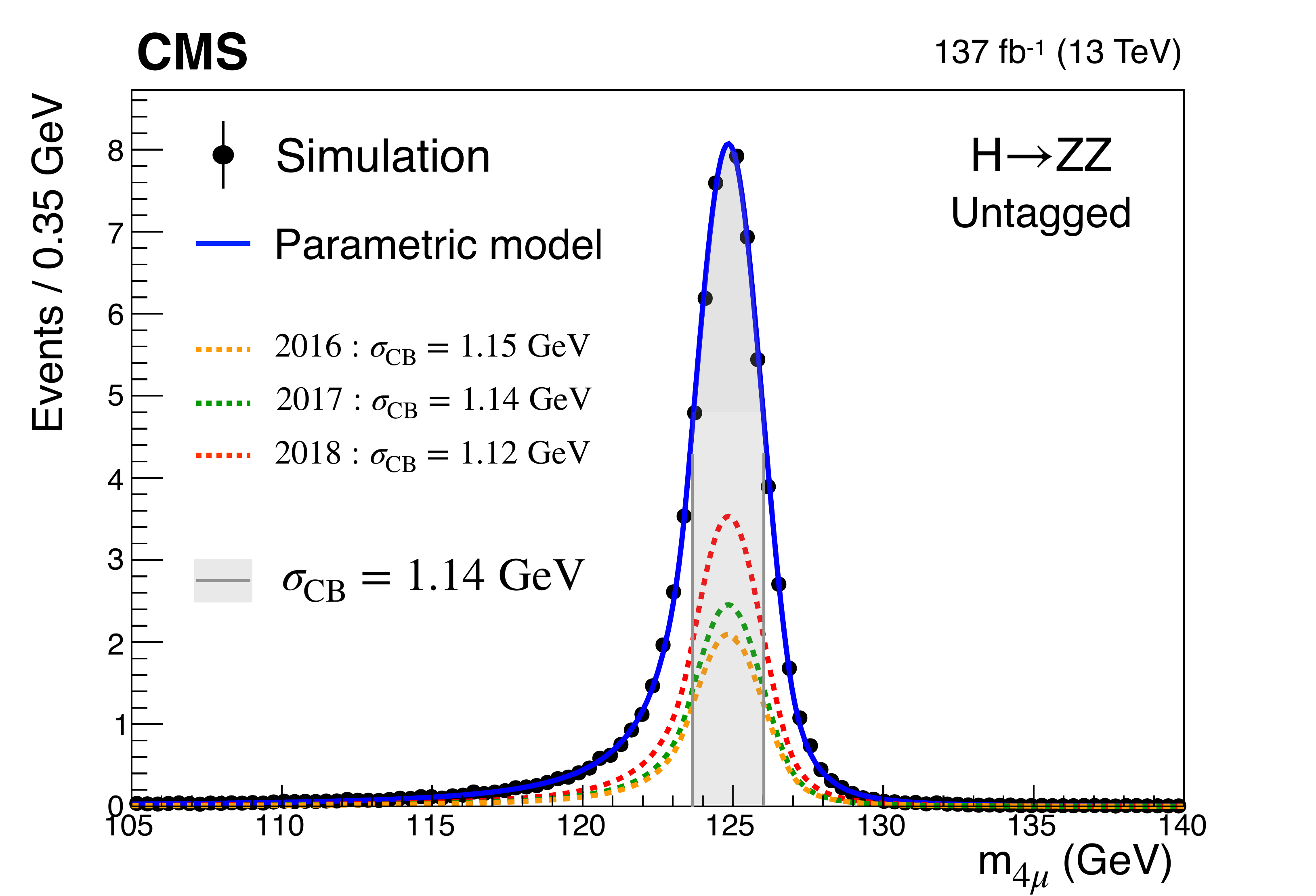}
		\caption{
		The shape of the parametric signal model for each year of simulated data, and for the sum of all years together.
    The black points represent weighted simulation events of the $\ggH$ production mechanism for $\mH=125\GeV$ and the blue line the corresponding model.
    Also shown is the $\sigma_{\text{CB}}$ value (half the width of the narrowest interval containing 68\% of the invariant mass distribution) in the gray shaded area.
    The contribution of the signal model from each year of data-taking is illustrated with the dotted lines.
    The models are shown for the $4\Pe$ (\cmsLeft) and $4\Pgm$ (\cmsRight) final states in the untagged event category.
			\label{fig:signal_fit}}
\end{figure}

The signal shape is parametrized by means of a double-sided Crystal Ball function~\cite{Oreglia:1980cs} around $\mH \approx 125\GeV$.
A Landau function is added in the total probability density function for the non-resonant part of the signal
for the case of $\WH$, $\ZH$ and $\ttH$ production modes. The signal shape is parametrized as a function of $\mH$
by performing a simultaneous fit of several mass points for all production modes in the 105 to 140\GeV mass range.
Each parameter of the double-sided Crystal Ball function has a linear dependence on $\mH$, for a total of 12 free parameters.
An examples of the fit is shown in Fig.~\ref{fig:signal_fit}.

\section{Systematic uncertainties}
\label{sec:systematics}

The systematic uncertainties are divided into experimental and theoretical.
The main experimental uncertainties originate from the imperfect knowledge of the detector; the dominant sources are the uncertainties in the luminosity measurement,
the lepton reconstruction and selection efficiency, the lepton and jet energy scale and resolution, the {\PQb} tagging efficiency, and the reducible background estimate.
The theoretical uncertainties account for the uncertainties in the modeling of the signal and background processes.

Both types of uncertainties can affect the signal selection,
cause migrations between the event categories, and affect the signal or background shapes used in the fit.
All the uncertainties affecting this analysis are modeled as nuisance parameters (NPs) that are profiled in the
maximum likelihood fit described in Section~\ref{sec:results}.

In the combination of the three data-taking periods, all theoretical uncertainties are treated as correlated across these periods.
The experimental uncertainties related to reconstruction and selection efficiency, the lepton energy scale and resolution, and the {\PQb}-tagging efficiency are also considered correlated across data-taking periods.
Luminosity uncertainty is treated as partially correlated.
All other experimental uncertainties are treated as uncorrelated.
Correlated sources of uncertainty are assigned the same NP and uncorrelated sources have a dedicated NP in the likelihood fit described in Section~\ref{sec:results}.

\begin{figure*}[!htb]
	\centering
		\includegraphics[width=0.8\textwidth]{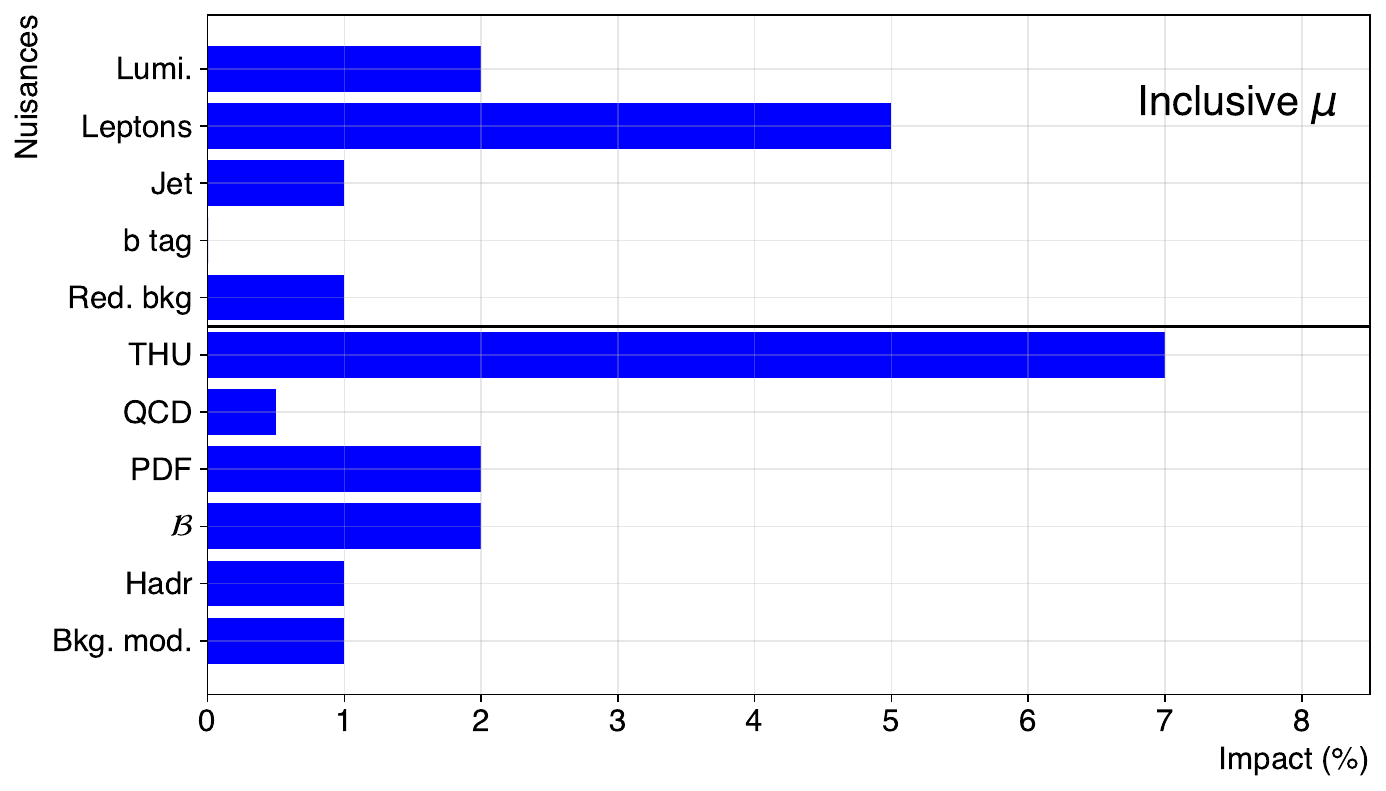}
		\includegraphics[width=0.8\textwidth]{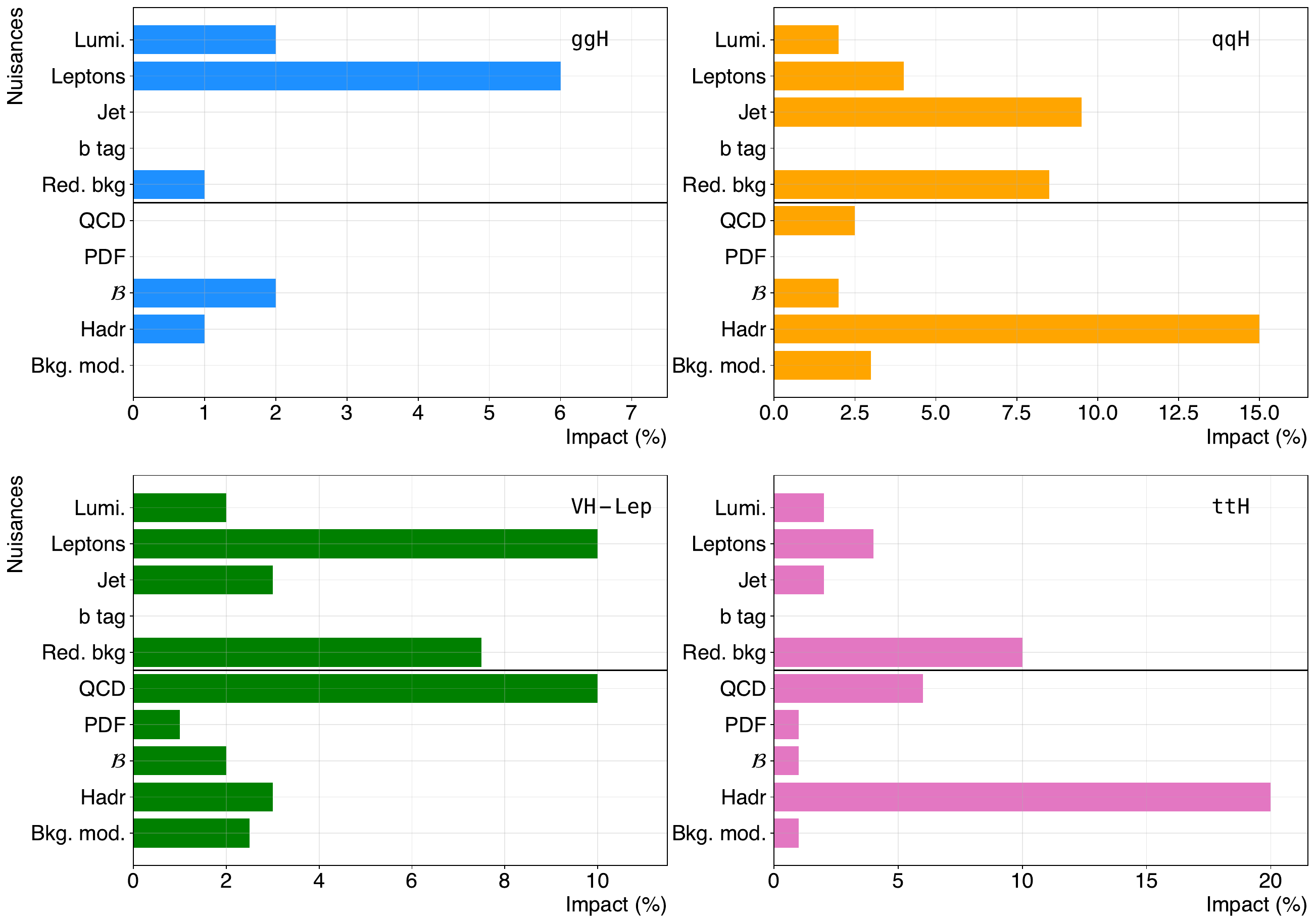}
		\caption{
		The impact of the dominant systematic uncertainties (in percent) on the inclusive signal strength $\mu$ and stage 0 production mode cross section described in Section~\ref{sec:results}.
    Impacts from different NPs are combined assuming no correlation.
		Only dominant experimental sources are presented: integrated luminosity uncertainty (Lumi.),
		lepton reconstruction and selection efficiency, scale and resolution (Leptons), jet energy scale and resolution (Jet), {\PQb}-tagging efficiency (B-tag),
		and reducible background estimation uncertainty (Red. bkg).
		Only dominant theoretical sources are presented: $\ggH$, $\VBF$, and $\VH$ cross section theoretical uncertainty scheme (THU), renormalization and factorization scale (QCD),
		choice of the PDF set (PDF), the branching fraction of $\Hllll$ ($\mathcal{B}$), modeling of hadronization and the underlying event (Hadr),
		and background modeling (Bkg. mod.). The THU uncertainty is not considered in the stage 0 cross section measurements. The uncertainties are rounded to the nearest 0.5\%.
			\label{fig:Impacts}}
\end{figure*}

The dominant sources of uncertainties and their effect on the analysis are discussed in detail in the following subsections.
The impact of a NP on a parameter of interest (POI) is defined as the shift induced on POI when NP is varied by a $\pm 1$ standard deviation from its post-fit value,
with all other parameters profiled as usual.
The relative impact of the dominant systematic uncertainties on some of the measurements discussed in Section~\ref{sec:results} is illustrated in Fig.~\ref{fig:Impacts}.

\subsection{Experimental uncertainties}
\label{subsec:Experimental_uncertainties}

The integrated luminosities of the 2016, 2017, and 2018 data-taking periods are individually known with uncertainties in the 2.3--2.5\% range~\cite{CMS-PAS-LUM-17-001,CMS-PAS-LUM-17-004,CMS-PAS-LUM-18-002},
while the total Run~2 (2016--2018) integrated luminosity has an uncertainty of 1.8\%, the improvement in precision reflecting the (uncorrelated) time evolution of some systematic effects.
The experimental uncertainty on the integrated luminosity measurement affects all final states, both signal and background.
Another experimental uncertainty common to all final states is the uncertainty in the lepton reconstruction and selection efficiency.
Here selection efficiency includes all the steps from trigger to impact parameter significance and finally identification and isolation requirements.
The uncertainty ranges from 1 to 2.3\% in the 4\Pgm channel and from 11 to 15.5\% in the 4\Pe channel.
While for muon efficiency measurements in the low $\pt^{\Pgm}$ regions we rely on low mass di-muon resonances,
the electron efficiency measurement relies solely on the \PZ boson resonance, resulting in a higher uncertainty in the low $\pt^{\Pe}$ region.

{\tolerance=800 Lepton momentum scale and resolution uncertainties are estimated from dedicated studies on the $\Zll$ mass distribution in data and simulation.
Events are classified according to the \pt and $\eta$ of one of the two leptons, determined randomly, and integrated over the other.
The dilepton mass distributions are then fit by a Breit-Wigner parameterization convolved with the double-sided Crystal Ball function described in Section~\ref{sec:signal}.
The scale uncertainty is found to be 0.04\% in the 4\Pgm channel and 0.3\% in the 4\Pe channel, while the resolution uncertainty is 20\% for both channels.
In both cases full correlation between the leptons in the event is assumed.
Both scale and resolution uncertainties alter the signal shape by allowing the corresponding parameters of the double-sided Crystal Ball function to vary.
The impact is found to be non-negligible only in the case of fiducial cross section measurements.\par}

The effects of the jet energy corrections are studied in a similar manner. While jet energy scale and smearing do not alter signal selection efficiency, they cause event migrations between the categories.
They can also alter the shape of the discriminants, but the effect on the shape is negligible.
The uncertainty in the jet energy scale ranges from 1\% in the high jet $\pt$ range and increases up to 5\% in the low jet $\pt$ range.
The uncertainty in jet energy resolution ranges from 1 to 2\%.
A detailed description of the determination of the jet energy scale and smearing uncertainties can be found in~\cite{Chatrchyan:2011ds}.
The effect on the analysis is studied in detail by propagating the uncertainties and estimating the effect on event migration in each of the 22 sub-categories.
Their impact on the inclusive measurements is found to be negligible.
However, the impact is significant in measurements of the $\VBF$ and $\VH$ production modes and differential cross section measurements as a function of jet kinematics,
where it is one of the leading sources of uncertainty.

The uncertainty in the {\PQb}-tagging efficiency is found to be 1\% in the high jet $\pt$ range and increases up to 3\% in the low jet \pt range.
The impact from the category migration is found to be negligible in all categories.

Finally, experimental uncertainties in the reducible background estimation, described in Section~\ref{sec:redbkgd},
originating from the background composition and misidentification rate uncertainties vary between 30 and 45\% depending on the final state and category.
However, the impact of this uncertainty on the measurements is found to be negligible.

Other sources of experimental uncertainties are also studied but their impact is negligible compared to the sources described above.

\subsection{Theoretical uncertainties}
\label{subsec:Theoretical_uncertainties}

Theoretical uncertainties that affect both the signal and background estimation include those related to the renormalization and factorization scales, and the choice of the PDF set.
The uncertainty from the renormalization and factorization scales is determined by varying these scales between 0.5 and 2 times their nominal value, while keeping their ratio between 0.5 and 2.
The uncertainty due to the PDF set is determined following the PDF4LHC recommendations by taking the root mean square of the variation of the results when using different replicas of the default NNPDF set~\cite{Botje:2011sn,Alekhin:2011sk}.
The uncertainties just described have an effect both on the signal and background yields, as well as on the migration of events between the categories.
An additional 10\% uncertainty in the K factor used for the $\ggZZ$ prediction is applied as described in Section~\ref{sec:irrbkgd}.
A systematic uncertainty of 2\%~\cite{deFlorian:2016spz} in the branching fraction of $\Hllll$ only affects the signal yield.

Theoretical uncertainties that affect the predictions of the STXS production bins are described in Ref.~\cite{deFlorian:2016spz}.
From here on we will refer to these uncertainties as the theoretical uncertainty scheme (THU).

The THU for the $\ggH$ process includes 9 NPs, which account for uncertainties in the cross section prediction for exclusive jet bins
(including the migration between the 0 and 1-jet, as well as between the 1 and $\geq$2-jet bins), the 2 jet and $\geq$3 jet $\VBF$ phase space,
migrations around the $\pt^{\PH}$ bin boundaries at 10, 60, and 120\GeV, and the uncertainty in the $\pt^{\PH}$ distribution due to missing higher order finite top quark mass corrections.

In the THU uncertainties for $\VBF$ and $\VH$ production, additional sources are introduced to account for the uncertainty in the modeling of the $\pt^{\PH}$,
$m_{\text{jj}}$ and $\pt^{{\PH\text{jj}}}$ distributions, as well as that of the number of jets in the event.
A total of 6 NPs account for the migrations of events across the $m_{\text{jj}}$ boundaries at 60, 120, 350, 700, 1000, and 1500\GeV.
Two additional NPs account for migrations across the $\pt^{\PH} = 200\GeV$ and $\pt^{\PH\text{jj}} = 25\GeV$ bin boundaries.
Finally, a single source is introduced to account for migrations between the zero and one jet, as well as the the two or more jet bins.
In each case, the uncertainty is computed by varying the renormalization and factorization scales and recalculating the fractional breakdown of the $\stxsqqH$ STXS stage 1.2 cross sections.

A set of THU uncertainties is considered as NPs in the likelihood fit when signal strength modifiers, rather than STXS, are measured.
In the STXS framework, THU uncertainties only enter at the interpretation step and are thus applied only to the SM cross section predictions.

Additional theoretical effects that only cause migration of signal and background events between categories originate from the modeling of the hadronization and the underlying event.
The underlying event modeling uncertainty is determined by varying initial- and final-state radiation scales between 0.25 and 4 times their nominal value.
The effects of the modeling of hadronization are determined by simulating additional events with the variation of the nominal \PYTHIA tune described in Section~\ref{sec:datasets}.

\section{Results}
\label{sec:results}

\begin{figure}[!htb]
	\centering
		\includegraphics[width=0.49\textwidth]{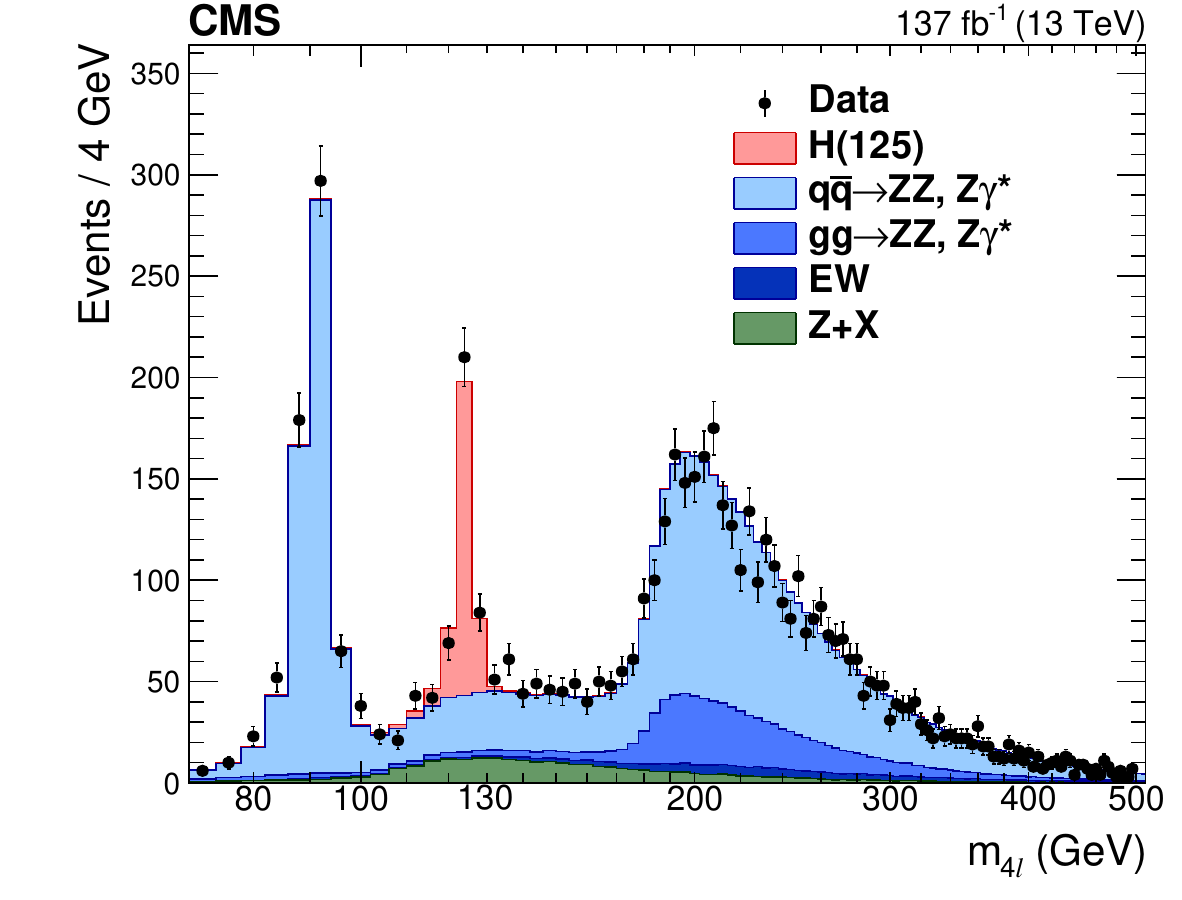}
		\includegraphics[width=0.49\textwidth]{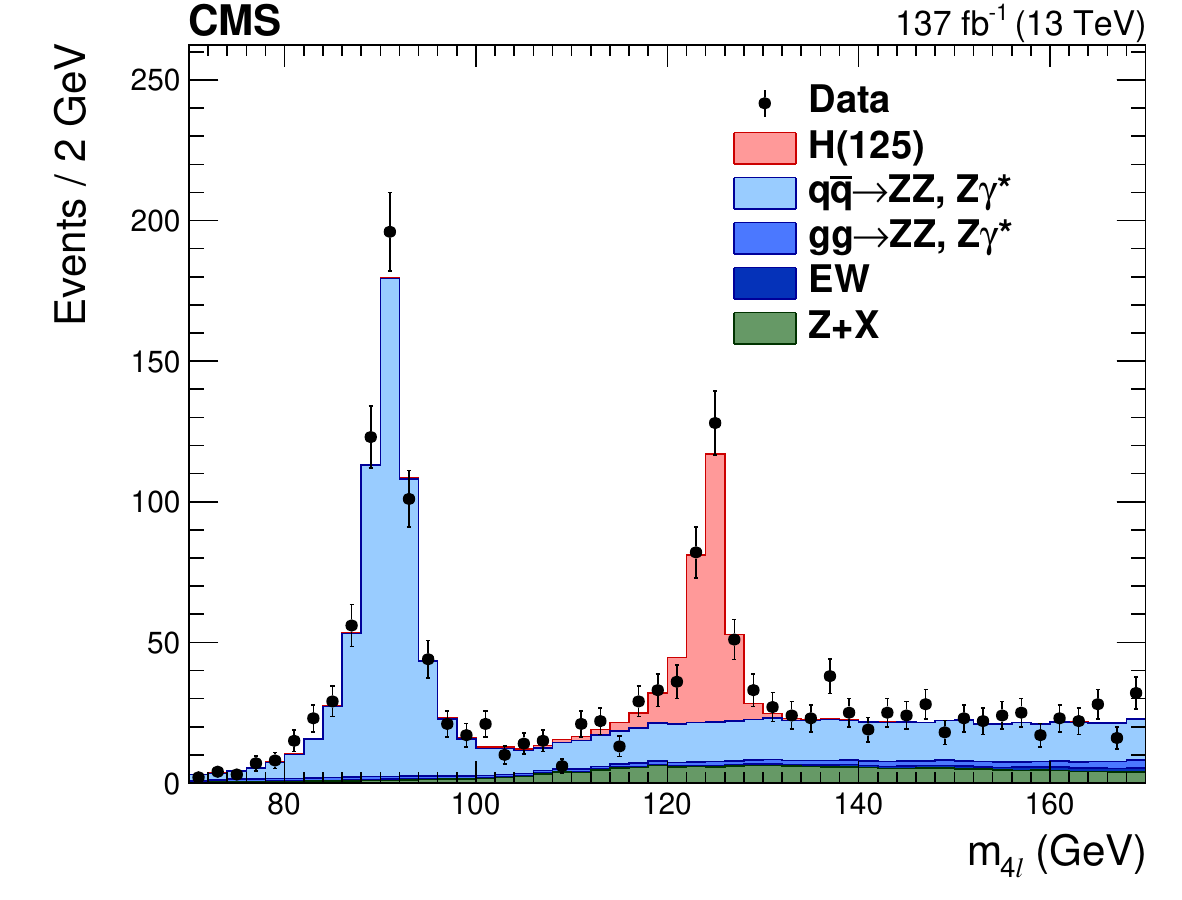}
		\caption{
		Four-lepton mass distribution, $\mllll$, up to $500\GeV$ with $4\GeV$ bin size (\cmsLeft) and in the low-mass range with $2\GeV$ bin size (\cmsRight).
		Points with error bars represent the data and stacked histograms represent the expected distributions for the signal and background processes.
		The SM Higgs boson signal with $\mH=125\GeV$, denoted as $\PH(125)$, the $\PZ\PZ$ and rare electroweak backgrounds are normalized to the SM expectation, the $\PZ$+X background to the estimation from data.
		\label{fig:combination}}
\end{figure}

The reconstructed four-lepton invariant mass distribution is shown in Fig.~\ref{fig:combination} for the $4\Pe$, $4\Pgm$ and $2\Pe2\Pgm$ events together,
and is compared with the expectations for signal and background processes.
The error bars on the data points correspond to the intervals at 68\% confidence level (\CL)~\cite{doi:GARWOOD}.
The observed distribution agrees with the expectation within the statistical uncertainties over the whole spectrum.

\begin{figure}[!htb]
	\centering
		\includegraphics[width=0.45\textwidth]{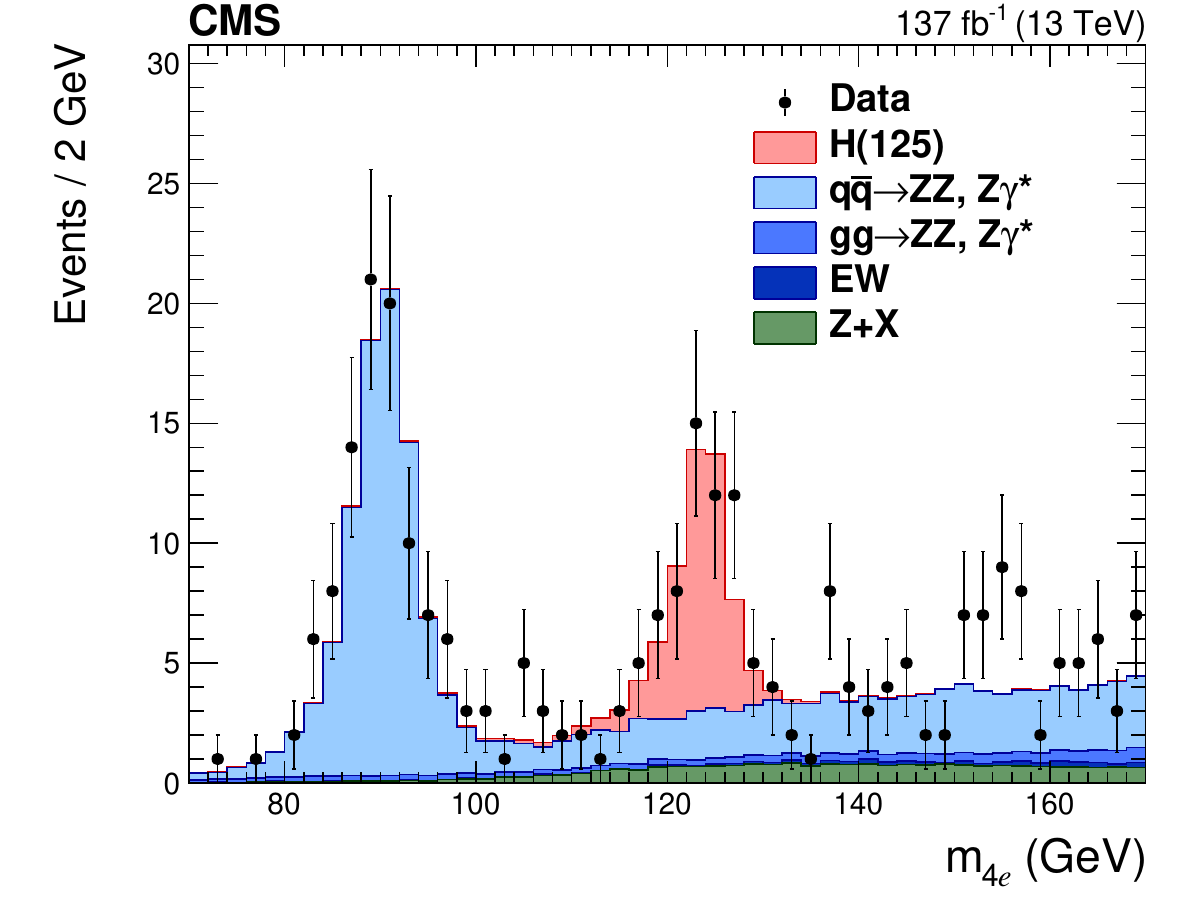}
		\includegraphics[width=0.45\textwidth]{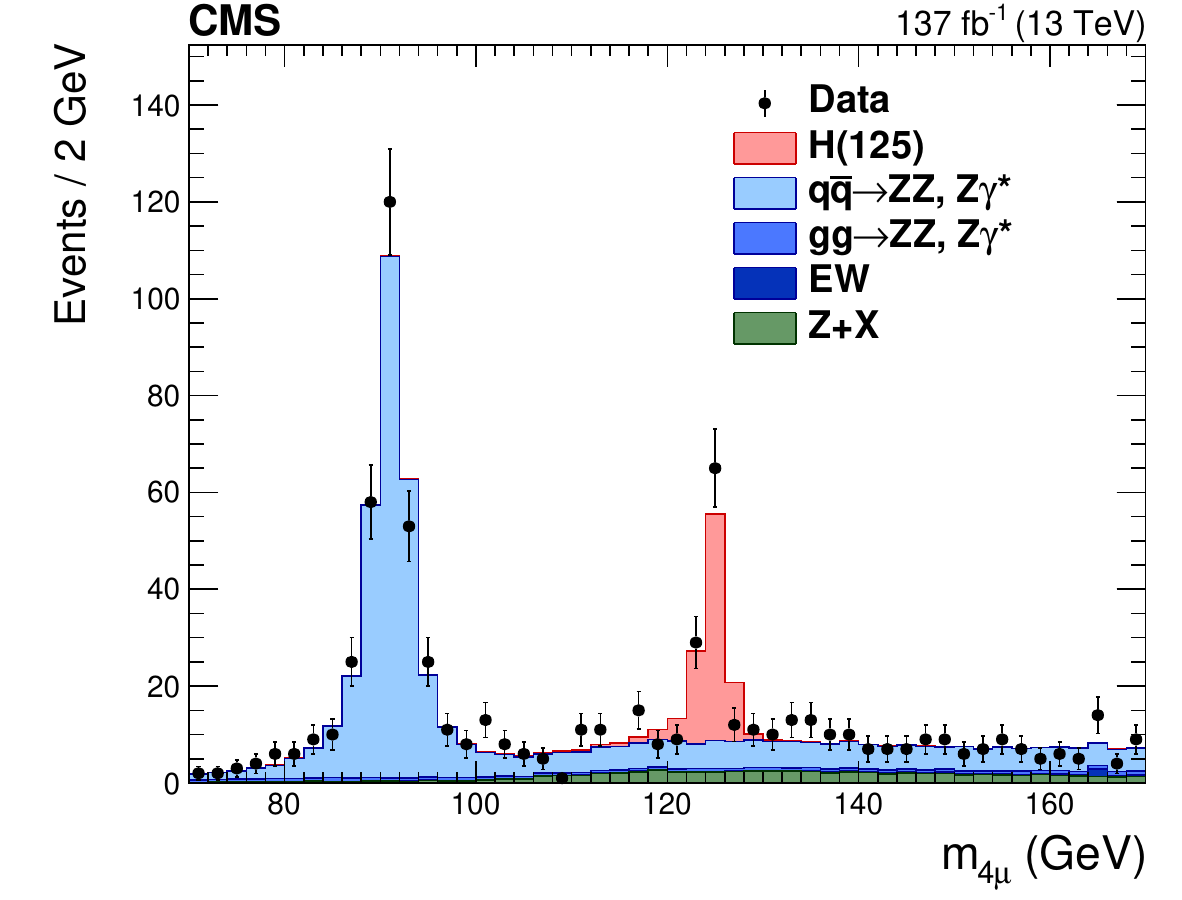}
		\includegraphics[width=0.45\textwidth]{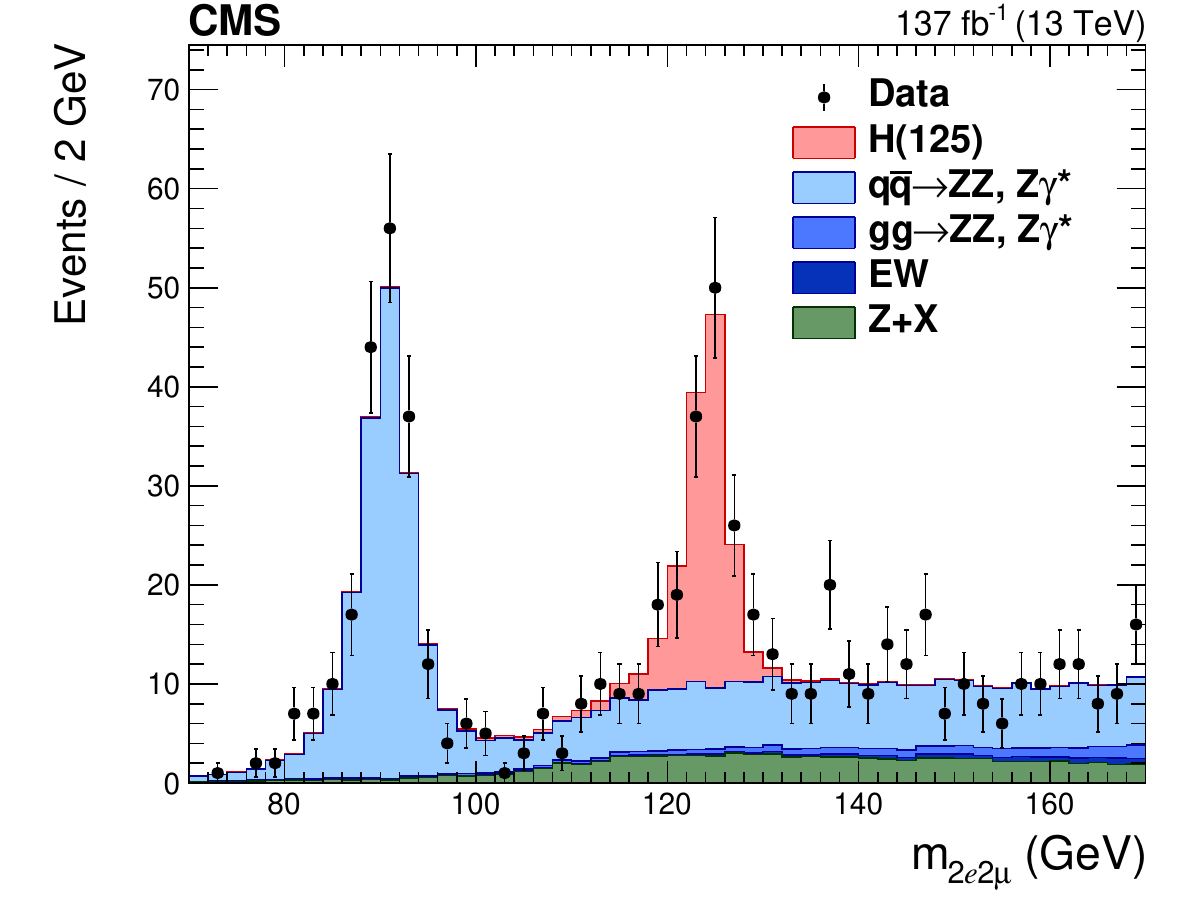}
		\caption{
		Four-lepton mass distribution in three final states: $4\Pe$ \cmsUpperLeft), $4\Pgm$ (\cmsUpperRight), and $2\Pe2\Pgm$ (lower).
		Points with error bars represent the data and stacked histograms represent the expected distributions for the signal and background processes.
		The SM Higgs boson signal with $\mH=125\GeV$, denoted as $\PH(125)$, the $\PZ\PZ$ and rare electroweak backgrounds are normalized to the SM expectation, the $\PZ$+X background to the estimation from data.
		\label{fig:final_states}}
\end{figure}

The reconstructed four-lepton invariant mass distribution is shown in Fig.~\ref{fig:final_states} for the three $4\ell$ final states and is compared with the expectations from signal and background processes.

\begin{figure*}[!htb]
	\centering
		\includegraphics[width=0.9\textwidth]{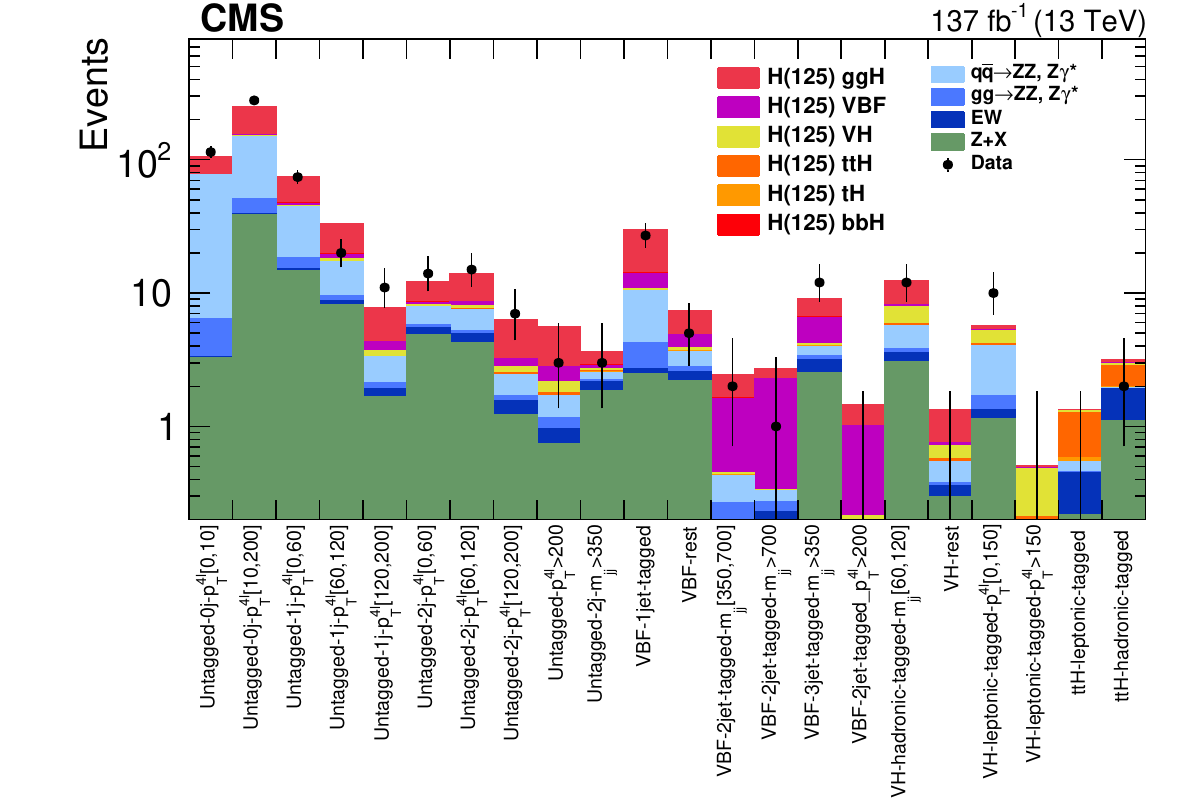}
		\caption{Distributions of the expected and observed number of events for the reconstructed event categories in the mass region $105<\mllll<140\GeV$.
		Points with error bars represent the data and stacked histograms represent the expected numbers of the signal and background events.
		The yields of the different \PH boson production mechanisms with $\mH=125\GeV$, denoted as $\PH(125)$,
		and those of the $\PZ\PZ$ and rare electroweak backgrounds are normalized to the SM expectations,
		while the $\PZ$+X background yield is normalized to the estimate from the data.
			\label{fig:STXS_Categorization}}
\end{figure*}

The number of candidates observed in the data and the expected yields for $\usedLumiABC$, for the backgrounds and \PH boson signal after
the full event selection, are given in Table~\ref{tab:EventYieldsPeakCateg} for each of the 22 reconstructed event categories (described in Section~\ref{subsec:Reco_Categories}) for the $105<\mllll<140\GeV$ mass window around the Higgs boson peak.
Figure~\ref{fig:STXS_Categorization} shows the number of expected and observed events for each of the categories.

\begin{table*}[htb]
	\centering
		\topcaption{Number of expected background and signal events and number of observed candidates after full analysis selection,
		for each event category, in the mass range $105<\mllll<140\GeV$ and for an integrated luminosity of $\usedLumiABC$.
		The yields are given for the different production modes. The uncertainties listed are statistical only.
		Signal is estimated from MC simulation at $\mH=125\GeV$, $\PZ\PZ$ and rare electroweak backgrounds are also estimated from MC simulation, and $\PZ$+X is estimated from data.
		\label{tab:EventYieldsPeakCateg}}
		\cmsTable
		{
		\begin{tabular}{cccccccc|cccc|cc|c}
			Reconstructed event & \multicolumn{7}{c|}{Signal} & \multicolumn{4}{c|}{Background} &  \multicolumn{2}{c|}{Expected} & Observed \\
			category             & $\ggH$ & $\VBF$    & $\WH$    & $\ZH$    & $\ttH$   & $\bbH$   & $\tH$   & $\qqZZ$  & $\ggZZ$  & EW & $\PZ$+X  &signal&total& \\
			\hline
			Untagged-0j-$\pt^{4\ell}[0,10]$ & 27.7 & 0.09 & 0.03 & 0.03 & 0.00 & 0.15 & 0.00 & 71.5 & 3.06 & 0.01 & 3.21 & 27.9$\pm$0.1 & 106$\pm$0 & 114
			\\
			Untagged-0j-$\pt^{4\ell}[10,200]$ & 96.2 & 1.69 & 0.60 & 0.77 & 0.01 & 1.01 & 0.00 & 98.1 & 11.6 & 0.35 & 37.8 & 100$\pm$0 & 248$\pm$1 & 278
			\\
			Untagged-1j-$\pt^{4\ell}[0,60]$ & 26.8 & 1.51 & 0.56 & 0.48 & 0.01 & 0.45 & 0.01 & 25.3 & 3.02 & 0.64 & 14.2 & 29.8$\pm$0.1 & 72.9$\pm$0.4 & 74
			\\
			Untagged-1j-$\pt^{4\ell}[60,120]$ & 13.5 & 1.31 & 0.51 & 0.41 & 0.02 & 0.11 & 0.01 & 7.81 & 0.82 & 0.62 & 7.95 & 15.9$\pm$0.1 & 33.1$\pm$0.3 & 20
			\\
			Untagged-1j-$\pt^{4\ell}[120,200]$ & 3.51 & 0.60 & 0.17 & 0.17 & 0.01 & 0.02 & 0.00 & 1.15 & 0.19 & 0.25 & 1.63 & 4.48$\pm$0.05 & 7.69$\pm$0.16 & 11
			\\
			Untagged-2j-$\pt^{4\ell}[0,60]$ & 3.45 & 0.29 & 0.15 & 0.14 & 0.08 & 0.09 & 0.02 & 2.14 & 0.32 & 0.63 & 4.75 & 4.20$\pm$0.06 & 12.1$\pm$0.2 & 14
			\\
			Untagged-2j-$\pt^{4\ell}[60,120]$ & 5.26 & 0.56 & 0.24 & 0.19 & 0.12 & 0.04 & 0.03 & 2.19 & 0.30 & 0.72 & 4.14 & 6.43$\pm$0.06 & 13.8$\pm$0.2 & 15
			\\
			Untagged-2j-$\pt^{4\ell}[120,200]$ & 3.07 & 0.40 & 0.16 & 0.13 & 0.07 & 0.01 & 0.02 & 0.75 & 0.14 & 0.34 & 1.19 & 3.86$\pm$0.05 & 6.28$\pm$0.14 & 7
			\\
			Untagged-$\pt^{4\ell}>200$ & 2.79 & 0.62 & 0.21 & 0.17 & 0.07 & 0.01 & 0.02 & 0.43 & 0.21 & 0.21 & 0.73 & 3.89$\pm$0.04 & 5.47$\pm$0.11 & 3
			\\
			Untagged-2j-$m_{\text{jj}}>350$ & 0.77 & 0.16 & 0.06 & 0.04 & 0.05 & 0.01 & 0.01 & 0.34 & 0.06 & 0.31 & 1.71 & 1.12$\pm$0.02 & 3.54$\pm$0.14 & 3
			\\
			$\VBF$-1jet-tagged & 15.5 & 3.29 & 0.22 & 0.16 & 0.00 & 0.13 & 0.01 & 6.85 & 1.53 & 0.20 & 2.44 & 19.3$\pm$0.1 & 30.3$\pm$0.2 & 27
			\\
			$\VBF$-2jet-tagged-$m_{\text{jj}}[350,700]$ & 0.83 & 1.19 & 0.01 & 0.01 & 0.00 & 0.01 & 0.00 & 0.19 & 0.07 & 0.11 & 0.14 & 2.05$\pm$0.03 & 2.55$\pm$0.05 & 2
			\\
			$\VBF$-2jet-tagged-$m_{\text{jj}}>700$ & 0.43 & 1.96 & 0.00 & 0.00 & 0.00 & 0.00 & 0.00 & 0.07 & 0.05 & 0.12 & 0.03 & 2.40$\pm$0.02 & 2.67$\pm$0.03 & 1
			\\
			$\VBF$-3jet-tagged-$m_{\text{jj}}>350$ & 2.52 & 2.35 & 0.06 & 0.06 & 0.03 & 0.03 & 0.05 & 0.62 & 0.21 & 0.64 & 2.43 & 5.11$\pm$0.05 & 9.01$\pm$0.17 & 12
			\\
			$\VBF$-2jet-tagged-$\pt^{4\ell}>200$ & 0.44 & 0.79 & 0.01 & 0.01 & 0.01 & 0.00 & 0.01 & 0.03 & 0.03 & 0.04 & 0.06 & 1.26$\pm$0.02 & 1.42$\pm$0.03 & 0
			\\
			VBF-rest & 2.48 & 0.94 & 0.13 & 0.09 & 0.04 & 0.04 & 0.01 & 0.98 & 0.20 & 0.39 & 2.18 & 3.74$\pm$0.05 & 7.49$\pm$0.17 & 5
			\\
			$\VH$-hadronic-tagged-$m_{\text{jj}}[60,120]$ & 4.11 & 0.25 & 1.09 & 0.96 & 0.13 & 0.06 & 0.02 & 1.69 & 0.22 & 0.52 & 2.93 & 6.62$\pm$0.06 & 12.0$\pm$0.2 & 12
			\\
			VH-rest & 0.57 & 0.03 & 0.09 & 0.06 & 0.03 & 0.01 & 0.00 & 0.16 & 0.02 & 0.06 & 0.33 & 0.79$\pm$0.02 & 1.36$\pm$0.06 & 0
			\\
			$\VH$-leptonic-tagged-$\pt^{4\ell}[0,150]$ & 0.33 & 0.04 & 0.85 & 0.26 & 0.10 & 0.03 & 0.03 & 2.16 & 0.36 & 0.19 & 1.11 & 1.64$\pm$0.02 & 5.47$\pm$0.13 & 10
			\\
			$\VH$-leptonic-tagged-$\pt^{4\ell}>150$ & 0.02 & 0.01 & 0.21 & 0.06 & 0.04 & 0.00 & 0.01 & 0.05 & 0.01 & 0.03 & 0.08 & 0.35$\pm$0.01 & 0.52$\pm$0.03 & 0
			\\
			$\ttH$-leptonic-tagged & 0.02 & 0.01 & 0.02 & 0.02 & 0.68 & 0.00 & 0.03 & 0.08 & 0.01 & 0.23 & 0.21 & 0.79$\pm$0.01 & 1.32$\pm$0.07 & 0
			\\
			$\ttH$-hadronic-tagged & 0.18 & 0.05 & 0.03 & 0.05 & 0.86 & 0.01 & 0.03 & 0.03 & 0.01 & 0.82 & 1.06 & 1.22$\pm$0.01 & 3.15$\pm$0.14 & 2
	\end{tabular}}
\end{table*}

\begin{figure*}[!htb]
	\centering
		\includegraphics[width=0.4\textwidth]{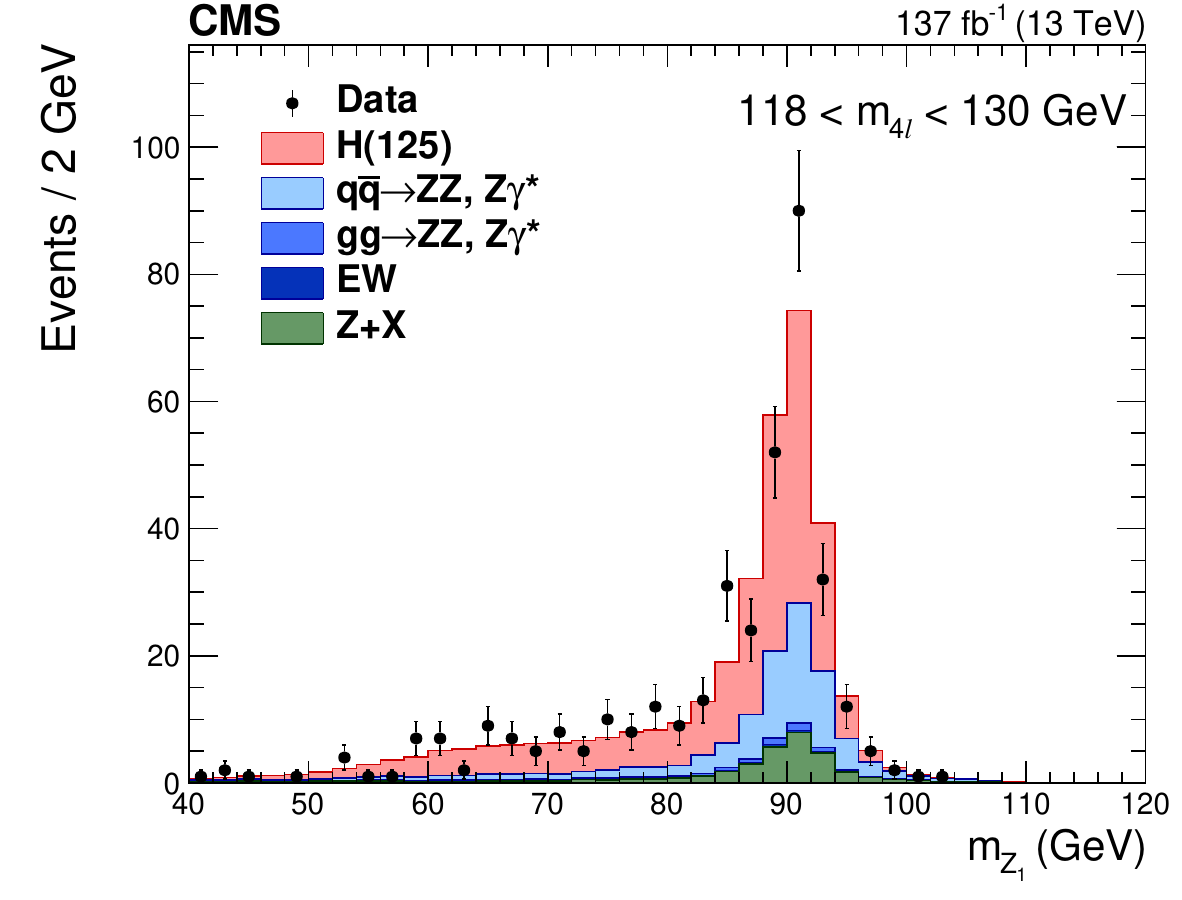}
		\includegraphics[width=0.4\textwidth]{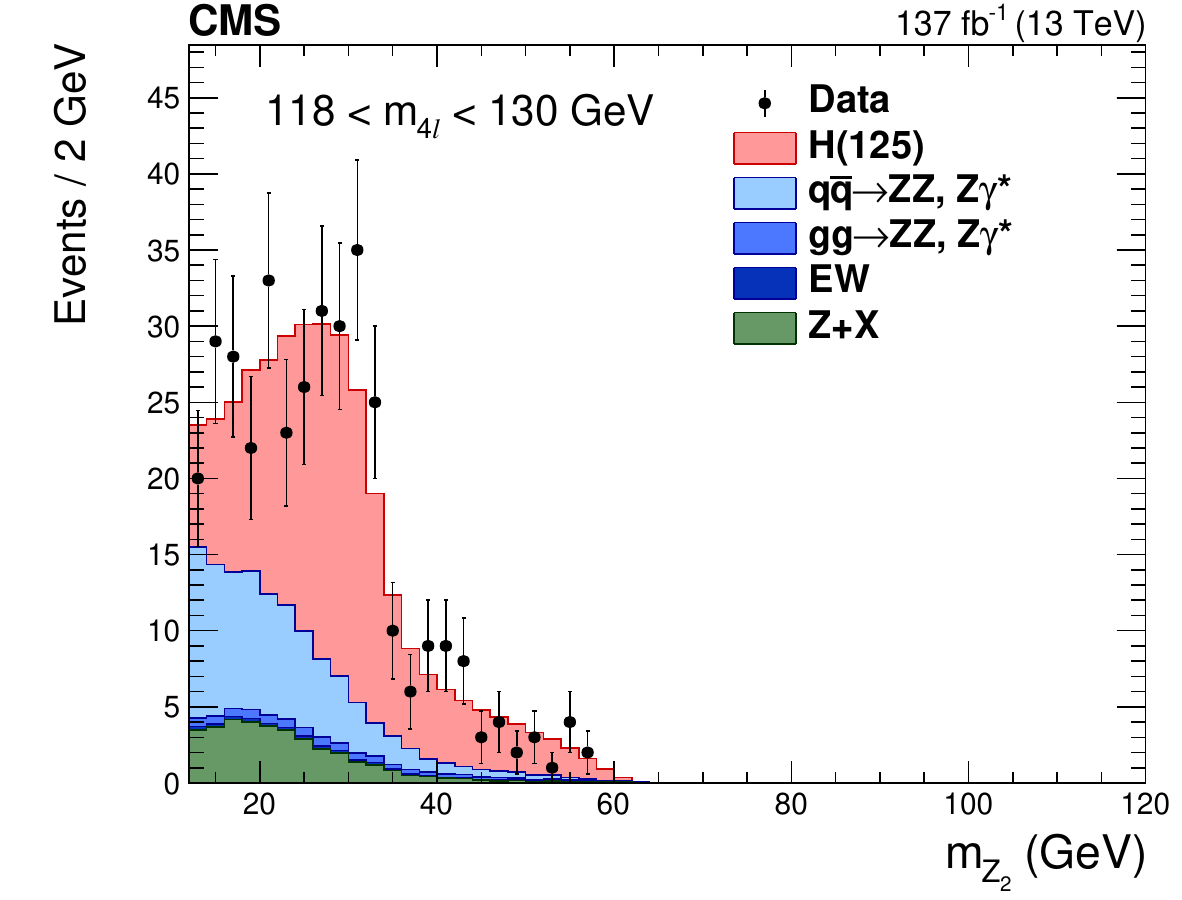}\\
		\includegraphics[width=0.6\textwidth]{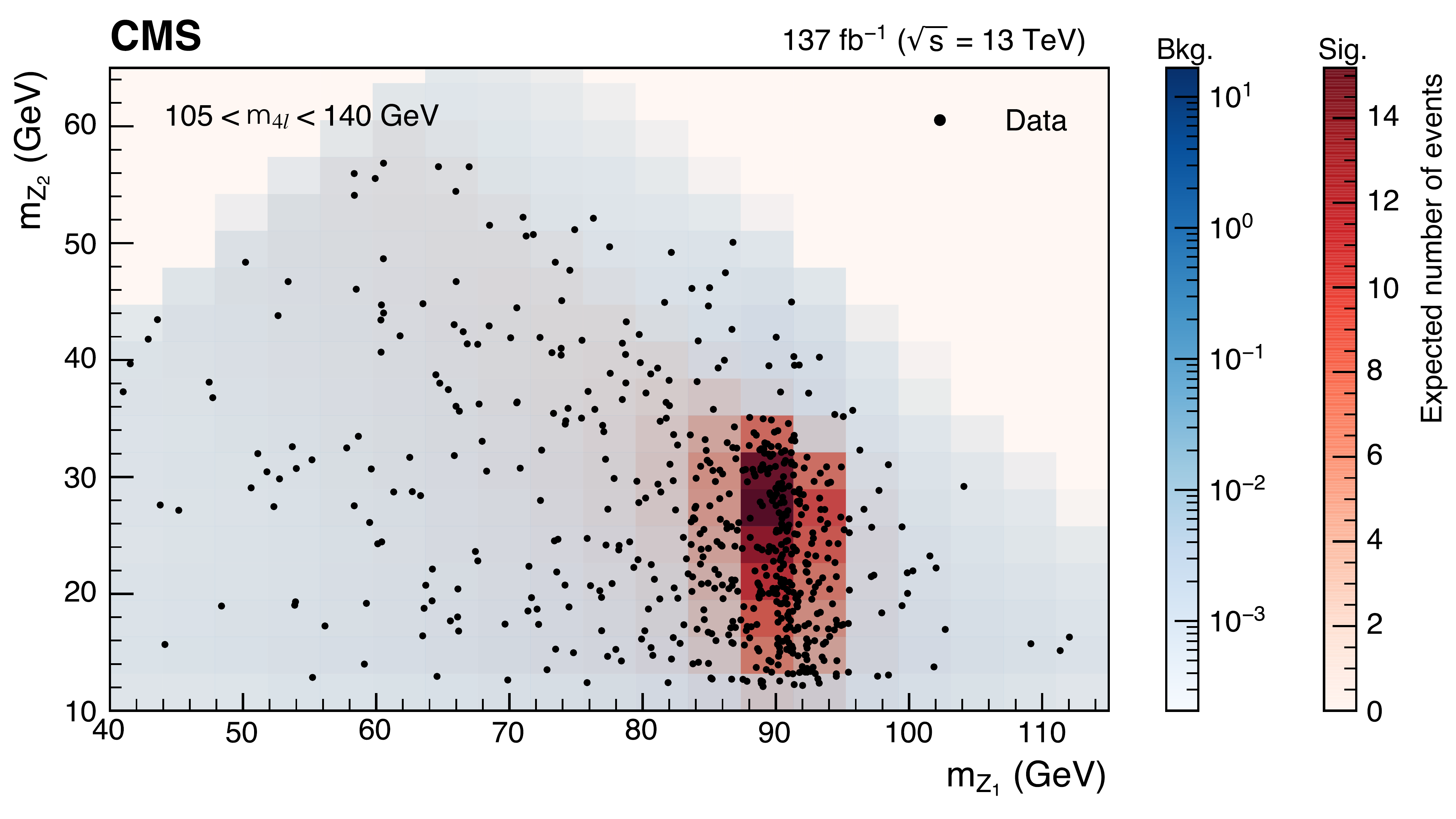}
		\caption{
		Distribution of the $\PZ_1$ (upper left) and $\PZ_2$ (upper right) reconstructed masses in the $118<\mllll<130\GeV$ mass region and their 2D distribution (lower) in the $105<\mllll<140\GeV$ mass region.
		The stacked histograms and the red and blue scales represent expected distributions of the signal and background processes and the points represent the data.
		The yields of the different \PH boson production mechanisms with $\mH=125\GeV$, denoted as $\PH(125)$,
		and those of the $\PZ\PZ$ and rare electroweak backgrounds are normalized to the SM expectations,
		while the $\PZ$+X background yield is normalized to the estimate from the data.
		\label{fig:MZ1MZ2}}
\end{figure*}

\begin{figure}[!htb]
	\centering
		\includegraphics[width=0.45\textwidth]{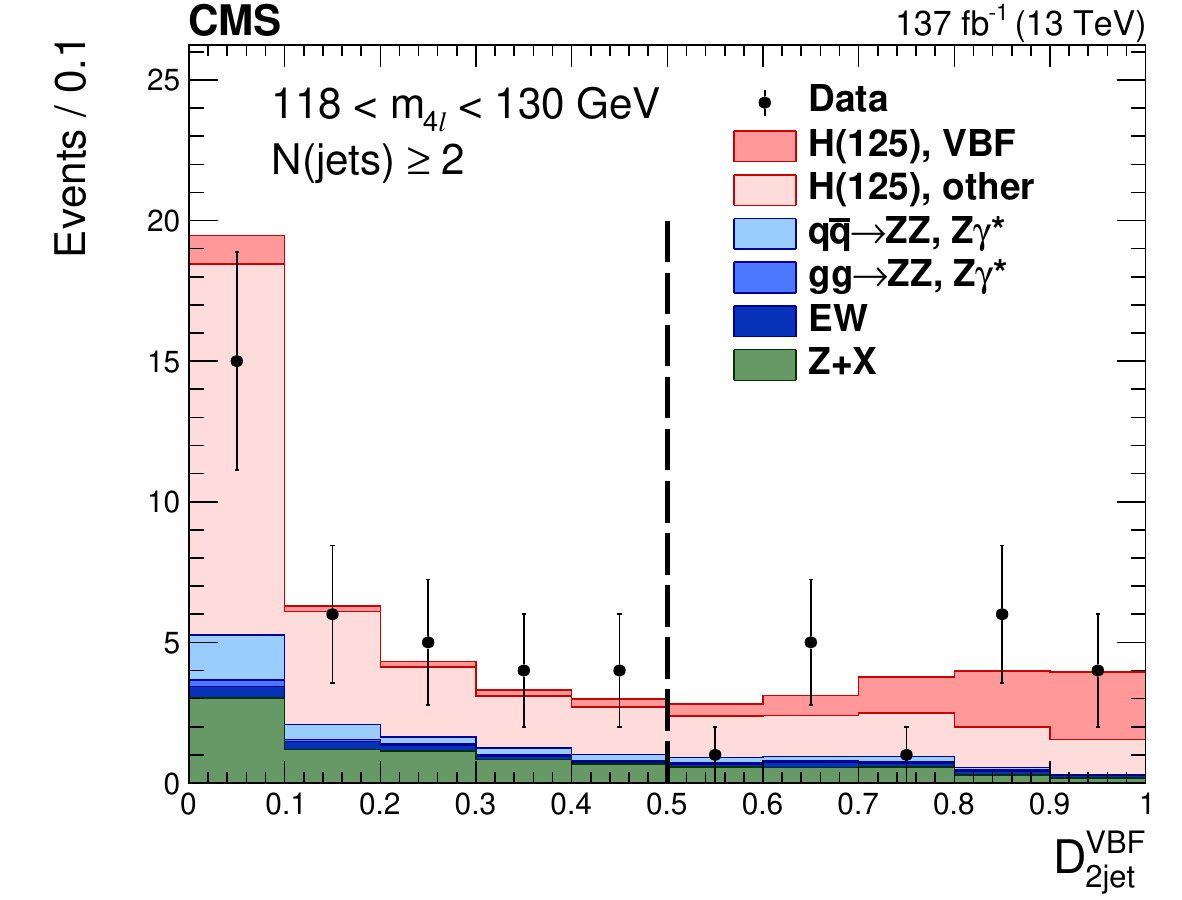}
		\includegraphics[width=0.45\textwidth]{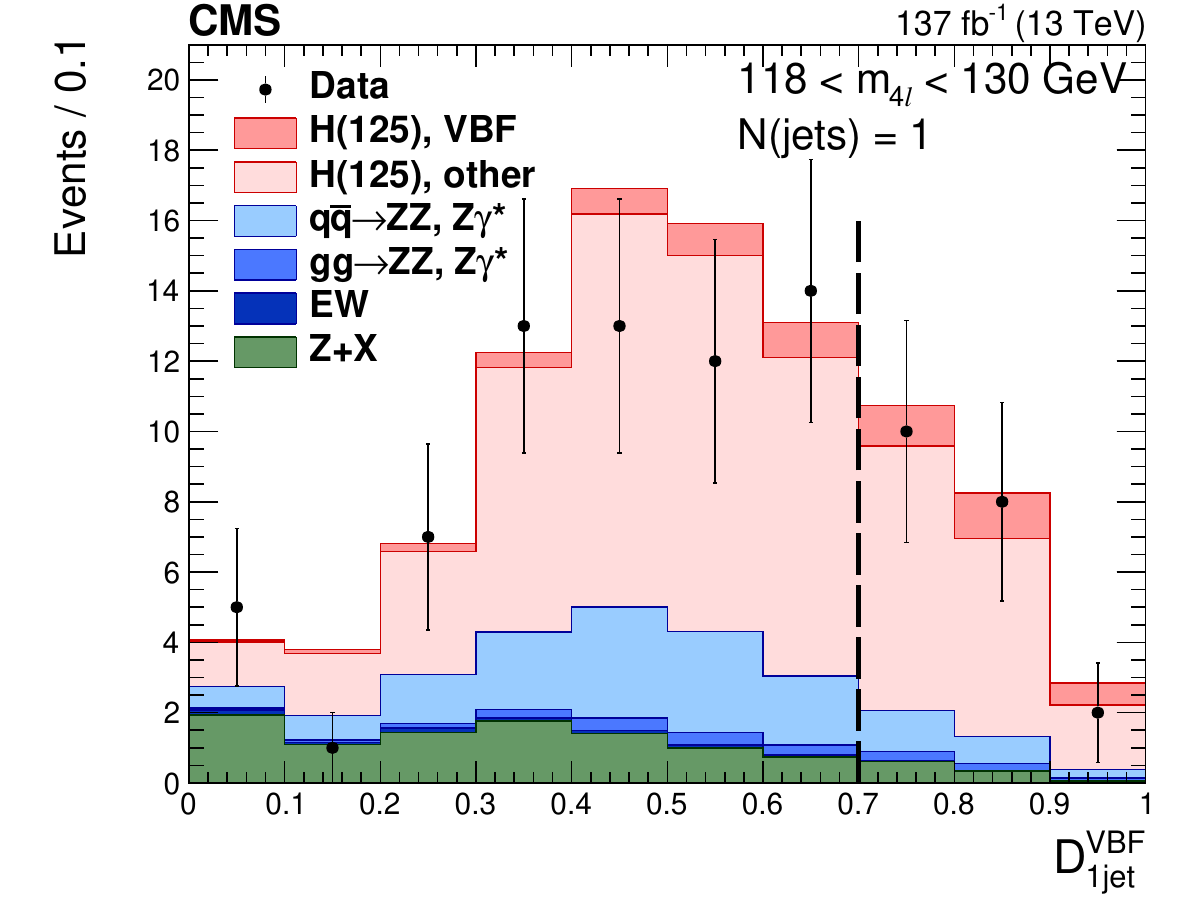}
		\includegraphics[width=0.45\textwidth]{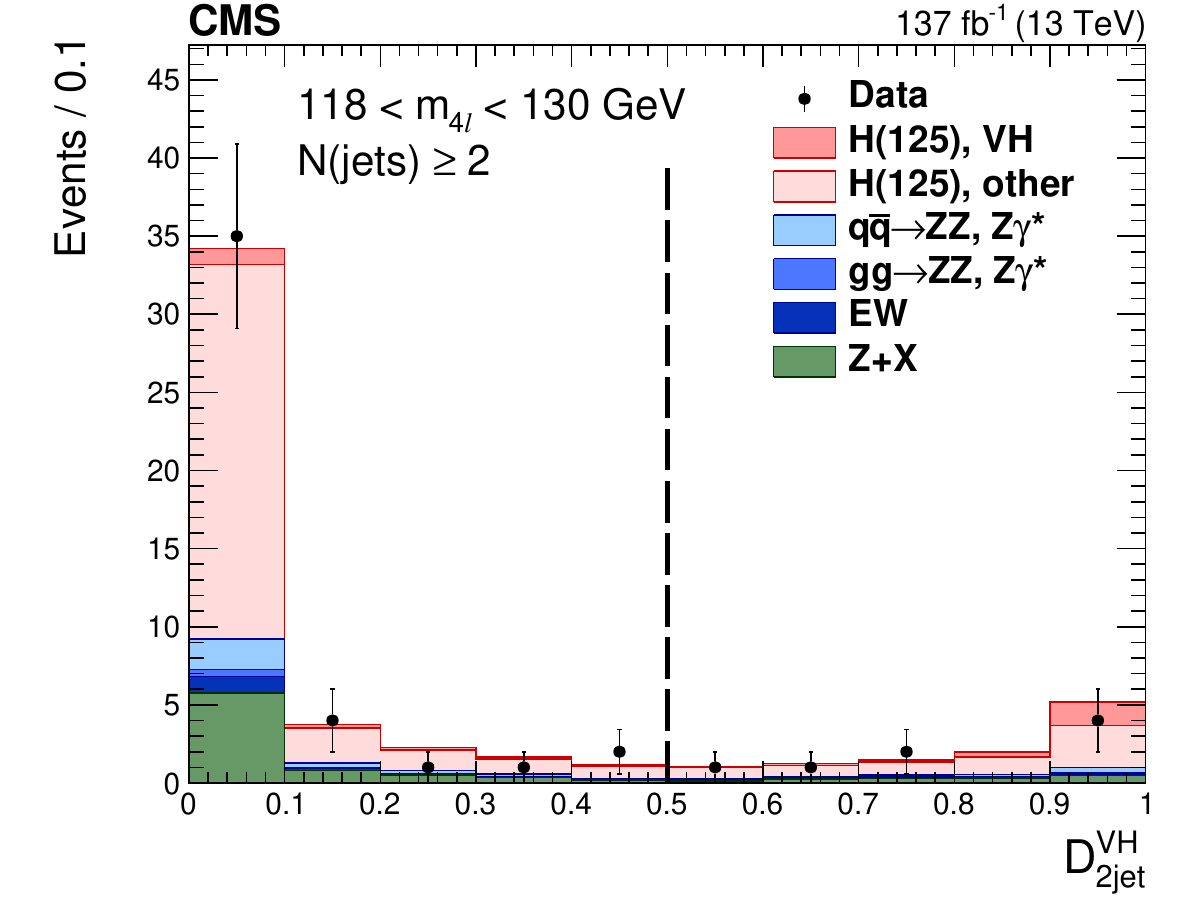}
		\caption{Distribution of categorization discriminants in the mass region $118<\mllll<130\GeV$: \DMeVbfjj (\cmsUpperLeft),
    \DMeVbfj (\cmsUpperRight), \DMeVh (lower) = max(\DMeWh,\DMeZh).
    Points with error bars represent the data and stacked histograms represent expected distributions of the signal and background processes.
    The SM Higgs boson signal with $\mH=125\GeV$, denoted as $\PH(125)$, and the $\PZ\PZ$ backgrounds and rare electroweak backgrounds are normalized to the SM expectation, the $\PZ$+X background to the estimation from data.
    The vertical dashed lines denote the working points used in the event categorization.
    The SM \PH boson signal is separated into two components: the production mode which is targeted by the specific discriminant, and other production modes, where the gluon fusion process dominates.
			\label{fig:Djet}}
\end{figure}

The reconstructed invariant masses of the $\PZ_1$ and $\PZ_2$ dilepton systems are shown in Fig.~\ref{fig:MZ1MZ2} for $118<\mllll<130\GeV$,
together with their 2D distribution in the $105<\mllll<140\GeV$ mass region.
The distribution of the discriminants used for event categorization along with the corresponding working point values are shown in Fig.~\ref{fig:Djet}.

The results presented in Sections~\ref{subsec:signalstrenghts} and \ref{subsec:stxs} are extracted with a two-dimensional likelihood fit that relies on two variables, the four-lepton invariant mass $\mllll$ and the matrix element kinematic discriminant $\mathcal{D}$.
The fiducial cross section measurements are extracted with a one-dimensional likelihood fit that relies only on the four-lepton invariant mass. The fit procedure and results are presented in Section~\ref{sec:crosssections}.
The fit is performed in the $105<\mllll<140\GeV$ mass region.
The parameters of interest (POIs) are estimated with their corresponding confidence intervals using a profile likelihood ratio test statistic~\cite{Cowan_2011,LHC-HCG},
in which the experimental and theoretical uncertainties are incorporated via NPs.
The choice of the POIs depends on the specific measurement under consideration,
while the remaining parameters are treated as NPs.
All the POIs considered in the analysis are forced to be greater than or equal to zero;
this reflects the fact that the signal yield is substantially larger than the background yield in the mass range studied.
Negative POIs would imply negative signal strength modifiers and a negative probability density function (pdf).
We define a two-dimensional pdf as the product of two one-dimensional pdfs:
\begin{linenomath*}
\begin{equation}
f(\mllll,\mathcal{D}) = \mathcal{P}(\mllll) \mathcal{P}(\mathcal{D}|\mllll).
\end{equation}
\end{linenomath*}
The first term, $\mathcal{P}(\mllll)$, is the unbinned analytical shape described in Section~\ref{sec:signal} for signals and Section~\ref{sec:bkgd} for backgrounds.
The second term, $\mathcal{P}(\mathcal{D}|\mllll)$, is a binned template of $\mathcal{D}$ that is conditional to $\mllll$.
This is achieved by creating a two-dimensional template of $\mllll$ vs. $\mathcal{D}$ and normalizing it to 1 for each bin of $\mllll$.

In almost all sub-categories we use a decay-only kinematic discriminant ($\mathcal{D}=\KD$) to separate the \PH boson signal from the background as defined in Eq.~(\ref{eq:ggmela}).
Conversely, in the sub-categories of the VBF-2jet-tagged, the $\mathcal{D}=\DbkgVBFdec$ discriminant (defined in Eq.~(\ref{eq:vbfmela})) is used, which is sensitive to the VBF production mechanism.
Similarly, in two sub-categories of the VH-hadronic-tagged category, the $\mathcal{D}=\DbkgVHdec$ discriminant (defined in Eq.~(\ref{eq:vhmela})) is used.

The $\ggH$, $\VBF$, $\WH$, $\ZH$ and $\ttH$ samples are used to build different signal templates for each of the nineteen STXS production bins described in Section~\ref{subsec:STXS_Categories}.
Irreducible background templates are built starting from $\qqZZ$ and $\ggZZ$ samples.
Finally, reducible background templates are built using data driven methods described in Section~\ref{sec:redbkgd}.
Following the described procedure, $\mathcal{P}(\mathcal{D}|\mllll)$ templates are obtained for the twenty-two event categories and the three final states ($4\mu$, $4\Pe$, $2\Pe2\mu$).

The unbinned likelihood function, $\mathcal{L}(\vec{\mu})$, is defined as the product over $N$ observed events:
\begin{linenomath*}
\ifthenelse{\boolean{cms@external}}
{
\begin{multline}
\mathcal{L}(\vec{\mu}) = \frac{1}{N} \prod_{\text{events}} \Bigg( \sum_{i=1}^{19} \mu_i S_i^{jk} f_{S}^{ijk}(\mllll,\mathcal{D}) +\\
 B^{jk} f_{B}^{jk}(\mllll,\mathcal{D})\Bigg)\re^{-\sum_{i} \mu_i S_i^{jk} + B^{jk}},
\end{multline}
}
{
\begin{equation}
\mathcal{L}(\vec{\mu}) = \frac{1}{N} \prod_{\text{events}} \left( \sum_{i=1}^{19} \mu_i S_i^{jk} f_{S}^{ijk}(\mllll,\mathcal{D}) + B^{jk} f_{B}^{jk}(\mllll,\mathcal{D})\right)\re^{-\sum_{i} \mu_i S_i^{jk} + B^{jk}},
\end{equation}
}
\end{linenomath*}
where $\mu_i$ is the signal strength modifier for the production bin $i$,
$S_i^{jk}$ are the predicted SM rates of events in the production bin $i$ that are observed in the reconstructed event category $j$ and final state $k$,
$B^{jk}$ are the predicted background rates in the reconstructed event category $j$ and final state $k$,
$f_{S}^{ijk} (\mllll,\mathcal{D})$ are the pdfs for the signal,
and $f_{B}^{jk}(\mllll,\mathcal{D})$ the pdfs for the background.

\begin{figure*}[!htbp]
\centering
\includegraphics[width=0.8\textwidth]{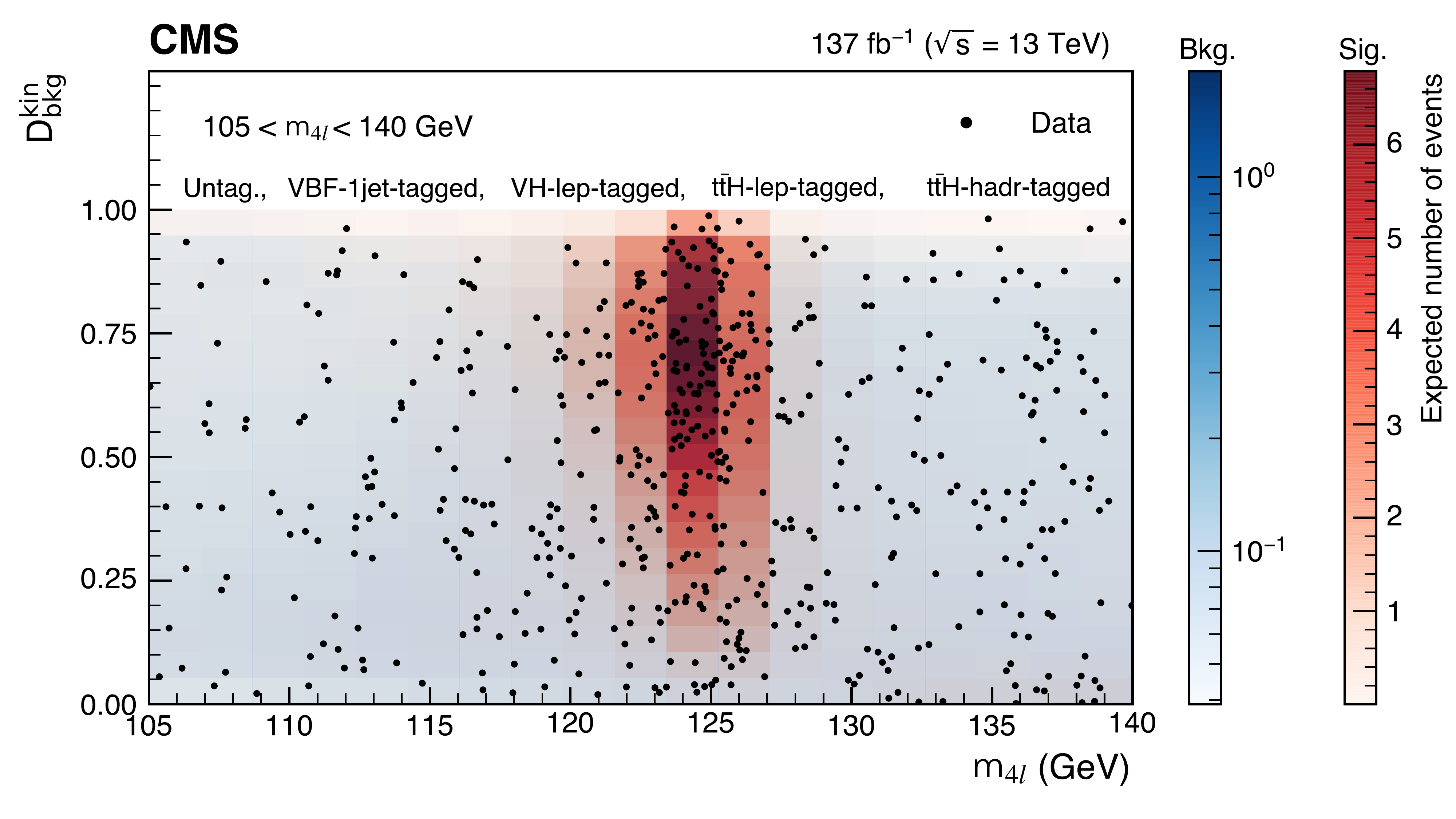} \\
\includegraphics[width=0.8\textwidth]{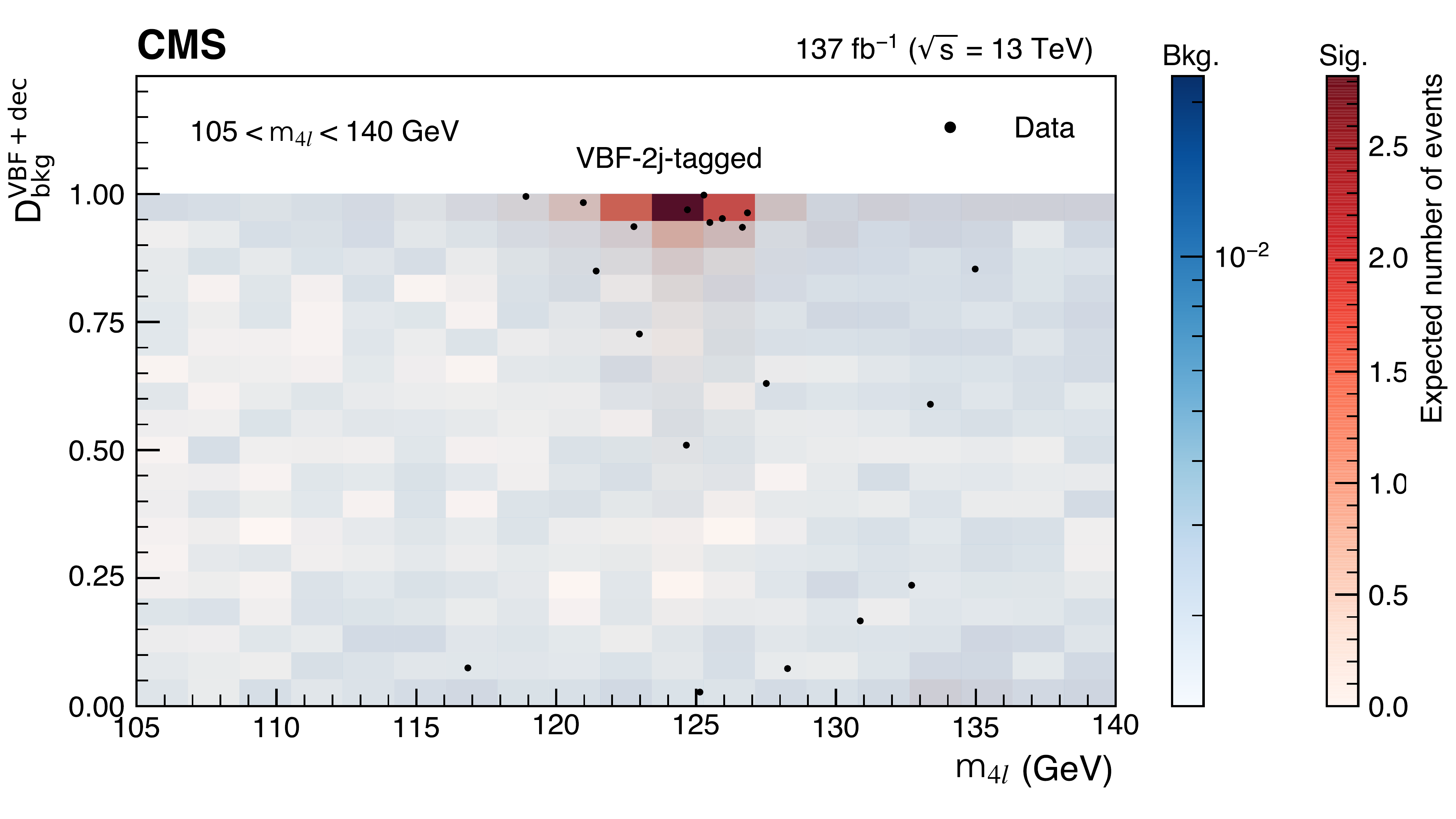}\\
\includegraphics[width=0.8\textwidth]{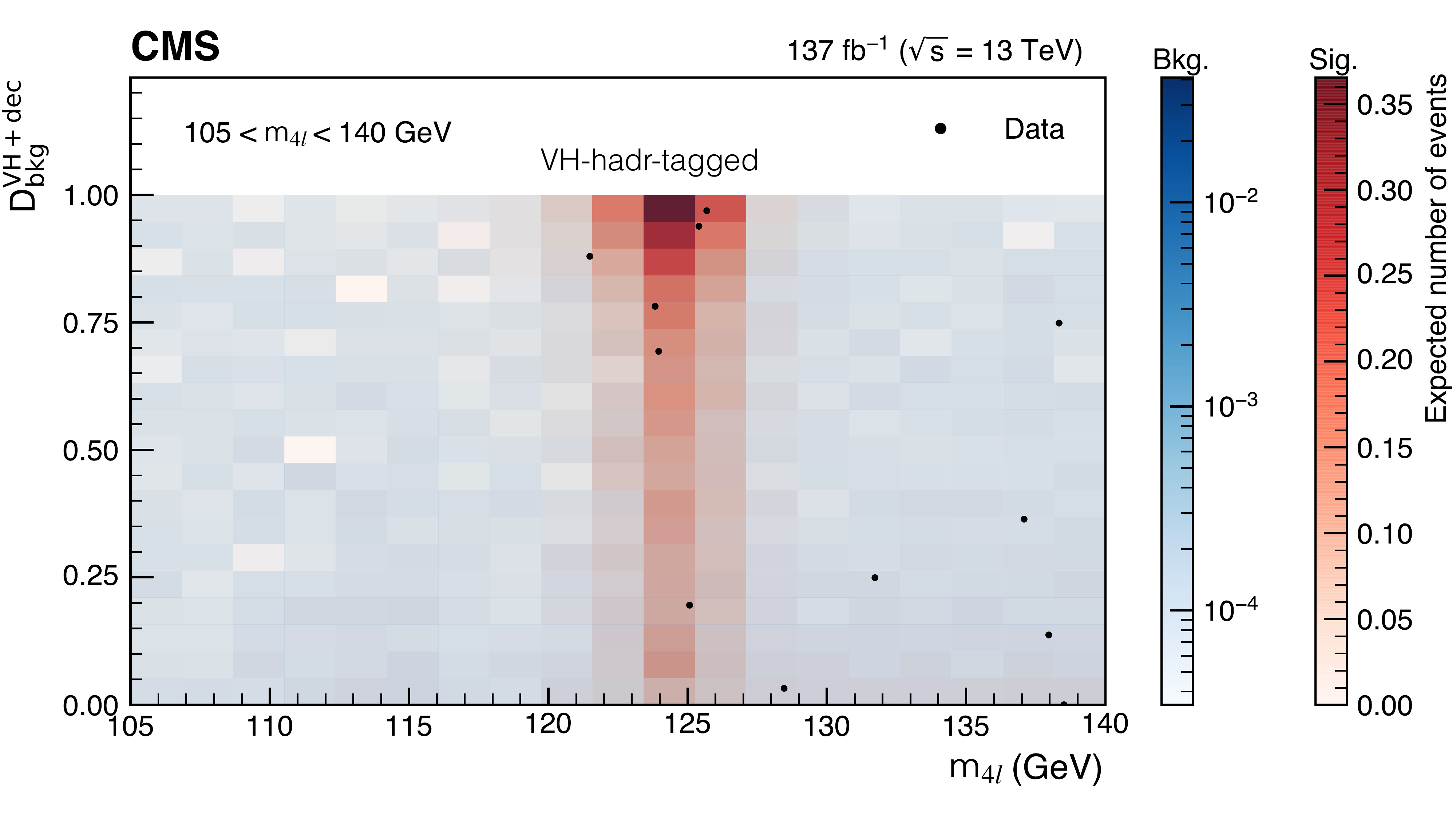}
\caption{
Distribution of three different kinematic discriminants versus $\mllll$: $\KD$ (upper), $\DbkgVBFdec$ (middle) and $\DbkgVHdec$ (lower) shown in the mass region $105<\mllll<140\GeV$.
The blue scale represents the expected total number of $\PZ\PZ$, rare electroweak, and $\PZ$+X background events.
The red scale represents the number of expected SM \PH boson signal events for $\mH=125\GeV$.
The points show the data from the categories listed in the legend.
\label{fig:KD}}
\end{figure*}

\begin{figure}[!htb]
\centering
\includegraphics[width=0.45\textwidth]{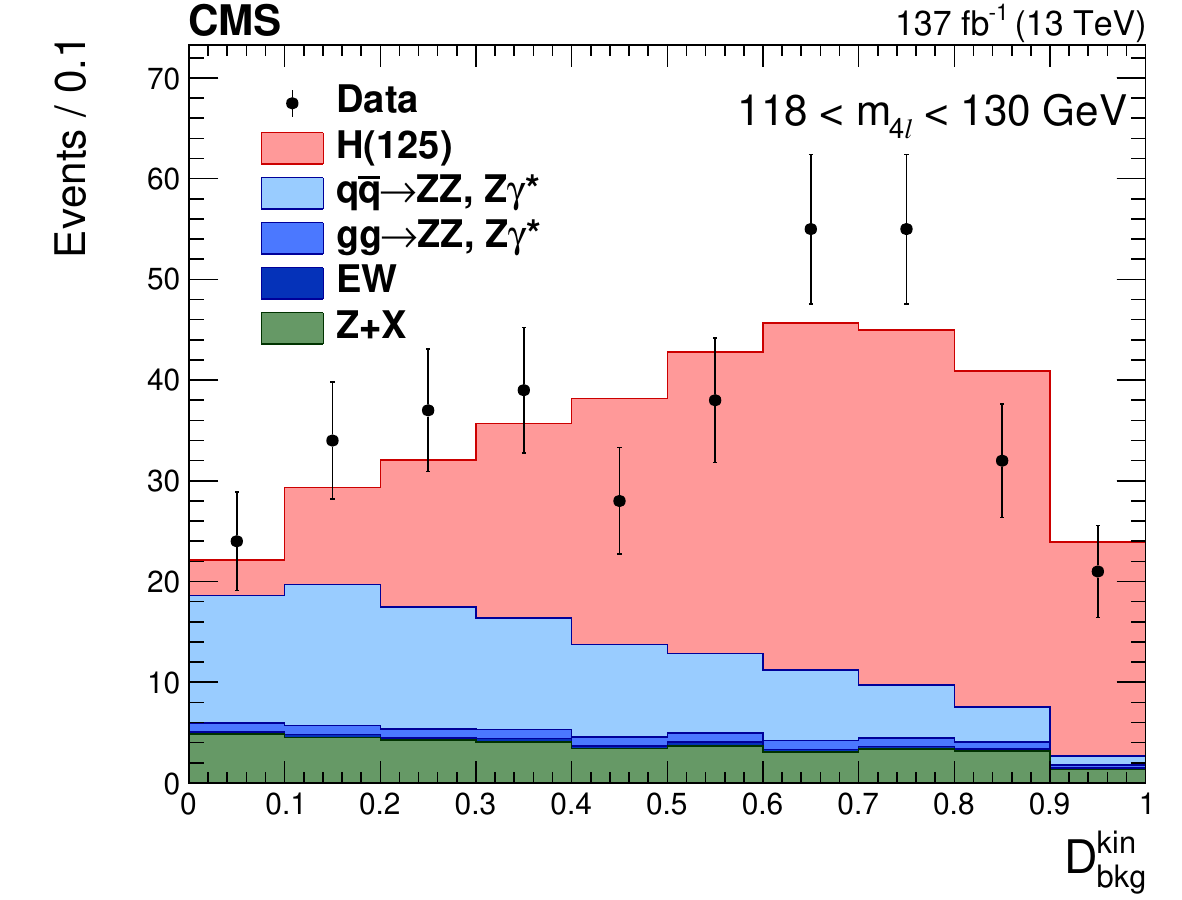}
\includegraphics[width=0.45\textwidth]{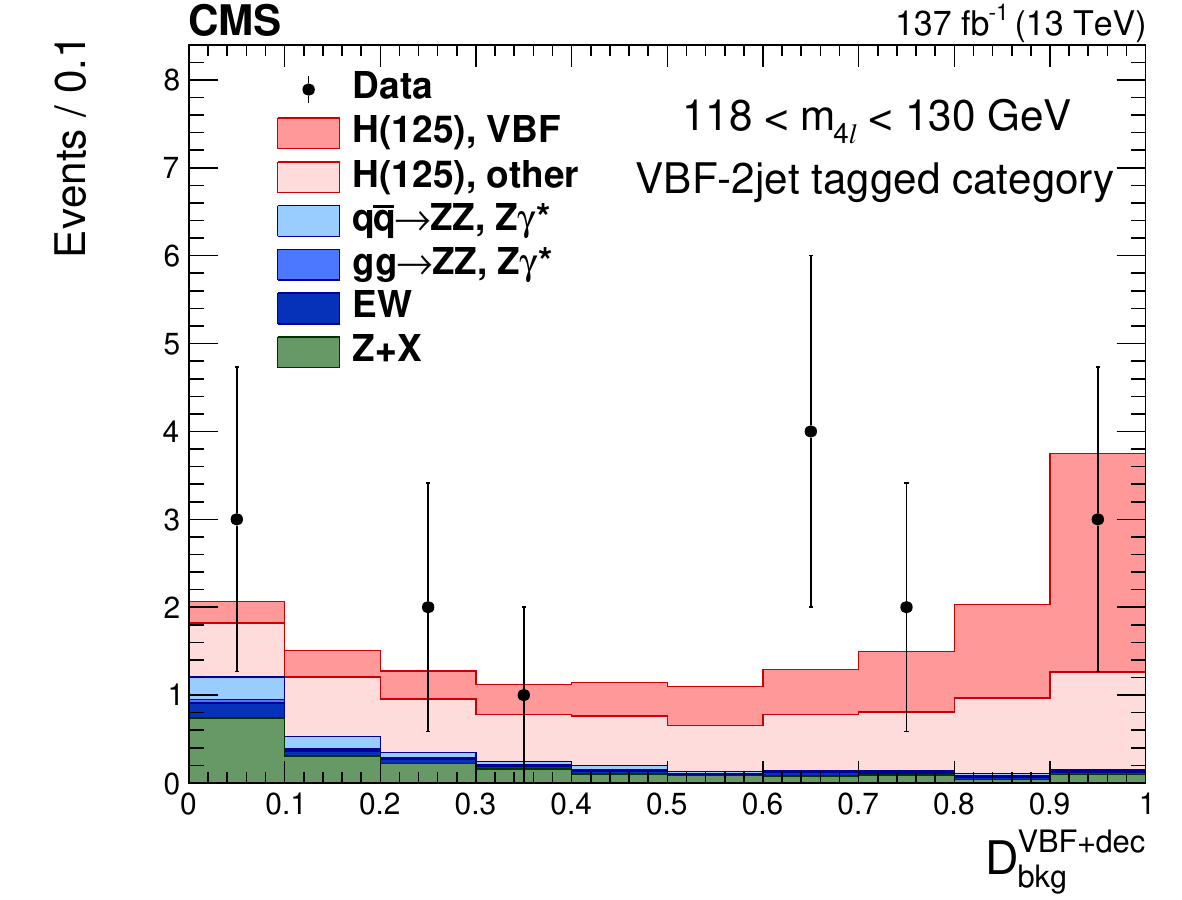}
\includegraphics[width=0.45\textwidth]{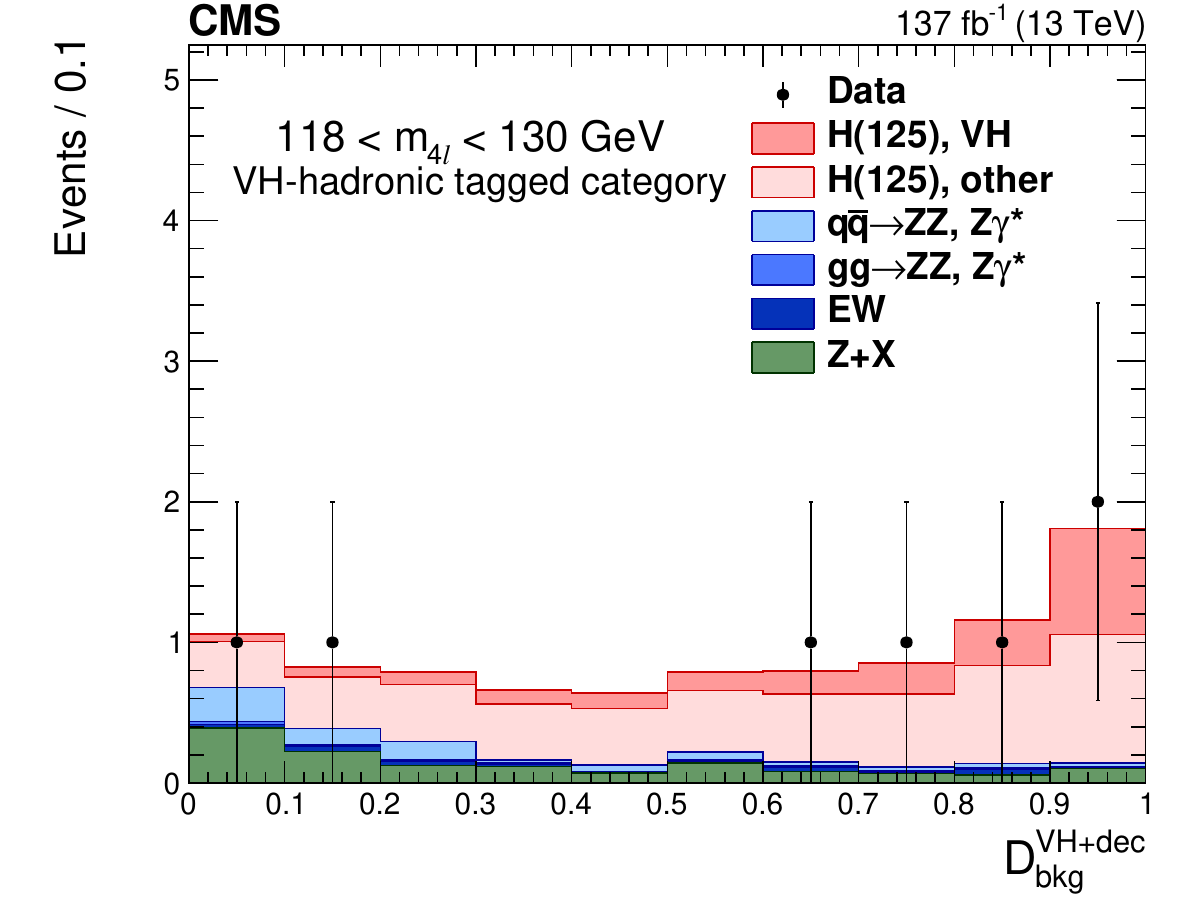}
\caption{Distribution of kinematic discriminants in the mass region $118<\mllll<130\GeV$:
(\cmsUpperLeft) \KD,
(\cmsUpperRight) \DbkgVBFdec,
(lower) \DbkgVHdec.
Points with error bars represent the data and stacked histograms represent expected distributions of the signal and background processes.
The yields of the different \PH boson production mechanisms with $\mH=125\GeV$, denoted as $\PH(125)$,
and those of the $\PZ\PZ$ and rare electroweak backgrounds are normalized to the SM expectations,
while the $\PZ$+X background yield is normalized to the estimate from the data.
In the middle and right figures the SM \PH boson signal is separated into two components: the production mode which is targeted by the specific discriminant,
and other production modes, where the gluon fusion process dominates.
\label{fig:fitDisc}}
\end{figure}

The correlation of the kinematic discriminants $\KD$, $\DbkgVBFdec$, and $\DbkgVHdec$ with the four-lepton invariant mass is shown in Fig.~\ref{fig:KD} for the mass interval $105<\mllll<140\GeV$.
Their distributions for the mass interval $118<\mllll<130\GeV$ are shown in Fig.~\ref{fig:fitDisc}.

\subsection{Signal strength modifier}
\label{subsec:signalstrenghts}

A simultaneous fit to all categories is performed to extract the signal strength modifier,
defined as the ratio of the observed \PH boson yield in the $\Hllll$ decay channel to the standard model expectation.

\begin{figure}[!htb]
\centering
\includegraphics[width=0.49\textwidth]{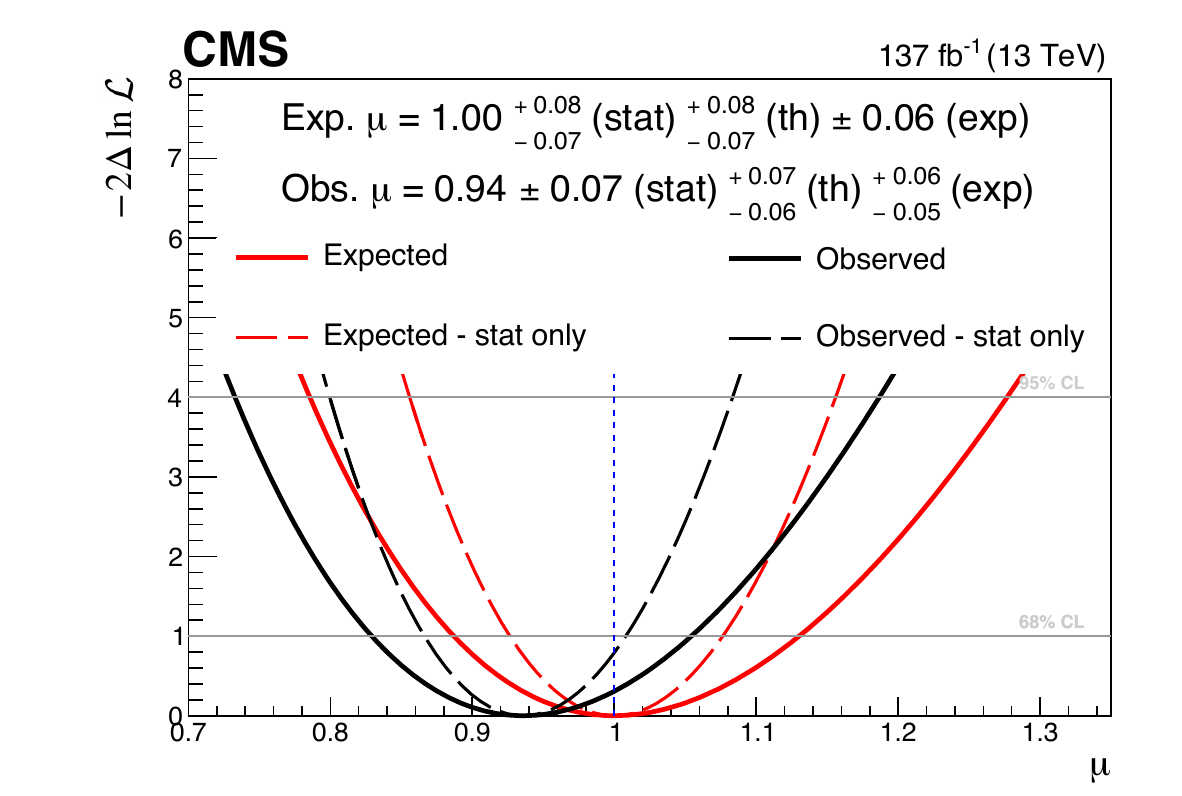}
\includegraphics[width=0.49\textwidth]{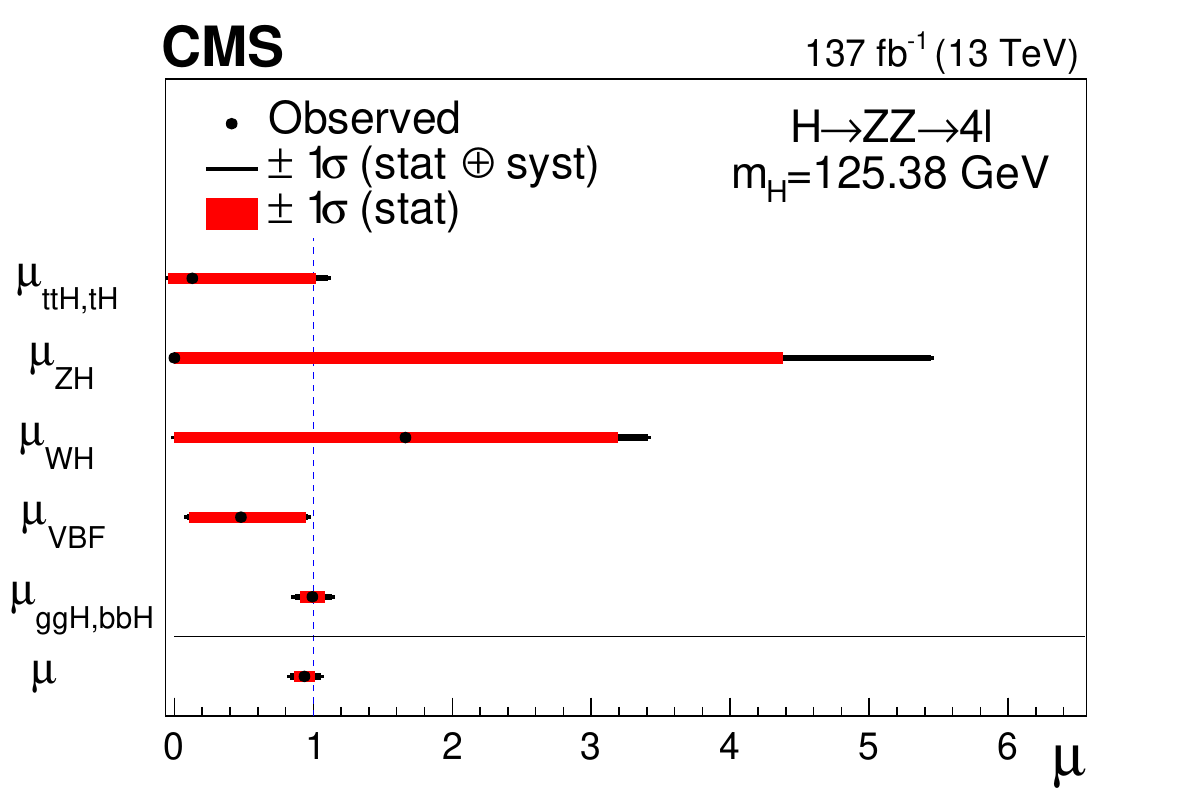}
\caption{
(\cmsLeft) The observed and expected profile likelihood scans of the inclusive signal strength modifier. The scans are shown both with (solid line) and without (dashed line) systematic uncertainties.
(\cmsRight) Results of likelihood scans for the signal strength modifiers corresponding to the five main SM \PH boson production mechanisms, compared to the SM prediction shown as a vertical dashed line.
The thick black lines indicate the one standard deviation confidence intervals including both statistical and systematic sources.
The thick red lines indicate the statistical uncertainties corresponding to the one standard deviation confidence intervals.
\label{fig:mucat}}
\end{figure}

The combined measurement of the inclusive signal strength modifier is measured to be $\mu = 0.94^{+0.12}_{-0.11}$ or $\mu = 0.94 \pm 0.07\stat^{+0.07}_{-0.06}\thy^{+0.06}_{-0.05}\,(\text{exp})$ at a fixed mass value $\mH = 125.38\GeV$,
which is the current most precise measurement of the \PH boson mass published by the CMS Collaboration~\cite{HggMass_Run2}.
In all subsequent fits, $\mH$ is fixed to this value.
The dominant experimental sources of systematic uncertainty are the uncertainties in the lepton identification efficiencies and luminosity measurement, while the dominant theoretical source is the uncertainty in the total gluon fusion cross section.
The contributions to the total uncertainty from experimental and theoretical sources are found to be similar in magnitude.
The signal strength modifiers are further studied in terms of the five main SM Higgs boson production mechanisms, namely $\ggH$, $\VBF$, $\ZH$, $\WH$, and $\ttH$.
The contributions of the $\bbH$ and $\tH$ production modes are also taken into account.
The relative normalizations of the $\bbH$ and the gluon fusion contributions are kept fixed in the fit, and so are the $\tH$ and $\ttH$ ones.
The results are shown in Fig.~\ref{fig:mucat} for the observed and expected profile likelihood scans of the inclusive signal strength modifier and those for the signal strength modifiers of the five main SM Higgs boson production mechanisms.
The corresponding numerical values, including the decomposition of the uncertainties into statistical and systematic components, as well as the expected uncertainties, are given in Table~\ref{tab:sigstr}.

The dependence of the measured signal strengths on the profiling of $\mH$ is checked and found to have a small impact both on the inclusive results and those in terms of the five main \PH boson production mechanisms, well within the measurement uncertainties.
The best fit signal value changes at most by 4\% and the profiled value of the mass is found to be $\mH=125.09^{+0.15}_{-0.14}\stat\GeV$.
It is important to note here that the precise determination of $\mH$ and the systematic uncertainties that enter its measurement are beyond the scope of this analysis.

\begin{table*}[hb]
	\centering
		\topcaption{
		Best fit values and $\pm 1$ standard deviation uncertainties for the expected and observed signal strength modifiers at $\mH=125.38\GeV$.
		The statistical and systematic uncertainties are given separately.
		\label{tab:sigstr}
			}
    \renewcommand{\arraystretch}{1.5}
    \begin{tabular}{ccc}
	& Expected & Observed \\
	\hline
	$\mu_{\ttH,\tH}$ & $1.00^{+1.23}_{-0.77}\stat^ {+0.51}_{-0.06}\syst$ & $0.17 ^{+0.88}_{-0.17}\stat^{+0.42}_{-0.00}\syst$ \\
	$\mu_{\WH}$ & $1.00^{+1.83}_{-1.00}\stat^{+0.75}_{-0.00}\syst$ & $1.66 ^{+1.52}_{-1.66}\stat^ {+0.85}_{-0.00}\syst$ \\
	$\mu_{\ZH}$ & $1.00^{+4.79}_{-1.00}\stat^ {+6.76}_{-0.00}\syst$ & $0.00 ^{+4.38}_{-0.00}\stat^ {+3.24}_{-0.00}\syst$ \\
  $\mu_{\mathrm{VBF}}$ & $1.00^{+0.53}_{-0.44}\stat^ {+0.18}_{-0.12}\syst$ & $0.48 ^{+0.46}_{-0.37}\stat^ {+0.14}_{-0.10}\syst$ \\
  $\mu_{\Pg\Pg\PH,\bbH}$ & $1.00 \pm 0.10\stat^ {+0.12}_{-0.10}\syst$ & $0.99 \pm 0.09\stat^ {+0.11}_{-0.09}\syst$ \\
	\hline
	$\mu$ & $1.00^{+0.08}_{-0.07}\stat^{+0.10}_{-0.08}\syst$ & $0.94 \pm 0.07\stat^{+0.09}_{-0.08}\syst$\\
\end{tabular}
\end{table*}

\begin{figure}[!htb]
\centering
\includegraphics[width=0.49\textwidth]{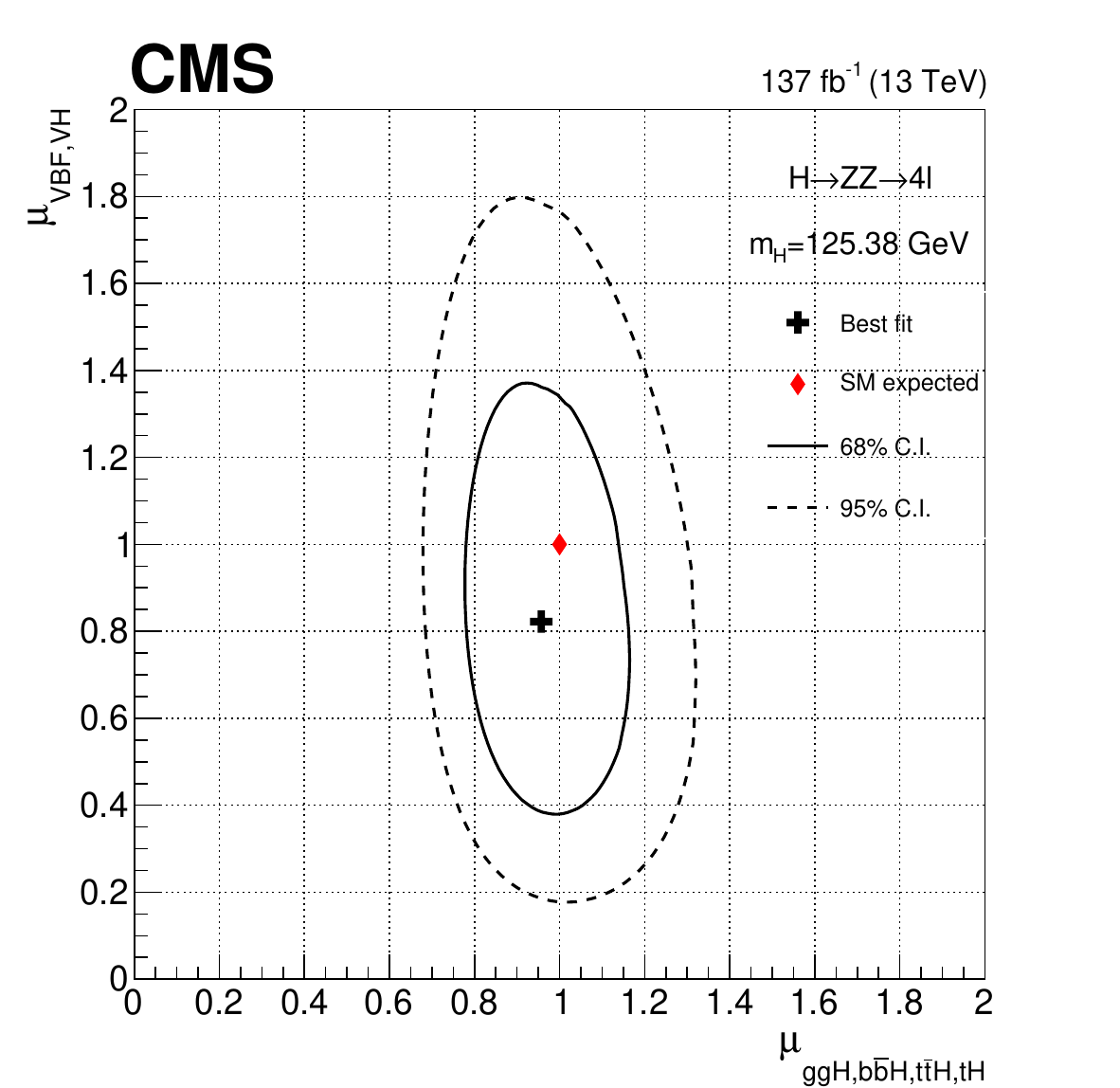}
\caption{
Result of the 2D likelihood scan for the $\muF\equiv\muFlong$ and $\muV\equiv\muVlong$ signal strength modifiers.
The solid and dashed contours show the 68 and 95\% \CL regions, respectively.
The cross indicates the best fit value, and the diamond represents the expected value for the SM Higgs boson.
\label{fig:mu2D}}
\end{figure}

Two signal strength modifiers, $\muF\equiv\muFlong$ and $\muV\equiv\muVlong$, are introduced for the fermion and vector-boson induced contributions to the expected SM cross section.
A two-parameter fit is performed simultaneously to the events reconstructed in all categories, leading to $\muF=\valMuF$ and $\muV=\valMuV$.
The expected values for $\mH=125.38\GeV$ are $\muF=1.00^{+0.15}_{-0.13}$ and $\muV=1.00^{+0.39}_{-0.33}$.
The 68 and 95\% \CL contours in the ($\muF,\muV$) plane are shown in Fig.~\ref{fig:mu2D} and the SM predictions lie within the 68\% \CL regions of this measurement.

\subsection{Simplified template cross section}
\label{subsec:stxs}

\begin{figure*}[!htbp]
\centering
\includegraphics[width=0.99\linewidth]{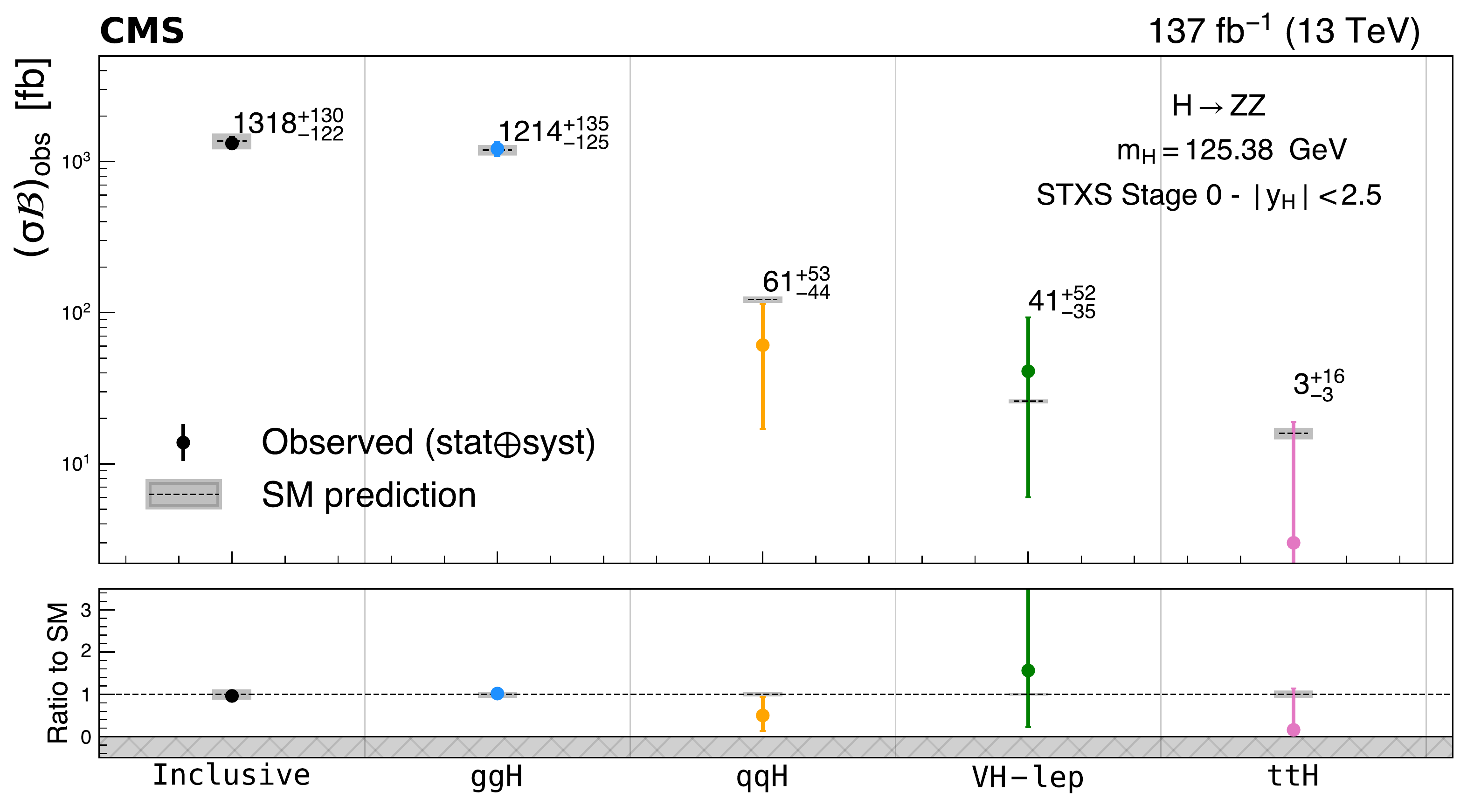}
\caption{
		The measured product of cross section times branching fraction for $\HZZ$ decay $\sigmaObs$ and the SM predictions $\sigmaSM$ for the stage 0 STXS production bins and the inclusive measurement at $\mH=125.38\GeV$.
		Points with error bars represent measured values and black dashed lines with gray uncertainty bands represent the SM predictions.
		In the bottom panel ratios of the measured cross sections and the SM predictions are shown along with the uncertainties for each of the bins and the inclusive measurement.
			\label{fig:stxs_0}
			}
\end{figure*}

\begin{figure*}[!htbp]
\centering
\includegraphics[width=0.85\linewidth]{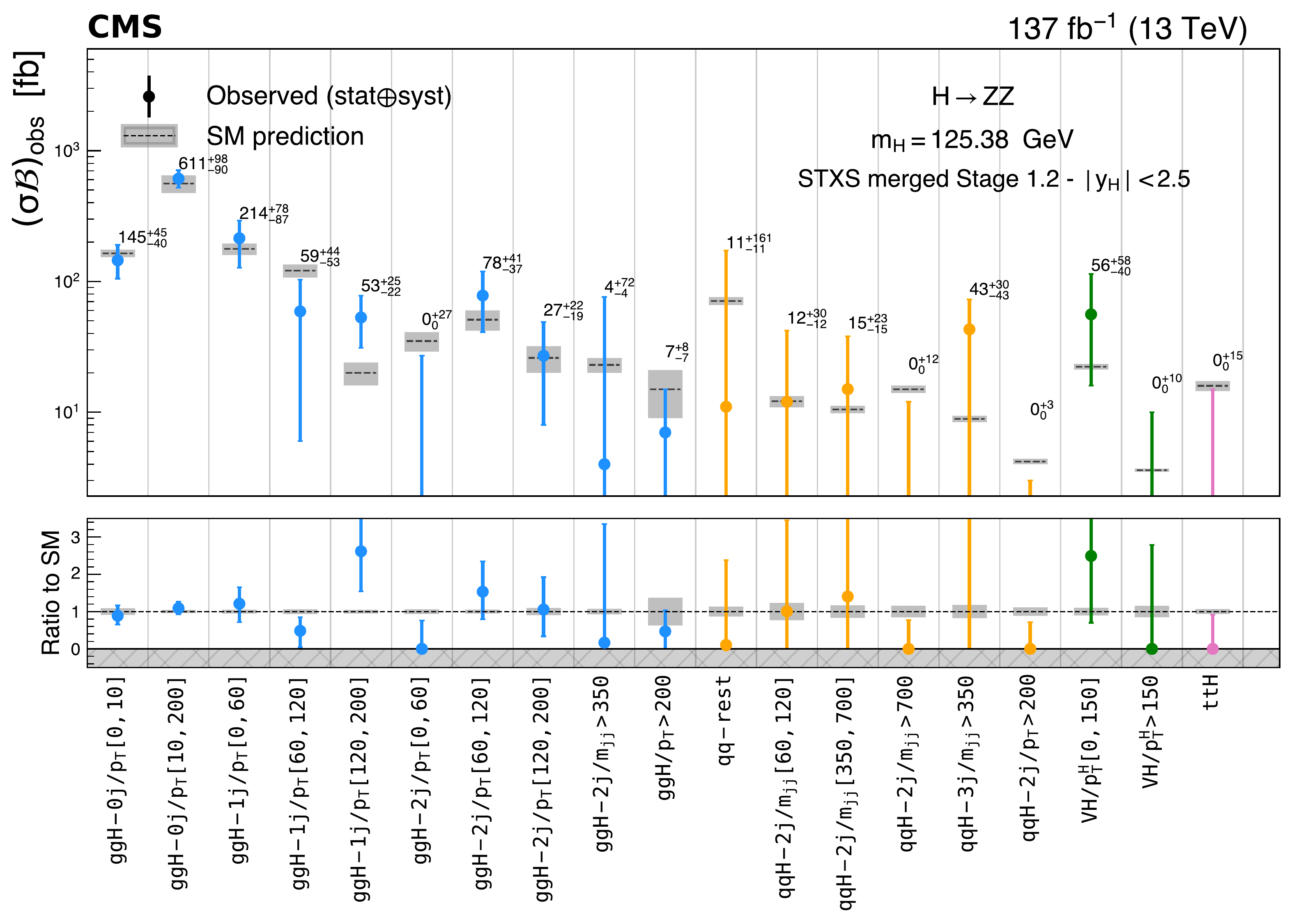}
\caption{
The measured cross sections $\sigmaObs$ and the SM predictions $\sigmaSM$ for $\HZZ$ decay and the merged stage 1.2 STXS production bins at $\mH=125.38\GeV$.
Points with error bars represent measured values and black dashed lines with gray uncertainty bands represent the SM predictions.
In the bottom panel ratios of the measured cross sections and the SM predictions are shown with corresponding uncertainties for each of the bins.
\label{fig:stxs_1}
}
\end{figure*}

The results for the \PH boson product of cross section times branching fraction for $\HZZ$ decay,
$\sigmaObs$, and comparisons with the SM expectation,
$\sigmaSM$,
for the stages of production bins defined in Section~\ref{subsec:STXS_Categories}, are shown in Fig.~\ref{fig:stxs_0} for the stage 0 and in Fig.~\ref{fig:stxs_1} for the merged stage 1.2.
The corresponding numerical values are given in Tables~\ref{tab:stage0} and~\ref{tab:stage1p2}.

\begin{figure*}[!htb]
	\centering
		\includegraphics[width=0.6\linewidth]{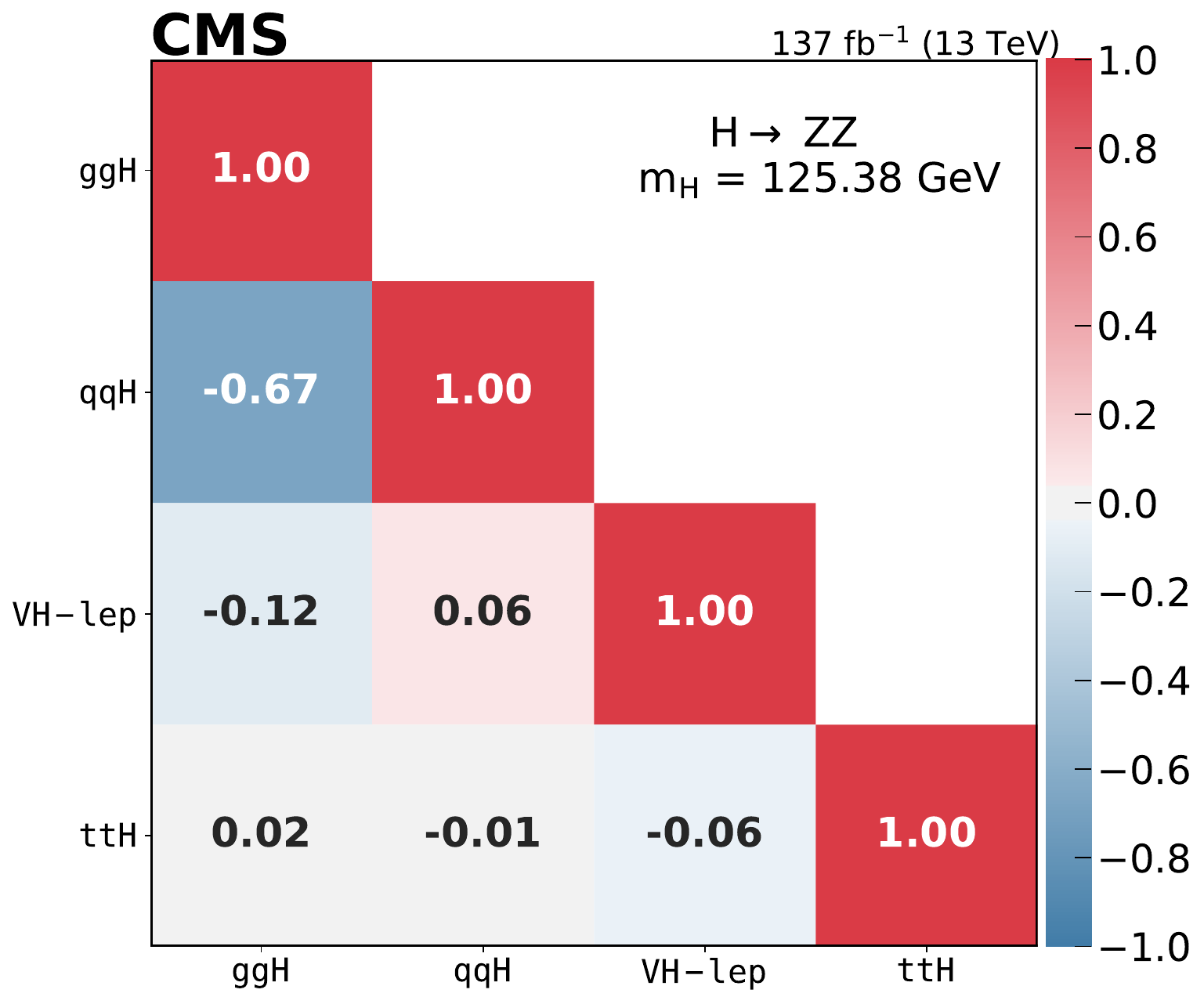}
		\includegraphics[width=0.98\linewidth]{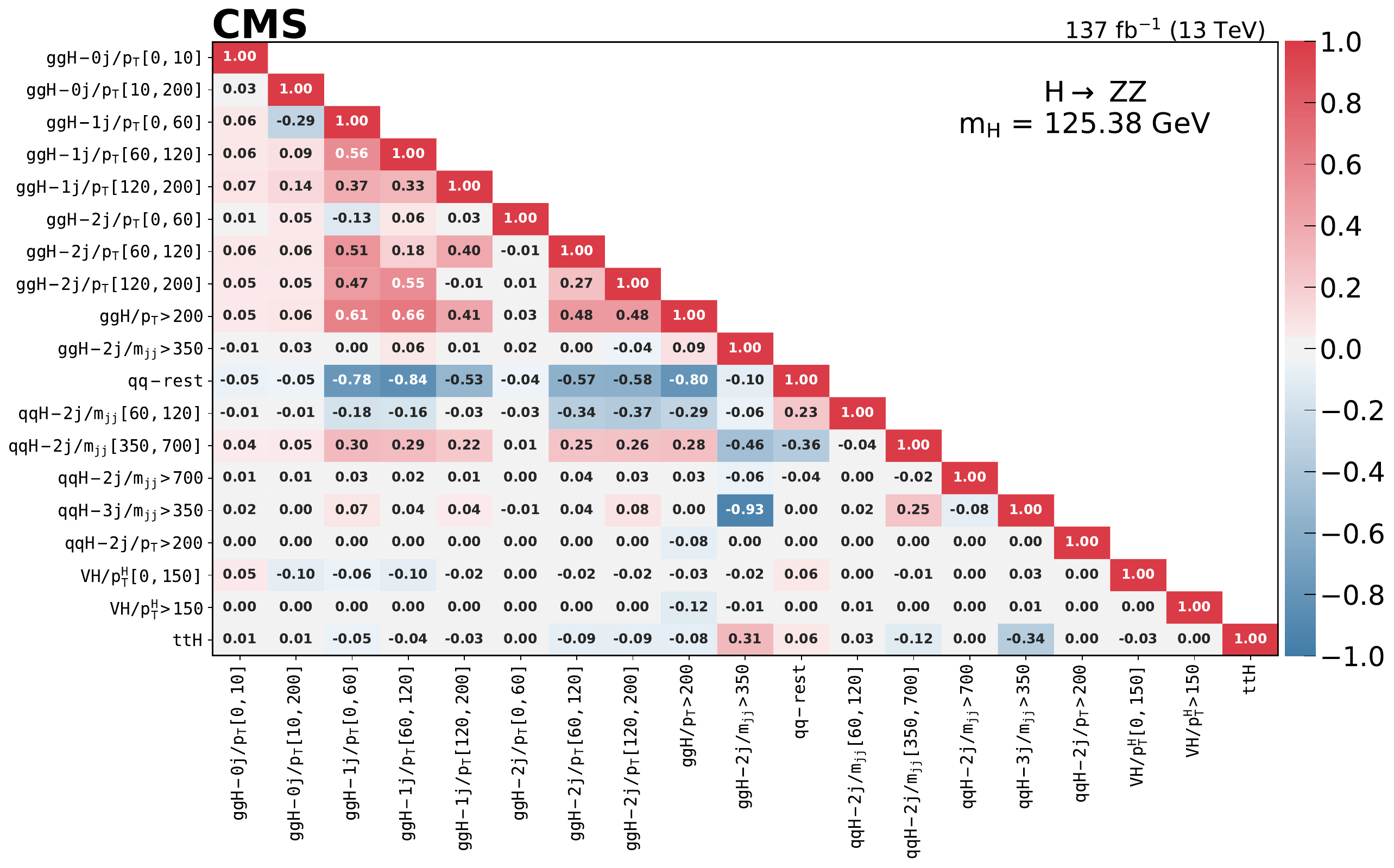}
		\caption{Correlation matrices between the measured cross sections for the stage 0 (upper) and the merged stage 1.2 (lower) for $\HZZ$ decay.
			\label{fig:corrmatrix}}
\end{figure*}

As discussed, the set of THU uncertainties described in Section~\ref{subsec:Theoretical_uncertainties} is not considered for the STXS measurements: THU uncertainties are model dependent and should be only considered in the interpretation of the results.
Therefore, the THU uncertainties are included in the SM predictions of the cross section.
The correlation matrices are shown in Fig.~\ref{fig:corrmatrix}.
The dominant experimental sources of systematic uncertainty are the same as for the signal strength modifiers measurement,
while the dominant theoretical source is the uncertainty in the category migration for the ggH process.

\begin{table}[hb!]
	\centering
		\topcaption{
		Best fit values and $\pm 1$ standard deviation uncertainties for the measured cross sections $\sigmaObs$,
		the SM predictions $\sigmaSM$, and their ratio for the stage 0 STXS production bins at $\mH=125.38 \GeV$ for $\HZZ$ decay.
		\label{tab:stage0}
			}
      \renewcommand{\arraystretch}{1.5}
    \begin{tabular}{cccc}
	& $\sigmaObs$~(fb) &  $\sigmaSM$~(fb) & $\sigmaObs / \sigmaSM$\\
	\hline
	$\stxsttH$ & $3^{+16}_{-3}$ & $15.9 \pm 1.4$ & $0.16^{+0.98}_{-0.16}$ \\
	$\stxsVHlep$ & $41^{+52}_{-35}$ & $25.9 \pm 0.8$ & $1.56^{+1.99}_{-1.34}$ \\
	$\stxsqqH$ & $61^{+53}_{-44}$ & $122 \pm 6$ & $0.50^{+0.44}_{-0.36}$ \\
	$\stxsggH$ & $1214^{+135}_{-125}$ & $1192 \pm 95$ & $1.02^{+0.11}_{-0.10}$ \\
  \hline
	$\mathrm{Inclusive}$ & $1318^{+130}_{-122}$ & $1369\pm 164$ & $0.96^{+0.10}_{-0.09}$ \\
	\end{tabular}
 \end{table}

\begin{table*}[hb]
	\centering
		\topcaption{
		Best fit values and $\pm 1$ standard deviation uncertainties for the measured cross sections $\sigmaObs$,
		the SM predictions $\sigmaSM$, and their ratio for the merged stage 1.2 STXS production bins at $\mH=125.38\GeV$ for $\HZZ$ decay.
		\label{tab:stage1p2}
			}
		\renewcommand{\arraystretch}{1.5}
		\begin{tabular}{cccc}
	& $\sigmaObs$~(fb) &  $\sigmaSM$~(fb) & $\sigmaObs / \sigmaSM$\\
	\hline
	$\stageoneggHbinone$ & $145^{+45}_{-40}$ & $164 \pm 11$ & $0.89^{+0.28}_{-0.24}$\\
	$\stageoneggHbintwo$ & $611^{+98}_{-90}$ & $561 \pm 87$ & $1.09^{+0.17}_{-0.16}$\\
	$\stageoneggHbinthree$ & $214^{+78}_{-87}$ & $177 \pm 18$ & $1.21^{+0.44}_{-0.49}$\\
  $\stageoneggHbinfour$ & $59^{+44}_{-53}$ & $121 \pm 14$ & $0.48^{+0.37}_{-0.44}$\\
	$\stageoneggHbinfive$ & $53^{+25}_{-22}$ & $20 \pm 4$ & $2.62^{+1.24}_{-1.08}$\\
	$\stageoneggHbinsix$ & $0^{+27}_{-0}$ & $35 \pm 6$ & $0.00^{+0.76}_{-0.00}$\\
	$\stageoneggHbinseven$ & $78^{+41}_{-37}$ & $51 \pm 9$ & $1.53^{+0.81}_{-0.73}$\\
	$\stageoneggHbineight$ & $27^{+22}_{-19}$ & $26 \pm 6$ & $1.06^{+0.87}_{-0.72}$\\
	$\stageoneggHbinnine$ & $4^{+72}_{-4}$ & $23 \pm 3$ & $0.17^{+3.2}_{-0.17}$\\
	$\stageoneggHbinten$ & $7^{+8}_{-7}$ & $15 \pm 6$ & $0.47^{+0.56}_{-0.47}$\\
	$\stageoneqqHbinone$ & $11^{+161}_{-11}$ & $71 \pm 5$ & $0.15^{+2.27}_{-0.15}$\\
	$\stageoneqqHbintwo$ & $12^{+30}_{-12}$ & $12.1 \pm 1.2$ & $1.01^{+2.45}_{-1.01}$\\
	$\stageoneqqHbinthree$ & $15^{+23}_{-15}$ & $10.5 \pm 0.7$ & $1.41^{+2.21}_{-1.41}$\\
	$\stageoneqqHbinfour$ &  $0^{+12}_{-0}$ & $15 \pm 1$ & $0.00^{+0.77}_{-0.00}$\\
	$\stageoneqqHbinfive$ & $43^{+30}_{-43}$ & $8.9 \pm 0.5$ & $4.84^{+3.38}_{-4.84}$\\
	$\stageoneqqHbinsix$ & $0^{+3}_{-0}$ & $4.2 \pm 0.2$ & $0.00^{+0.72}_{-0.00}$\\
	$\stageoneVHbinone$ & $56^{+58}_{-40}$ & $22.3 \pm 1.1$ & $2.49^{+2.60}_{-1.79}$\\
	$\stageoneVHbintwo$ & $0^{+10}_{-0}$ & $3.6 \pm 0.1$ & $0.00^{+2.79}_{-0.00}$\\
	$\stxsttH$ & $0^{+15}_{-0}$ & $15.9 \pm 1.4$ & $0.00^{+0.91}_{-0.00}$\\
	\end{tabular}
 \end{table*}

\subsection{Fiducial cross section}
\label{sec:crosssections}

In this section the cross section measurement for the process $\Pp\Pp\to\Hllll$ within a fiducial volume that closely matches the reconstruction level selection is presented.
In particular, the integrated fiducial cross section is measured as well as differential cross sections as a function of the transverse momentum of the \PH boson ($\pt^{\PH}$),
its rapidity ($\abs{y^{\PH}}$), the number of associated jets ($N^{\text{j}}$), and the transverse momentum of the leading jet ($\pt^{\text{j}}$).
These measurements are largely independent of the assumptions on the relative fractions and kinematic distributions of the individual production modes.
The definition of the fiducial volume is based on generator-level quantities and is identical to that in Ref.~\cite{CMSH4l2016}.
In order to reduce the experimental uncertainties, only jets with $\pt^{\text{j}}>30\GeV$ and $\abs{\eta^{\text{j}}}<2.5$ are considered for
the differential cross sections as a function of jet observables. An increase in model dependence compared to Ref.~\cite{CMSH4lFiducial8TeV} is observed when
using the $\ZZ$ candidate selection at reconstruction level where the candidate with the best $\KD$ discriminant value is chosen.
Therefore, the fiducial cross section measurement is performed using the event selection algorithm in Ref.~\cite{CMSH4lFiducial8TeV}.
Specifically, the $\PZ_{1}$ candidate is chosen to be the one with $m(\PZ_1)$ closest to the nominal $\PZ$ boson mass, and in cases where
multiple $\PZ_{2}$ candidates satisfy all criteria, the pair of leptons with the largest sum of the transverse momenta magnitudes is chosen.
The full fiducial volume definition is detailed in Table~\ref{tab:FidDef} and the acceptance for various SM production modes is given in Table~\ref{tab:summarySM}.

\begin{table*}[!htb]
	\centering
		\topcaption{
			Summary of requirements used in the definition of the fiducial phase space for the $\Hllll$ cross section measurements.
			\label{tab:FidDef}
		}
		\cmsTable
		{
		\begin{tabular}{lc}
			\multicolumn{2}{c}{Requirements for the $\Hllll$ fiducial phase space} \\
			\hline
			\multicolumn{2}{c}{Lepton kinematics and isolation} \\[\cmsTabSkip]
			\vspace{-0.4cm} & \\
			Leading lepton $\pt$ & $\pt > 20\GeV$ \\
			\vspace{-0.4cm} & \\
			Next-to-leading lepton $\pt$ & $\pt > 10\GeV$ \\
			\vspace{-0.4cm} & \\
			Additional electrons (muons) $\pt$ & $\pt > 7 (5)\GeV$ \\
			\vspace{-0.4cm} & \\
			Pseudorapidity of electrons (muons) & $\abs{\eta} <$ 2.5 (2.4) \\
			\vspace{-0.4cm} & \\
			Sum of scalar $\pt$ of all stable particles within $\Delta R < 0.3$ from lepton & $<0.35 \pt$ \\[\cmsTabSkip]
			\multicolumn{2}{c}{Event topology} \\[\cmsTabSkip]
			\multicolumn{2}{l}{Existence of at least two same-flavor OS lepton pairs, where leptons satisfy criteria above} \\
			Inv. mass of the $\PZ_1$ candidate & $40 < m_{\PZ_{1}} < 120 \GeV$ \\
			\vspace{-0.4cm} & \\
			Inv. mass of the $\PZ_2$ candidate & $12 < m_{\PZ_{2}} < 120 \GeV$ \\
			\vspace{-0.4cm} & \\
			Distance between selected four leptons & $\Delta R(\ell_{i},\ell_{j})>0.02$ for any $i\neq j$  \\
			\vspace{-0.4cm} & \\
			Inv. mass of any opposite sign lepton pair & $m_{\ell^{+}\ell'^{-}}>4 \GeV$ \\
			\vspace{-0.4cm} & \\
			Inv. mass of the selected four leptons & $105 < \mllll < 140 \GeV$  \\
		\end{tabular}}
		\normalsize
\end{table*}

A maximum likelihood fit of the signal and  background parameterizations to the observed $4\ell$
mass distribution, $N_{\mathrm{obs}}(\mllll)$, is performed to extract
the integrated fiducial cross section for the process $\Pp\Pp\to\PH\to4\ell$ ($\sigma_{\mathrm{fid}}$).
The fit is carried out inclusively (\ie, without any event categorization) and does not use the $\KD$ observable
in order to minimize the model dependence.
The fit is performed simultaneously in all final states and assumes a \PH boson mass $\mH=125.38\GeV$, while
the branching fractions of the \PH boson to different final states ($4\Pe,4\Pgm,2\Pe2\Pgm$) are free parameters in the fit.
The systematic uncertainties described in Section~\ref{sec:systematics} are included in the form of NPs and the results are obtained using an
asymptotic approach~\cite{LHC-HCG} with a test statistic based on the profile likelihood ratio~\cite{Cowan_2011}.
This procedure accounts for the unfolding of detector effects from the observed distributions and is the same as in
Refs.~\cite{CMSH4lFiducial8TeV}~and~\cite{CMSHggFiducial8TeV}.

The number of expected events in each final state $f$ and in each bin $i$ of a given observable is expressed as a
function of $\mllll$ as:
\begin{linenomath}
\ifthenelse{\boolean{cms@external}}
{
\begin{equation}
\label{eqn:m4l}
\begin{aligned}
N_{\mathrm{exp}}^{f,i}(m_{4\ell}) &= N_{\mathrm{fid}}^{f,i}(m_{4\ell})+N_{\mathrm{nonfid}}^{f,i}(m_{4\ell})+N_{\mathrm{nonres}}^{f,i}(m_{4\ell})\\
&+N_{\text{bkg}}^{f,i}(m_{4\ell}) \\
&=\sum_j\epsilon_{i,j}^{f}  \left(1+f_{\mathrm{nonfid}}^{f,i} \right)\sigma_{\mathrm{fid}}^{f,j}  \mathcal{L}\mathcal{P}_{\mathrm{res}}(m_{4\ell}) \\
&\,\,\,+ N_{\mathrm{nonres}}^{f,i}\mathcal{P}_{\mathrm{nonres}}(m_{4\ell})+N_{\text{bkg}}^{f,i}\mathcal{P}_{\text{bkg}}(m_{4\ell}).
\end{aligned}
\end{equation}
}
{
\begin{equation}
\label{eqn:m4l}
\begin{aligned}
N_{\mathrm{exp}}^{f,i}(m_{4\ell}) &= N_{\mathrm{fid}}^{f,i}(m_{4\ell})+N_{\mathrm{nonfid}}^{f,i}(m_{4\ell})+N_{\mathrm{nonres}}^{f,i}(m_{4\ell})+N_{\text{bkg}}^{f,i}(m_{4\ell}) \\
&=\sum_j\epsilon_{i,j}^{f}  \left(1+f_{\mathrm{nonfid}}^{f,i} \right)\sigma_{\mathrm{fid}}^{f,j}  \mathcal{L}\mathcal{P}_{\mathrm{res}}(m_{4\ell}) \\
&\,\,\,+ N_{\mathrm{nonres}}^{f,i}\mathcal{P}_{\mathrm{nonres}}(m_{4\ell})+N_{\text{bkg}}^{f,i}\mathcal{P}_{\text{bkg}}(m_{4\ell}).
\end{aligned}
\end{equation}
}
\end{linenomath}
The shape of the resonant signal contribution, $\mathcal{P}_{\mathrm{res}}(m_{4\ell})$, is described by a double-sided Crystal Ball function as discussed in Section~\ref{sec:signal}, and the normalization is used to extract the fiducial cross section.
The non-resonant signal function, $\mathcal{P}_{\mathrm{nonres}}(m_{4\ell})$, is determined by the $\WH$, $\ZH$, and $\ttH$ contributions where one of the leptons from the \PH boson decay is lost or not selected.
It is modeled by a Landau distribution with shape parameters constrained in the fit to be within a range determined from simulation.
This contribution is referred to as the ``combinatorial signal'' and is treated as a background in this measurement.

The quantity $\epsilon_{i,j}^{f}$ represents the detector response matrix that maps the number of expected events in bin \textit{j} of a given observable at the fiducial level to the number of expected events in bin \textit{i} at the reconstruction level.
This response matrix is determined using simulated signal samples and includes corrections for residual differences between data and simulation.
In the case of the integrated fiducial cross section measurement, the response matrices become single numbers,
which are listed in Table~\ref{tab:summarySM} for different SM production mechanism.

{\tolerance=800 An  additional resonant contribution arises from events which are accepted but do not originate from the fiducial phase space. These events are due to detector effects that cause differences between
the quantities used for the fiducial phase space definition and the
corresponding quantities at the reconstruction level.
This contribution is treated as background and is referred to as the ``non-fiducial signal'' contribution.
A simulated sample is used to verify that the shape of the distribution for these events is identical to that of the fiducial signal,
and its normalization is fixed to the corresponding fraction of the fiducial signal.
The value of this fraction, which we denote as $f_{\mathrm{nonfid}}$, is determined from simulation for each of the signal models studied.
The value of $f_{\mathrm{nonfid}}$ for different signal models is shown in Table~\ref{tab:summarySM}.\par}
\begin{table*}[!h!tb]
	\centering
		\topcaption{
			Summary of the fraction of signal events for different SM signal production modes within the fiducial phase space (acceptance $\mathcal{A}_{\mathrm{fid}}$),
			reconstruction efficiency ($\epsilon$) for signal events in the fiducial phase space,
			and ratio of the number of reconstructed events outside the fiducial phase space to that of the reconstructed events in the fiducial phase space ($f_{\mathrm{nonfid}}$).
			For all production modes the values given are for $\mH = 125\GeV$.
			Also shown in the last column is the factor $(1+f_{\mathrm{nonfid}})\epsilon$ which regulates the signal yield for a given fiducial cross section, as shown in Eq.~(\ref{eqn:m4l}).
			The uncertainties listed are statistical only. The theoretical uncertainty in $\mathcal{A}_{\mathrm{fid}}$ for the SM is less than 1\%.
			\label{tab:summarySM}
		}
		\begin{tabular}{lcccc}
			Signal process & $\mathcal{A}_{\mathrm{fid}}$ & $\epsilon$ & $f_{\mathrm{nonfid}}$  & $(1+f_{\mathrm{nonfid}})\epsilon$ \\
			\hline
			$\ggH$ (\POWHEG) & 0.402 $\pm$ 0.001 & 0.598 $\pm$ 0.002 & 0.054 $\pm$ 0.001 & 0.631 $\pm$ 0.002 \\
 		  $\VBF$ & 0.445 $\pm$ 0.002 & 0.615 $\pm$ 0.002 & 0.043 $\pm$ 0.001 & 0.641 $\pm$ 0.003 \\
 		  $\WH$ & 0.329 $\pm$ 0.002 & 0.604 $\pm$ 0.003 & 0.078 $\pm$ 0.002 & 0.651 $\pm$ 0.004 \\
 		  $\ZH$  & 0.340 $\pm$ 0.003 & 0.613 $\pm$ 0.005 & 0.082 $\pm$ 0.004 & 0.663 $\pm$ 0.006 \\
 		  $\ttH$ & 0.315 $\pm$ 0.004 & 0.588 $\pm$ 0.007 & 0.181 $\pm$ 0.009 & 0.694 $\pm$ 0.010 \\
		\end{tabular}
\end{table*}

\begin{figure}[!htb]
	\centering
	\includegraphics[width=0.48\textwidth]{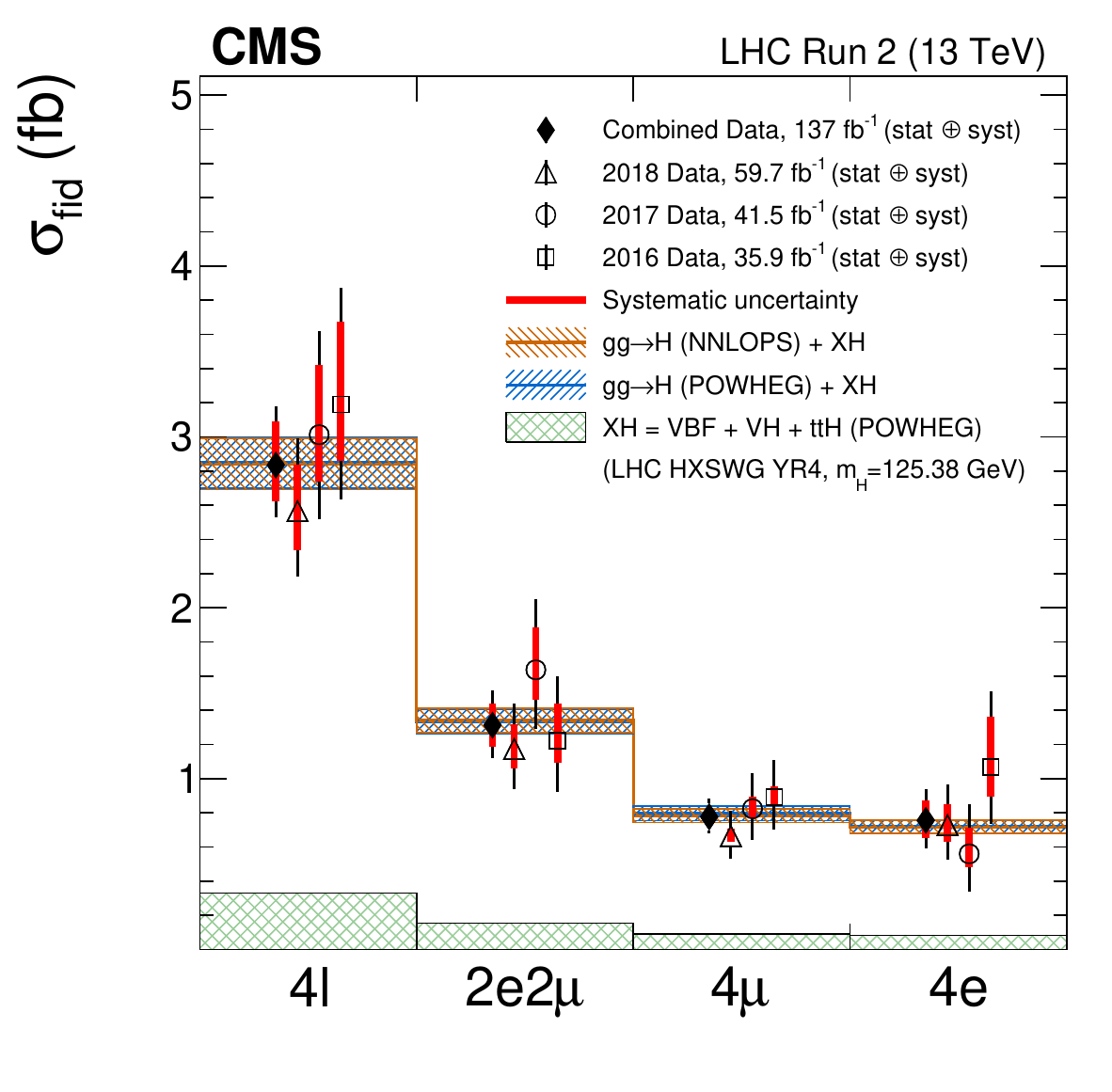}
	\includegraphics[width=0.48\textwidth]{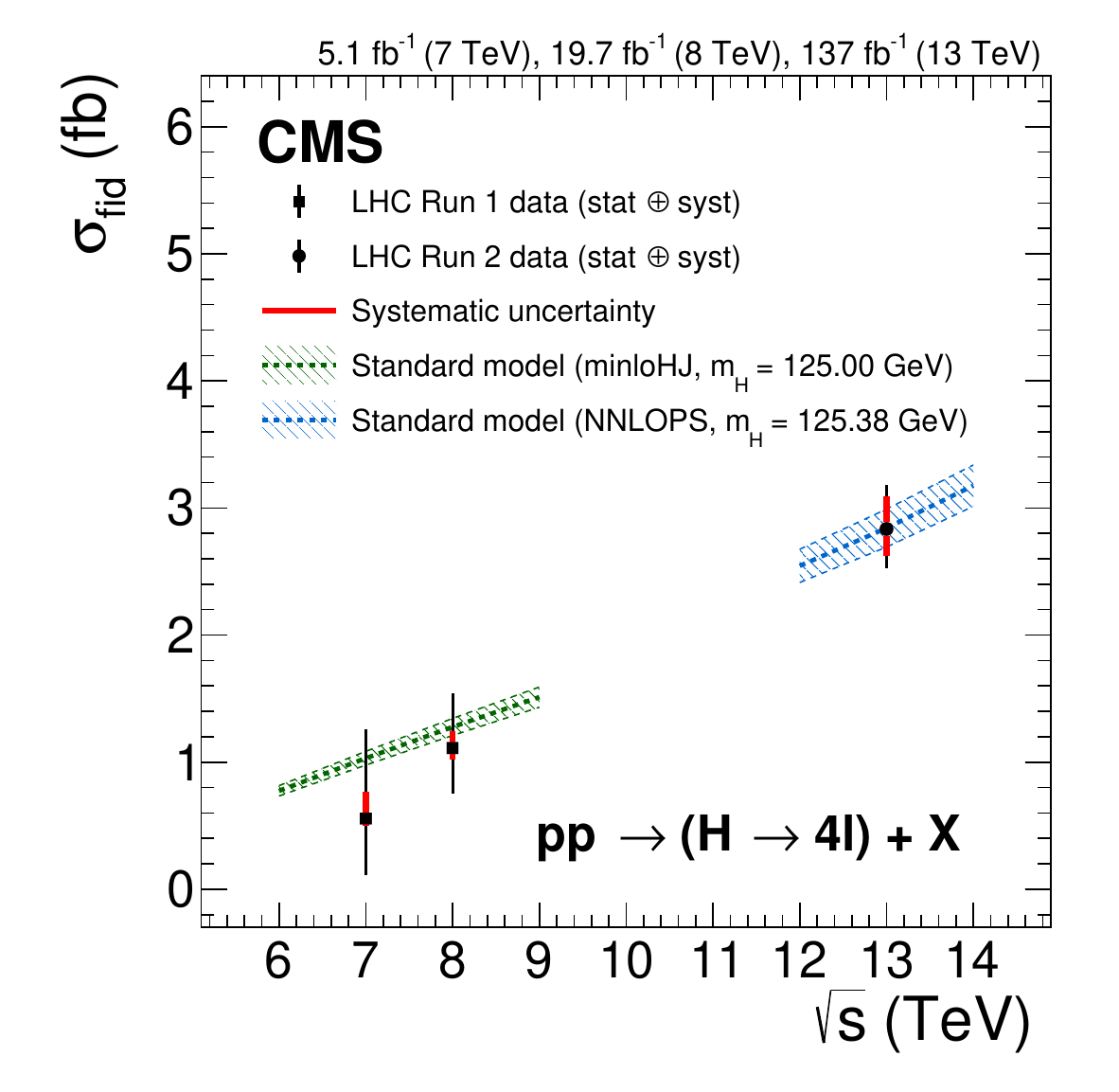}
	\caption{
	    The measured inclusive fiducial cross section in different final states (\cmsLeft) and integrated as a function of $\sqrt{s}$ (\cmsRight).
		The acceptance is calculated using \POWHEG at  $\sqrt{s}=13\TeV$ and \textsc{HRes}~\cite{Grazzini:2013mca,deFlorian:2012mx}
		at  $\sqrt{s}=$~7 and 8\TeV, and the total gluon fusion cross section and uncertainty are taken from
		Ref.~\cite{Anastasiou2016}. The fiducial volume for $\sqrt{s}=6$--$9\TeV$ uses the lepton isolation definition from
		Ref.~\cite{CMSH4lFiducial8TeV} and the SM predictions and measurements are calculated at $\mH=125.0\GeV$,
		while for $\sqrt{s}=12$--$14\TeV$ the definition described in the text is used and SM predictions and measurements are calculated at $\mH = 125.38\GeV$.
		\label{fig:fiducial_inclusive}}
\end{figure}

The integrated fiducial cross section is measured to be
$$\sigma_{{\mathrm{fid}}}=2.84^{+0.34}_{-0.31}=2.84^{+0.23}_{-0.22}\stat^{+0.26}_{-0.21}\syst \unit{fb}$$ at $\mH = 125.38\GeV$.
This can be compared to the SM expectation $\sigma_{{\mathrm{fid}}}^{\mathrm{SM}}=2.84\pm0.15 \unit{fb}$.
The measured inclusive fiducial cross sections in different final states and integrated as a function of center-of-mass energy are shown in Fig.~\ref{fig:fiducial_inclusive}.
The corresponding numerical values, including the decomposition of the uncertainties into statistical and systematic components, and the corresponding expected uncertainties, are given in Table~\ref{tab:fiducial_inclusive}.

\begin{table*}[!htb]
	\centering
		\topcaption{
    The measured inclusive fiducial cross section and $\pm 1$ standard deviation uncertainties for different final states and data-taking periods at $\mH=125.38 \GeV$.
    The statistical and systematic uncertainties are given separately for the inclusive measurements.
			\label{tab:fiducial_inclusive}
		}
    \renewcommand{\arraystretch}{1.5}
    \begin{tabular}{ccccc}
	    & $2\Pe 2\Pgm$~(fb) & $4\Pgm$~(fb) & $4\Pe$~(fb) & Inclusive (fb) \\
    \hline
	   2016 & $1.22^{+0.38}_{-0.30}$ & $0.89^{+0.22}_{-0.19}$ & $1.07^{+0.44}_{-0.33}$ & $3.19^{+0.68}_{-0.56}=3.19^{+0.48}_{-0.45}\stat^{+0.48}_{-0.33}\syst$   \\
	   2017 & $1.64^{+0.41}_{-0.35}$ & $0.82^{+0.21}_{-0.18}$ & $0.56^{+0.29}_{-0.22}$ & $3.01^{+0.60}_{-0.50}=3.01^{+0.44}_{-0.41}\stat^{+0.41}_{-0.27}\syst$   \\
	   2018 & $1.17^{+0.27}_{-0.24}$ & $0.66^{+0.15}_{-0.13}$ & $0.73^{+0.24}_{-0.20}$ & $2.57^{+0.42}_{-0.38}=2.57^{+0.33}_{-0.31}\stat^{+0.27}_{-0.23}\syst$   \\
	   2016--2018 & $1.31^{+0.20}_{-0.19}$ & $0.78^{+0.10}_{-0.10}$ & $0.76^{+0.18}_{-0.16}$ & $2.84^{+0.34}_{-0.31}=2.84^{+0.23}_{-0.22}\stat^{+0.26}_{-0.21}\syst$   \\
\end{tabular}
\end{table*}

\begin{figure}[!htb]
	\centering
	\includegraphics[width=0.48\textwidth]{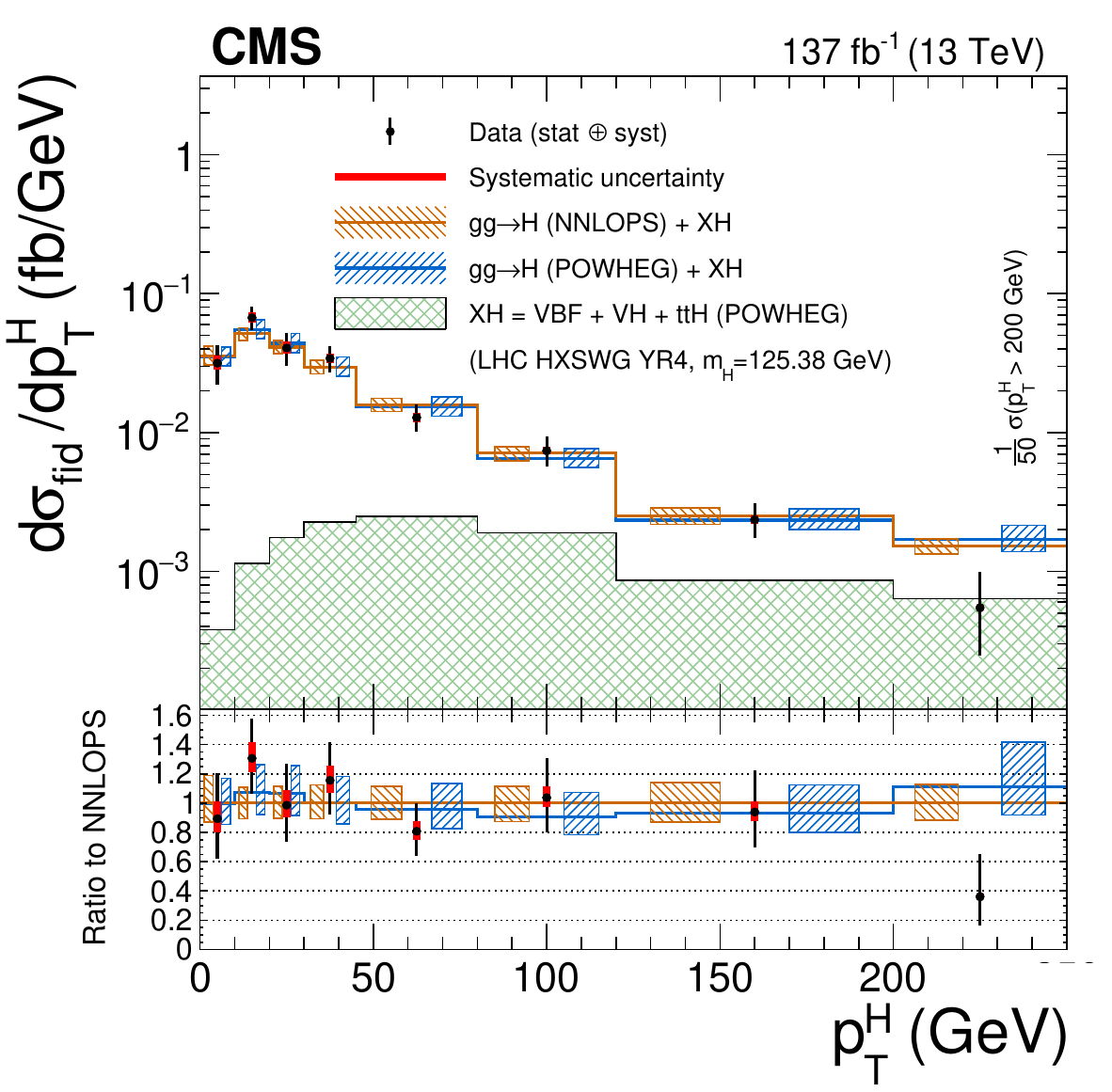}
	\includegraphics[width=0.48\textwidth]{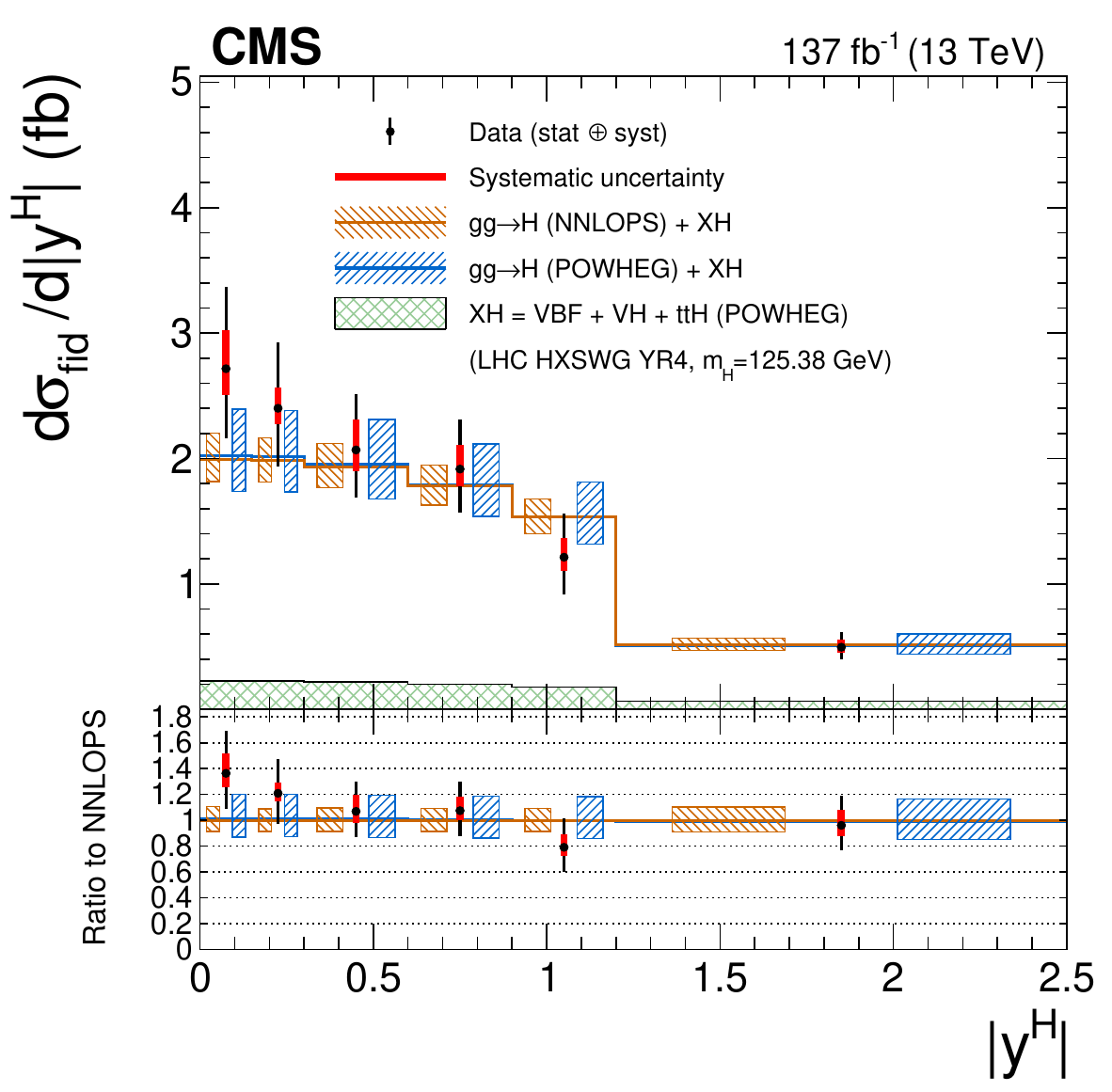} \\
	\caption{
		Differential cross sections as a function of $\pt^{\PH}$ (\cmsLeft) and $\abs{y^{\PH}}$ (\cmsRight).
		The acceptance and theoretical uncertainties in the differential bins are calculated using \POWHEG.
		The sub-dominant component of the signal ($\VBF + \VH + \ttH$) is denoted as XH.
		\label{fig:fiducial_diff_Hboson}}
\end{figure}

\begin{figure}[!htb]
	\centering
	\includegraphics[width=0.48\textwidth]{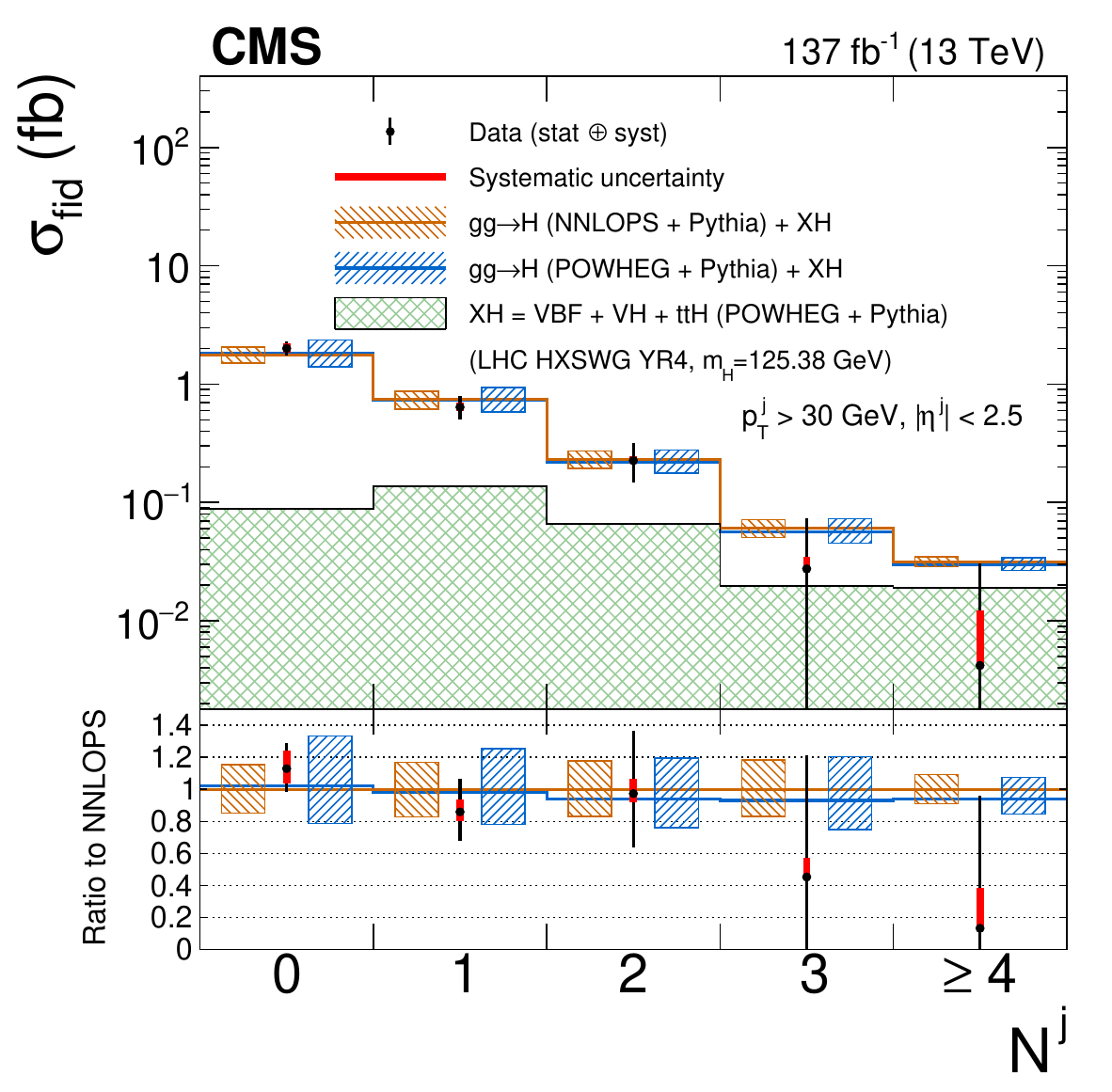}
	\includegraphics[width=0.48\textwidth]{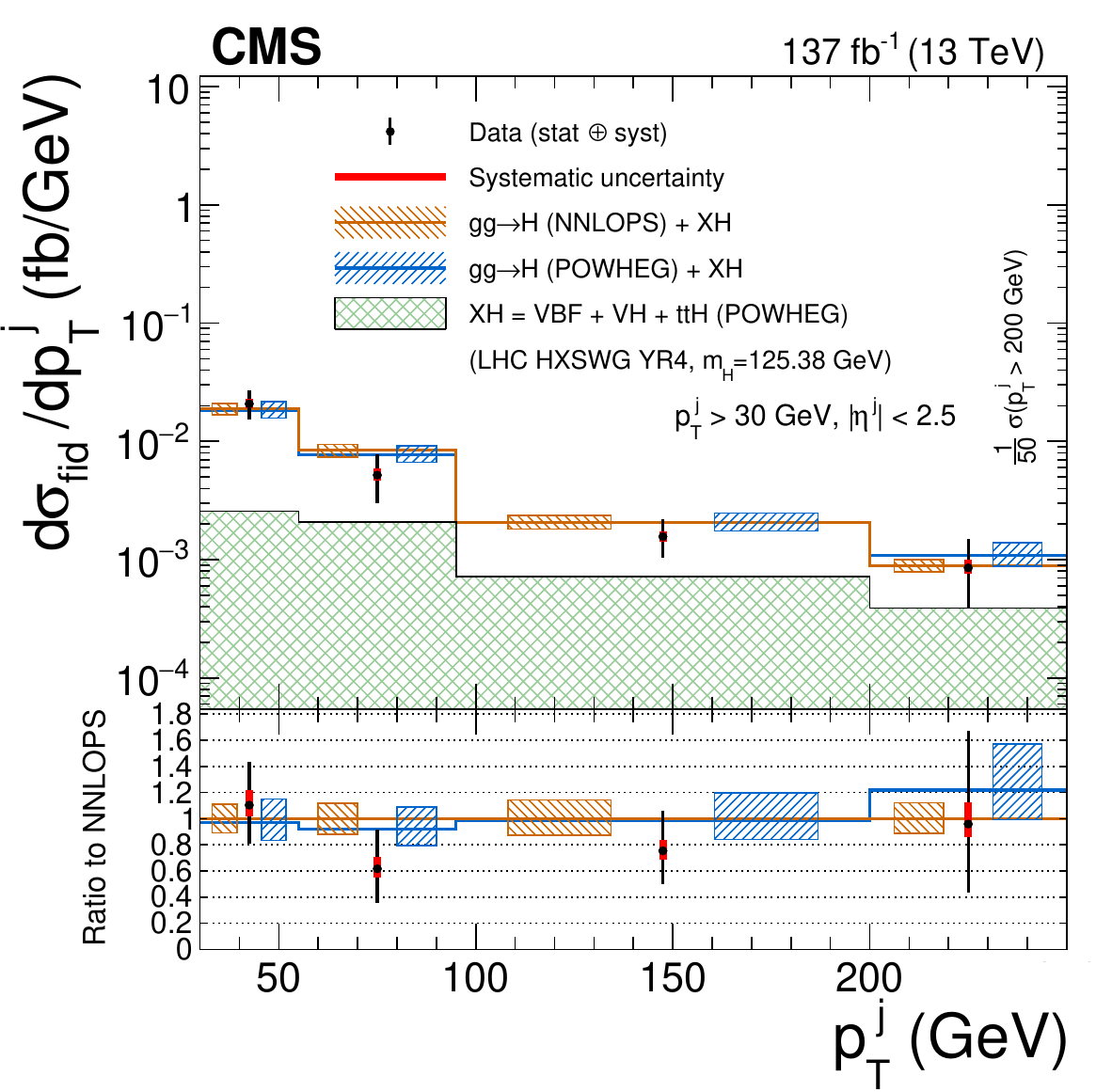}
	\caption{
		Differential cross sections as a function of the number of associated jets (left), and \pt of the leading jet (right).
		The acceptance and theoretical uncertainties in the differential bins are calculated using \POWHEG.
		The sub-dominant component of the signal ($\VBF + \VH + \ttH$) is denoted as XH.
		\label{fig:fiducial_diff_jets}}
\end{figure}

The measured differential cross sections as a function of the \PH boson transverse momentum and rapidity are shown in Fig.~\ref{fig:fiducial_diff_Hboson}.
The corresponding numerical values are given in Tables~\ref{tab:fiducial_ptH} and~\ref{tab:fiducial_yH}.
Finally, the measured differential cross sections as a function of the number of associated jets and the transverse momentum of the leading jet are shown in Fig.~\ref{fig:fiducial_diff_jets}.
The corresponding numerical values are given in Tables~\ref{tab:fiducial_Nj} and~\ref{tab:fiducial_ptj}.

\begin{table}[!htb]
	\centering
		\topcaption{
    The measured differential fiducial cross section and $\pm 1$ standard deviation uncertainties for the $\pt^{\PH}$ observable at $\mH=125.38 \GeV$.
    The breakdown of the total uncertainty (unc.) into statistical and systematic components is given.
			\label{tab:fiducial_ptH}
		}
    \renewcommand{\arraystretch}{1.5}
    \begin{tabular}{ccccc}
    Bin range ({\GeVns}) & $\rd\sigma_{\mathrm{fid}}$ (fb) & unc. & \stat  & \syst \\ \hline
    0--10 &  0.32 &  $^{+0.11}_{-0.10}$ &  $^{+0.10}_{-0.09}$ &  $^{+0.04}_{-0.03}$ \\
    10--20 &  0.67 &  $^{+0.14}_{-0.13}$ &  $^{+0.13}_{-0.12}$ &  $^{+0.06}_{-0.05}$ \\
    20--30 &  0.41 &  $^{+0.12}_{-0.10}$ &  $^{+0.11}_{-0.10}$ &  $^{+0.04}_{-0.04}$ \\
    30--45 &  0.51 &  $^{+0.12}_{-0.10}$ &  $^{+0.11}_{-0.10}$ &  $^{+0.04}_{-0.04}$ \\
    45--80 &  0.45 &  $^{+0.10}_{-0.09}$ &  $^{+0.10}_{-0.09}$ &  $^{+0.04}_{-0.03}$ \\
    80--120 &  0.30 &  $^{+0.08}_{-0.07}$ &  $^{+0.07}_{-0.07}$ &  $^{+0.02}_{-0.02}$ \\
    120--200 &  0.19 &  $^{+0.06}_{-0.05}$ &  $^{+0.06}_{-0.05}$ &  $^{+0.01}_{-0.01}$ \\
    200--13000 &  0.03 &  $^{+0.02}_{-0.02}$ &  $^{+0.02}_{-0.01}$ &  $^{+0.00}_{-0.00}$ \\
    \end{tabular}
\end{table}

\begin{table}[!htb]
	\centering
		\topcaption{
    The measured differential fiducial cross section and $\pm 1$ standard deviation uncertainties for the $\abs{y^{\PH}}$ observable at $\mH=125.38 \GeV$.
    The breakdown of the total uncertainty (unc.) into statistical and systematic components is given.
			\label{tab:fiducial_yH}
		}
    \renewcommand{\arraystretch}{1.5}
    \begin{tabular}{ccccc}
    Bin range  & $\rd\sigma_{\mathrm{fid}}$ (fb) & unc. & \stat  & \syst \\ \hline
    0.0--0.15 &  0.41 &  $^{+0.10}_{-0.08}$ &  $^{+0.09}_{-0.08}$ &  $^{+0.05}_{-0.03}$ \\
    0.15--0.3 &  0.36 &  $^{+0.08}_{-0.07}$ &  $^{+0.07}_{-0.07}$ &  $^{+0.03}_{-0.02}$ \\
    0.3--0.6 &  0.62 &  $^{+0.13}_{-0.11}$ &  $^{+0.11}_{-0.10}$ &  $^{+0.07}_{-0.05}$ \\
    0.6--0.9 &  0.57 &  $^{+0.12}_{-0.10}$ &  $^{+0.10}_{-0.10}$ &  $^{+0.06}_{-0.04}$ \\
    0.9--1.2 &  0.36 &  $^{+0.10}_{-0.09}$ &  $^{+0.09}_{-0.08}$ &  $^{+0.05}_{-0.03}$ \\
    1.2--2.5 &  0.64 &  $^{+0.15}_{-0.13}$ &  $^{+0.13}_{-0.12}$ &  $^{+0.08}_{-0.05}$ \\
    \end{tabular}
\end{table}

\begin{table}[!htb]
	\centering
		\topcaption{
    The measured differential fiducial cross section and $\pm 1$ standard deviation uncertainties for the $N^{\text{j}}$ observable at $\mH=125.38 \GeV$.
    The breakdown of the total uncertainty (unc.) into statistical and systematic components is given.
			\label{tab:fiducial_Nj}
		}
    \renewcommand{\arraystretch}{1.5}
    \begin{tabular}{ccccc}
    Bin range & $\rd\sigma_{\mathrm{fid}}$ (fb) & unc. & \stat  & \syst \\ \hline
    0 &  2.00 &  $^{+0.29}_{-0.26}$ &  $^{+0.21}_{-0.20}$ &  $^{+0.20}_{-0.17}$ \\
    1 &  0.64 &  $^{+0.15}_{-0.14}$ &  $^{+0.14}_{-0.13}$ &  $^{+0.06}_{-0.04}$ \\
    2 &  0.23 &  $^{+0.09}_{-0.08}$ &  $^{+0.09}_{-0.08}$ &  $^{+0.02}_{-0.01}$ \\
    3 &  0.03 &  $^{+0.05}_{-0.03}$ &  $^{+0.05}_{-0.03}$ &  $^{+0.01}_{-0.00}$ \\
    $\geq$4 &  0.00 &  $^{+0.03}_{-0.00}$ &  $^{+0.03}_{-0.00}$ &  $^{+0.01}_{-0.00}$ \\
    \end{tabular}
\end{table}

\begin{table}[!htb]
	\centering
		\topcaption{
    The measured differential fiducial cross section and $\pm 1$ standard deviation uncertainties for the $\pt^\text{j}$ observable at $\mH=125.38 \GeV$.
    The breakdown of the total uncertainty (unc.) into statistical and systematic components is given.
			\label{tab:fiducial_ptj}
		}
    \renewcommand{\arraystretch}{1.5}
    \begin{tabular}{ccccc}
    Bin range ({\GeVns})  & $\rd\sigma_{\mathrm{fid}}$ (fb) & unc. & \stat  & \syst \\ \hline
    30--55 &  0.52 &  $^{+0.16}_{-0.14}$ &  $^{+0.15}_{-0.13}$ &  $^{+0.05}_{-0.04}$ \\
    55--95 &  0.21 &  $^{+0.10}_{-0.09}$ &  $^{+0.10}_{-0.09}$ &  $^{+0.03}_{-0.02}$ \\
    95--200 &  0.16 &  $^{+0.07}_{-0.06}$ &  $^{+0.06}_{-0.05}$ &  $^{+0.02}_{-0.01}$ \\
    200--13000 &  0.04 &  $^{+0.03}_{-0.02}$ &  $^{+0.03}_{-0.02}$ &  $^{+0.01}_{-0.01}$ \\
    \end{tabular}
\end{table}

For all the fiducial measurements the dominant systematic uncertainties are those on the lepton identification efficiencies and luminosity measurement,
while the theoretical uncertainties are smaller.
In order to assess the model dependence of the measurement, the unfolding procedure is repeated
using different response matrices created by varying the relative fraction of each SM production mode
within its experimental constraints. The uncertainty is negligible with respect to the experimental systematic uncertainties.

\clearpage

\section{Summary}
\label{sec:summary}

{\tolerance=800
Several measurements of the Higgs boson production in the four-lepton final state at $\sqrt{s} = 13 \TeV$ have been presented,
using data samples corresponding to an integrated luminosity of $\usedLumiABC$.
Thanks to a large signal-to-background ratio and the complete reconstruction of the final state decay products,
this channel enables a detailed study of the Higgs boson production properties.
The measured signal strength modifier is $\mu=0.94 \pm 0.07\stat^{+0.07}_{-0.06}\thy^{+0.06}_{-0.05}\,(\text{exp})$
and the integrated fiducial cross section is measured to be $\sigma_{\text{fid}}=2.84^{+0.23}_{-0.22}\stat^{+0.26}_{-0.21}\syst\unit{fb}$ with a standard model prediction of $2.84 \pm 0.15 \unit{fb}$ for the same fiducial region..
The signal strength modifiers for the main Higgs boson production modes are also reported.
A new set of measurements, designed to quantify the different Higgs boson production processes in specific kinematical regions of phase space, have also been presented.
The differential cross sections as a function of the  transverse momentum and rapidity of the Higgs boson, the number of associated jets,
and the transverse momentum of the leading associated jet are determined.
All results are consistent, within their uncertainties, with the expectations for the standard model Higgs boson.
\par}

\begin{acknowledgments}
   We gratefully acknowledge the LHC Higgs Working Group for its role in developing stage 1.2 of the simplified template cross section framework.
   We congratulate our colleagues in the CERN accelerator departments for the excellent performance of the LHC and thank the technical and administrative staffs at CERN and at other CMS institutes for their contributions to the success of the CMS effort. In addition, we gratefully acknowledge the computing centers and personnel of the Worldwide LHC Computing Grid and other centers for delivering so effectively the computing infrastructure essential to our analyses. Finally, we acknowledge the enduring support for the construction and operation of the LHC, the CMS detector, and the supporting computing infrastructure provided by the following funding agencies: BMBWF and FWF (Austria); FNRS and FWO (Belgium); CNPq, CAPES, FAPERJ, FAPERGS, and FAPESP (Brazil); MES (Bulgaria); CERN; CAS, MoST, and NSFC (China); COLCIENCIAS (Colombia); MSES and CSF (Croatia); RIF (Cyprus); SENESCYT (Ecuador); MoER, ERC PUT and ERDF (Estonia); Academy of Finland, MEC, and HIP (Finland); CEA and CNRS/IN2P3 (France); BMBF, DFG, and HGF (Germany); GSRT (Greece); NKFIA (Hungary); DAE and DST (India); IPM (Iran); SFI (Ireland); INFN (Italy); MSIP and NRF (Republic of Korea); MES (Latvia); LAS (Lithuania); MOE and UM (Malaysia); BUAP, CINVESTAV, CONACYT, LNS, SEP, and UASLP-FAI (Mexico); MOS (Montenegro); MBIE (New Zealand); PAEC (Pakistan); MSHE and NSC (Poland); FCT (Portugal); JINR (Dubna); MON, RosAtom, RAS, RFBR, and NRC KI (Russia); MESTD (Serbia); SEIDI, CPAN, PCTI, and FEDER (Spain); MOSTR (Sri Lanka); Swiss Funding Agencies (Switzerland); MST (Taipei); ThEPCenter, IPST, STAR, and NSTDA (Thailand); TUBITAK and TAEK (Turkey); NASU (Ukraine); STFC (United Kingdom); DOE and NSF (USA).

	\hyphenation{Rachada-pisek} Individuals have received support from the Marie-Curie program and the European Research Council and Horizon 2020 Grant, contract Nos.\ 675440, 724704, 752730, and 765710 (European Union); the Leventis Foundation; the Alfred P.\ Sloan Foundation; the Alexander von Humboldt Foundation; the Belgian Federal Science Policy Office; the Fonds pour la Formation \`a la Recherche dans l'Industrie et dans l'Agriculture (FRIA-Belgium); the Agentschap voor Innovatie door Wetenschap en Technologie (IWT-Belgium); the F.R.S.-FNRS and FWO (Belgium) under the ``Excellence of Science -- EOS" -- be.h project n.\ 30820817; the Beijing Municipal Science \& Technology Commission, No. Z191100007219010; the Ministry of Education, Youth and Sports (MEYS) of the Czech Republic; the Deutsche Forschungsgemeinschaft (DFG), under Germany's Excellence Strategy -- EXC 2121 ``Quantum Universe" -- 390833306, and under project number 400140256 - GRK2497; the Lend\"ulet (``Momentum") Program and the J\'anos Bolyai Research Scholarship of the Hungarian Academy of Sciences, the New National Excellence Program \'UNKP, the NKFIA research grants 123842, 123959, 124845, 124850, 125105, 128713, 128786, and 129058 (Hungary); the Council of Science and Industrial Research, India; the Ministry of Science and Higher Education and the National Science Center, contracts Opus 2014/15/B/ST2/03998 and 2015/19/B/ST2/02861 (Poland); the National Priorities Research Program by Qatar National Research Fund; the Ministry of Science and Higher Education, project no. 0723-2020-0041 (Russia); the Programa Estatal de Fomento de la Investigaci{\'o}n Cient{\'i}fica y T{\'e}cnica de Excelencia Mar\'{\i}a de Maeztu, grant MDM-2015-0509 and the Programa Severo Ochoa del Principado de Asturias; the Thalis and Aristeia programs cofinanced by EU-ESF and the Greek NSRF; the Rachadapisek Sompot Fund for Postdoctoral Fellowship, Chulalongkorn University and the Chulalongkorn Academic into Its 2nd Century Project Advancement Project (Thailand); the Kavli Foundation; the Nvidia Corporation; the SuperMicro Corporation; the Welch Foundation, contract C-1845; and the Weston Havens Foundation (USA).
\end{acknowledgments}

\bibliography{auto_generated}
\cleardoublepage \appendix\section{The CMS Collaboration \label{app:collab}}\begin{sloppypar}\hyphenpenalty=5000\widowpenalty=500\clubpenalty=5000\vskip\cmsinstskip
\textbf{Yerevan Physics Institute, Yerevan, Armenia}\\*[0pt]
A.M.~Sirunyan$^{\textrm{\dag}}$, A.~Tumasyan
\vskip\cmsinstskip
\textbf{Institut f\"{u}r Hochenergiephysik, Wien, Austria}\\*[0pt]
W.~Adam, J.W.~Andrejkovic, T.~Bergauer, S.~Chatterjee, M.~Dragicevic, A.~Escalante~Del~Valle, R.~Fr\"{u}hwirth\cmsAuthorMark{1}, M.~Jeitler\cmsAuthorMark{1}, N.~Krammer, L.~Lechner, D.~Liko, I.~Mikulec, F.M.~Pitters, J.~Schieck\cmsAuthorMark{1}, R.~Sch\"{o}fbeck, M.~Spanring, S.~Templ, W.~Waltenberger, C.-E.~Wulz\cmsAuthorMark{1}
% ...
\vskip\cmsinstskip
\dag: Deceased\\
1:  Also at Vienna University of Technology, Vienna, Austria\\
% ...
\end{sloppypar}
\end{document}